\documentclass[final,times,3p,12pt,authoryear]{elsarticle}

\makeatletter
\def\ps@pprintTitle{%
    \let\@oddhead\@empty
    \let\@evenhead\@empty
    \def\@oddfoot{\footnotesize\itshape
         {} \hfill }%
    \let\@evenfoot\@oddfoot
    }
\makeatother

\usepackage{amsfonts,amssymb,amsmath,amsthm,autobreak,cancel,
    dsfont,mathtools,mathrsfs,siunitx,upgreek}

\usepackage{booktabs,longtable,dcolumn,makecell,
    multicol,multirow,tabularx,xltabular,rotating}

\usepackage{lmodern}                    

\usepackage[activate={true,nocompatibility}]{microtype}
\usepackage[linesnumbered,ruled,vlined]{algorithm2e}         
\usepackage[]{algpseudocode}
\usepackage[title,toc,titletoc]{appendix}
\usepackage{autobreak}
\usepackage[english]{babel}             
\usepackage{caption}                    
\usepackage{comment}                    
\usepackage[mathlines]{lineno}
\usepackage{enumitem}                   
\usepackage{float}                      
\usepackage{forest}
\usepackage[T1]{fontenc}                
\usepackage[bottom]{footmisc}           
\usepackage{graphicx,wrapfig}           
\usepackage[hidelinks,colorlinks=true,urlcolor=blue,linkcolor=black,citecolor=blue]{hyperref}
\usepackage{lscape}                     
\usepackage{listings,minted}            
\usepackage{csquotes}                   
\usepackage{marvosym}	                
\usepackage[version=3]{mhchem}
\usepackage{parskip}                    
\usepackage{subcaption}                 
\usepackage[dvipsnames]{xcolor}

\numberwithin{equation}{section}
\allowdisplaybreaks[1]
\DeclareMathOperator{\R}{\mathrm{R}}

\DeclareMathOperator{\T}{{\top}}
\newcommand{\vect}[1]{\mathbf{#1}}

\newcommand{\mat}[1]{\mathbf{#1}}
\DeclareMathOperator{\tr}{tr}
\DeclareMathOperator{\sym}{sym}
\DeclareMathOperator{\skw}{skw}

\DeclareMathOperator{\divg}{\nabla_{\mathit{x}} \cdot}
\DeclareMathOperator{\Divg}{\nabla \cdot}
\DeclareMathOperator{\grad}{\nabla_{\mathit{x}}}
\DeclareMathOperator{\Grad}{\nabla}
\DeclareMathOperator{\curl}{\nabla_{\mathit{x}} \times}
\DeclareMathOperator{\Curl}{\nabla \times}

\makeatletter
\DeclareRobustCommand\bigop[2][1]{%
  \mathop{\vphantom{\sum}\mathpalette\bigop@{{#1}{#2}}}\slimits@
}
\newcommand{\bigop@}[2]{\bigop@@#1#2}
\newcommand{\bigop@@}[3]{%
  \vcenter{%
    \sbox\z@{$#1\sum$}%
    \hbox{\resizebox{\ifx#1\displaystyle#2\fi\dimexpr\ht\z@+\dp\z@}{!}{$\m@th#3$}}%
  }%
}
\makeatother
\newcommand{\bigA}{\DOTSB\bigop[1]{\mathsf{A}}}

\makeatletter
\renewcommand*\env@matrix[1][*\c@MaxMatrixCols c]{%
  \hskip -\arraycolsep
  \let\@ifnextchar\new@ifnextchar
  \array{#1}}
\makeatother

\newdefinition{remark}{Remark}

\SetKwFor{While}{while}{}{end while}%
\SetArgSty{textnormal}

\SetCommentSty{mycommfont}

\begin{document}

\begin{frontmatter}

\title{A transient nonlinear finite element framework and implementation of coupled electro-chemo-mechanics of polyelectrolyte hydrogels}

\author[1]{Bibekananda Datta}       
\corref{corr-auth}
\ead{bdatta1@jhu.edu}
\cortext[corr-auth]{Corresponding author. The author has moved to Los Alamos National Laboratory.}

\author[2]{Brandon K. Zimmerman}

\author[1]{Thao D. Nguyen}

\affiliation[1]{organization={Department of Mechanical Engineering, Johns Hopkins University},
        city={Baltimore},
        state={MD 21218},
        country={USA}}
           
\affiliation[2]{organization={Lawrence Livermore National Laboratory}, 
        city={Livermore},
        state={CA 94550},
        country={USA}}

\begin{abstract}
Polyelectrolyte  (PE) hydrogels exhibit complex behavior characterized by large mechanical deformations, nonlinear stress response, solvent transport, and ion diffusion. The interplay between these mechanisms can lead to unexpected swelling dynamics, deformation patterns, and stress response.  As such, advanced computational tools are needed for the efficient design of PE hydrogel-based devices, such as actuators and sensors for soft robotics, microfluidic valves, and drug delivery systems. In this work, we develop a numerical framework to simulate the coupled electro-chemo-mechanical behavior of PE hydrogels using finite element analysis. Applying this framework, an electro-chemo-mechanical model for PE hydrogels in a dilute ionic solution is implemented as a user element (UEL) subroutine in Abaqus/Standard. The model and UEL implementation are validated by comparing to experiments in the literature for transient free-swelling of a DMAEA gel in a solution of varying ionic strengths, then applied to study the consolidation behavior under confined compression and the transient bending behavior of a hydrogel bilayer. The simulations show that the ionic strength of the external solution, fixed charge density, and Flory-Huggins parameter play significant roles in the magnitude of the transient swelling and consolidation behavior.  
\end{abstract}


\end{frontmatter}

\section{Introduction}      
\label{sec:introduction}

\begin{figure}[ht]
\begin{center}
    \includegraphics[width=0.5\textwidth, trim={8.5cm 6cm 8.5cm 6cm}, clip] {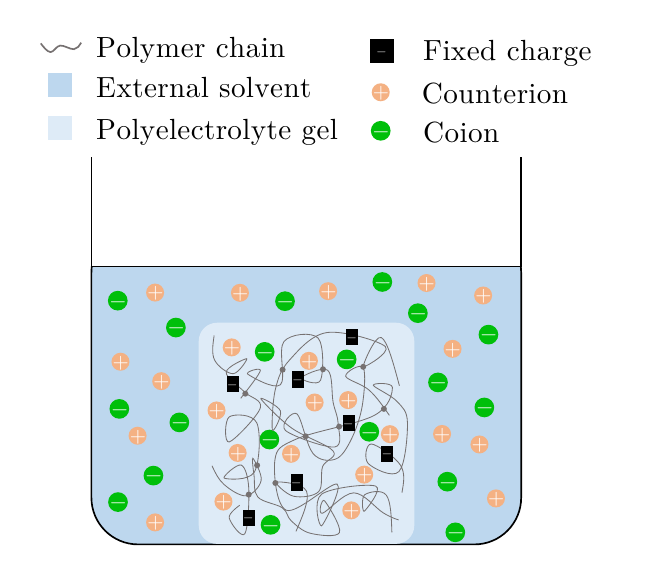}
    \caption{Schematic of a swollen anionic polyelectrolyte hydrogel immersed in a solvent bath with coions and counterions.}
    \label{fig:pe-gel-schematic}
\end{center}
\end{figure}

Polyelectrolyte (PE) hydrogels are crosslinked elastomeric networks in which the functional side groups of the polymer chains contain ionizable fixed charges. Based on the type of fixed charges, positive or negative, PE gels can be categorized as either cationic or anionic. As shown in Figure \ref{fig:pe-gel-schematic}, coions are the free-moving ions that have the same type of charge as the polymer fixed charge, whereas counterions are the ions that have the opposite type of charge. In comparison to non-ionic gels, PE gels show a higher degree of swellability because of their unique fixed charge characteristics. The ionic strength and pH of the external chemical environment can also control the extent and kinetics of swelling and deswelling of PE gels \citep{katchalskyPolyelectrolyteGelsSalt1955,siegelPHdependentEquilibriumSwelling1988,rydzewskiSwellingShrinkingPolyelectrolyte1990,brannon-peppasEquilibriumSwellingBehavior1991}. These diverse stimuli-responsive behavior makes PE hydrogels attractive for a wide variety of engineering applications, including drug deliver systems \citep{hoareHydrogelsDrugDelivery2008,liDesigningHydrogelsControlled2016}, soft sensors and actuators \citep{beebeFunctionalHydrogelStructures2000,gerlachChemicalPHSensors2005,yewAnalysisPHElectrically2007,panPolyelectrolyteHydrogelVersatile2022,sunRecentAdvancesFlexible2022}, wearable electronics \citep{keplingerStretchableTransparentIonic2013,yangIonicCable2015,kimHighlyStretchableTransparent2016}, and tissue scaffolds \citep{kwonNegativelyChargedPolyelectrolyte2006,wuChitosanbasedPolyelectrolyteComplex2012,bendersExtracellularMatrixScaffolds2013}.

Early continuum modeling efforts for polyelectrolyte gels include multiphasic mixture theories developed for soft biological tissues, such as cartilage (e.g., \cite{laiTriphasicTheorySwelling1991,ateshianTheoryReactiveMixtures2007}). In these models, mass and momentum balances are formulated separately for each species in the mixture. More recent efforts have applied a chemo-mechanical approach, where a hydrogel is treated as a single-phase continuum through which water and other species are transported. The chemo-mechanical approach can be traced to the pioneering work of \cite{floryStatisticalMechanicsCrosslinked1943a}, who developed a theory for the equilibrium swelling of hydrogels by accounting for the elastic energy of the polymer network and the mixing entropy between polymers and solvent. \cite{hongTheoryCoupledDiffusion2008} generalized this approach by developing a thermodynamically consistent framework coupling the finite deformation elastic behavior of the network with the transient effects of transport, where the solvent flux is driven by gradients in chemical potential. Soon after, \cite{hongLargeDeformationElectrochemistry2010} extended their theory to include ionic species within a dielectric medium to analyze the equilibrium electro-chemo-mechanical response of polyelectrolyte gels. The approach has been applied most commonly to model the equilibrium swelling response of pH-responsive anionic and cationic gels
\citep{marcombeTheoryConstrainedSwelling2010,liModelIdealElastomeric2014,drozdovSwellingPHresponsiveCationic2015,drozdovModelingEffectsPH2015} and salt-induced volume collapse of PE gel \citep{yuSaltinducedSwellingVolume2017}. 

In recent years, there has been growing interest in further extending the coupled electro-chemo-mechanical theory to analyze transient behaviors of PE gels for more advanced applications.  \cite{zhangKineticsPolyelectrolyteGels2020} proposed a non-equilibrium thermodynamic model for PE gels, which included cross-diffusion of the solvent and ions based on the Stefan-Maxwell approach \citep{maxwellIVDynamicalTheory1867}. Following a similar approach, \cite{narayanCoupledElectrochemomechanicalTheory2022} studied stimuli-responsive behaviors of PE gels for soft actuator-related applications, and \cite{celoraKineticModelPolyelectrolyte2022} studied the interfacial behaviors of PE gels undergoing a volume phase transition. Recently, we developed a nonequilibrium electro-chemo-mechanical model for copolymerized DNA hydrogels, which described the growth and remodeling of polymer chains driven via dynamic polymerization of charged DNA hairpins into DNA crosslinks \citep{zimmermanReactiveElectrochemomechanicalTheory2024}. While these works include numerical implementations for finite element analysis and illustrative numerical examples, a generalized finite element framework for the fully coupled electro-chemo-mechanical theory for PE gels, similar to those available for non-ionic gels \citep{zhangFiniteElementMethod2009, chesterFiniteElementImplementation2015, bouklasNonlinearTransientFinite2015}, has not been presented. In our previous study, we implemented a coupled electro-chemo-mechanical model for copolymerized DNA hydrogels in the open-source finite element software package, FEBio\footnote{Available as \textsf{Reactive Hydrogel} material plug-in for download on \url{https://febio.org/plugins/}.}, by leveraging the equivalence between governing equations of reactive multiphasic mixture theory, which FEBio \citep{maasFEBioFiniteElements2012,ateshianMultiphasicFiniteElement2013,ateshianComputationalModelingChemical2014} is based around, and those derived by \cite{zimmermanReactiveElectrochemomechanicalTheory2024} for reactive electro-chemo-mechanics.

In this work, we present a generalized finite element framework for a fully coupled electro-chemo-mechanical theory for PE gels described in Section \ref{sec:theory-summary}. The theory is similar to that developed by \cite{narayanCoupledElectrochemomechanicalTheory2022}, except that we treat the PE gel as an electroneutral medium instead of a dielectric medium, following our earlier work \citep{zimmermanReactiveElectrochemomechanicalTheory2024}. This is a reasonable simplifying assumption for structural applications that do not involve electroactuation \citep{tanakaCollapseGelsElectric1982,zimmermanDirectOsmoticPressure2021,liSpontaneousRapidElectroactuated2022,stewartElectrochemomechanicalTheoryHydrogel2023}. 
Treating the PE gel as an electroneutral medium allows the electric potential to be computed by enforcing the electroneutrality condition rather than by solving Maxwell's equation  \citep{zhangKineticsPolyelectrolyteGels2020, narayanCoupledElectrochemomechanicalTheory2022,celoraKineticModelPolyelectrolyte2022}. In addition, we adopt a hydrated reference state to describe the as-manufactured conditions of the hydrogel. The material parameters, swelling volume change, and deformation are defined for the hydrated reference state to facilitate model calibration and validation.  

A semi-discrete Galerkin method is presented in Section \ref{sec:finite-element} for solving the mechanical equilibrium equation and the transient transport equations for solvent and ionic species using a monolithic scheme. The constitutive model for PE gels involves a nonlinear relationship between chemical potential and concentration. These equations, together with the electroneutrality condition, are solved locally at the integration points. In Section \ref{sec:constitutive-model}, we specify constitutive models for the free energy density of a PE gel in a dilute ion solution and for solvent and ion diffusion. We adopt a quasi-incompressible approach to describe the mechanical behavior of the polymer network, which employs a large bulk modulus to penalize volumetric deformations and approximate incompressibility. The model is implemented as a user element (UEL) subroutine in Abaqus/Standard \citep{dassaultsystemesSIMULIAUserAssistance2023a}, and applied to examples of transient free swelling, consolidation under confined compression, and transient shape change of a hydrogel bilayer\footnote{The source code for the UEL subroutine and user instructions can be downloaded from \url{https://github.com/bibekanandadatta/Abaqus-UEL-Polyelectrolyte_Gel} or \url{https://github.com/NguyenLabJHU/Abaqus-UEL-Polyelectrolyte_Gel}.}. To mitigate volumetric locking due to incompressibility, we utilize the F-bar element formulation proposed by \cite{netoDesignSimpleLow1996}. The free swelling simulations are validated using experimental results from the literature and are further used to study the convergence behavior of the algorithm. The finite element framework provides an efficient and robust computational tool for the design and analysis of polyelectrolyte hydrogel structures.

\section{Summary of coupled electro-chemo-mechanical theory}    
\label{sec:theory-summary}

In this section, we briefly summarize a coupled electro-chemo-mechanical theory for PE hydrogels. The detailed derivation of the coupled continuum theory, including thermodynamic considerations, is given in Appendix \ref{sec:appendix-theory}. Following our earlier works \citep{yoonFunctionalStimuliResponsive2014, zimmermanReactiveElectrochemomechanicalTheory2024}, we consider the hydrated reference state for the hydrogel, which is more representative of the as-manufactured hydrogel than the dry reference state. The theory also treats the PE hydrogel as an electroneutral medium, which simplifies the calculation of the electric potential in the implementation.

\subsection{Multiplicative deformation kinematics}    
\label{sec:kinematics}

\begin{figure}[ht]
\begin{center}
    \includegraphics[width=0.8\textwidth, trim={5cm 5cm 5cm 5cm}, clip] {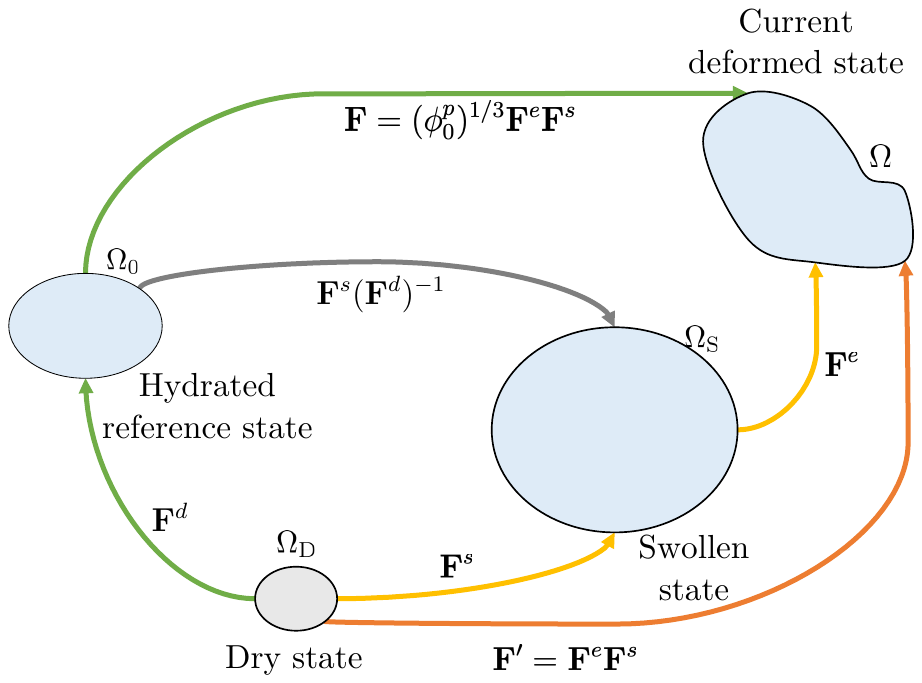}
    \caption{Multiplicative decomposition of elastic and swelling deformation for a pre-swollen polyelectrolyte gel.}
    \label{fig:kinematics}
\end{center}
\end{figure}

Hydrogels are typically synthesized by crosslinking a polymer solution, which produces a partially swollen gel composed of solvent and polymers. We introduce  $\Omega_0(\vect{X}) \in \mathbb{R}^3$ with an initial polymer volume fraction, $\phi^p_0$, as the hydrated reference state of an as-prepared PE gel, as shown in Figure \ref{fig:kinematics}. The material point coordinate, $\vect{X}$, is mapped to the current deformed state, $\Omega(\vect{x}) \in \mathbb{R}^{3}$, with spatial point coordinate, $\vect{x}$, via smooth one-to-one mapping $\vect{x} = \varphi(\vect{X},t)$. As is standard, the deformation gradient, $\vect{F}$, is defined as,
\begin{equation} \label{eq:deformation}
    \vect{F} = \frac{\partial \vect{x}}{\partial \vect{X}}.
\end{equation}
Let $\vect{F}^d = \lambda^d \mathds{1}$ be the swelling deformation from the dry state to the hydrated reference state, where $\lambda^d$ is the isotropic swelling stretch between these two states. Thus, the volume change from the dry state to the hydrated reference state is,
\begin{equation} \label{eq:pre-swollen-vol}
    J^d = \det (\vect{F}^d) = \left( \lambda^d \right)^3 = \frac{dV}{dV_{d}} = \frac{dV}{dV^p} = \frac{1}{\phi^p_0},
\end{equation}
where $V$ and $V_{d}$ are the hydrated reference state volume and the dry state volume of the gel, respectively. Since the mass and volume of the polymer network do not change from the dry state to the hydrated reference state, $ V^p = V_{d} $ represents the total volume of the polymers at the hydrated reference state. Following \cite{yoonFunctionalStimuliResponsive2014} and \cite{zimmermanReactiveElectrochemomechanicalTheory2024}, we can now decompose the deformation gradient, $\vect{F}$, as follows,
\begin{equation} \label{eq:def-grad}
    \vect{F} 
    = \vect{F}'  (\vect{F}^d)^{-1} 
    = \left( \phi^p_0 \right)^{1/3} \vect{F}^e \vect{F}^s,
\end{equation}
where $\vect{F}' = \vect{F}^e \vect{F}^s$ represents the total deformation of the gel from the dry state to the current deformed state. Here, $\vect{F}^s$ represents the swelling deformation gradient from the dry state to the swollen state, and $\vect{F}^e$ represents the elastic deformation gradient between the swollen state and the current state. We assume that the swelling deformation is isotropic, $\vect{F}^s = \lambda^s \mathds{1}$, where $\lambda^s$ is the swelling stretch from the dry state to the swollen state. It follows from Eq. \eqref{eq:def-grad} that the volumetric deformation, $J = \det(\vect{F})$, represents the volume change from the hydrated reference state to the current deformed state, and can be expressed as,
\begin{equation} \label{eq:total-volume-detF}
    J = \phi^p_0 J^e J^s,
\end{equation}
where $J^e = \det(\vect{F}^e)$ and $J^s = \det(\vect{F}^s)$ represent the volume change due to elastic deformation and swelling, respectively. We further assume the swelling volume change in general occurs from the uptake of solvent and ions such that,
\begin{equation} \label{eq:swelling-vol}
    J^s = 
    \frac{1}{\phi^p_0} \left( C^w \mathcal{V}^w 
    + C^p \mathcal{V}^p 
    + \sum_{\beta_k} C^{\beta_k} \mathcal{V}^{\beta_k} \right),
\end{equation}
where $C^w$, $C^p$, and $C^{\beta_k}$ are the referential concentration of the solvent, polymer, and the $ k$-th ion, $\beta_k$, respectively, which is measured as the moles per reference volume. Additionally, $\mathcal{V}^w$, $\mathcal{V}^w$, and $\mathcal{V}^{\beta_k}$ are the molar volumes of the solvent, polymer, and the $k-$th ion, $\beta_k$, respectively.

\subsection{Governing equations, initial and boundary conditions} 
\label{sec:governing-eqs}

For a time interval $t \in [0, \tau]$, physical processes within the PE gel domain, $\Omega_0$, are described by a set of governing partial differential equations with appropriate initial conditions and boundary conditions. The boundary of the PE gel in the hydrated reference state, $\partial \Omega_0$, can be divided into non-overlapping portions with essential boundary conditions for the displacement, $\vect{g}$, on $\Gamma_{\vect{g}}$, solvent chemical potential, $h$, on $\Gamma_{h}$, and ion electrochemical potential, $q^{\beta_k}$, on $\Gamma_{q^{\beta_k}}$, and natural boundary conditions for the first Piola-Kirchhoff traction, $\vect{T}$, on $\Gamma_{\vect{T}}$, solvent molar flux, $I^w$, on $\Gamma_{I^w}$, and ion molar flux, $I^{\beta_k}$, on $\Gamma_{I^{\beta_k}}$.

Ignoring the inertial effect, the governing equation for local linear momentum balance, as well as the corresponding boundary conditions, are given by,
\begin{equation} \label{eq:momentum-balance}
\begin{aligned}
    \Divg \vect{P} + \rho_R \vect{B}  & = 0 && \mathrm{in} \  \Omega_0,  \\
    \vect{u} & = \vect{g} && \mathrm{on} \ \Gamma_{\vect{g}}, \\
    \vect{P} \cdot \vect{N} & = \vect{T} && \mathrm{on} \  \Gamma_{\vect{T}},
\end{aligned}
\end{equation}
where $\vect{P}$ is the first Piola-Kirchhoff stress tensor, $\rho_{\R}$ is the referential mass density, $\vect{B}$ is the body force per unit of mass in the reference configuration, and $\vect{N}$ is the normal to the surface $\Gamma_{\vect{T}}$. Using the relation $\vect{P} = \vect{FS}$, where $\vect{S}$ is the second Piola-Kirchhoff stress tensor, the local form of the linear momentum balance Eq. \eqref{eq:momentum-balance}$_1$ can be expressed as,
\begin{equation}
    \Divg (\vect{FS}) +  \rho_R \vect{B} = 0 \quad \mathrm{in} \ \Omega_0.
\end{equation}
The governing equation for the solvent mass balance, the initial condition, and the boundary conditions are given by,
\begin{equation} \label{eq:solvent-mass-balance}
\begin{aligned}
    \dot{C}^{w} & = - \Divg \vect{J}^{w}      && \mathrm{in} \ \Omega_0, \\
    \mu^w(\vect{X}, t=0)  & =  \mu_0^{w}        && \mathrm{in} \ \Omega_0, \\
    \mu^w & = h                                 && \mathrm{on} \ \Gamma_{h} , \\
    - \vect{J}^{w} \cdot \vect{N} &= I^w        && \mathrm{on} \ \Gamma_{I^w},
\end{aligned}
\end{equation}
where $\vect{J}^w$ is the referential molar flux of the solvent, measured as moles per unit area in the reference configuration, per unit time, and $\vect{N}$ is the normal to the surface $\Gamma_{{I}^{w}}$.

Finally, the governing equation for the mass balance of any ion, $\beta_k$, the initial condition, and the boundary conditions are given by,
\begin{equation} \label{eq:ion-mass-balance}
\begin{aligned}
    \dot{C}^{\beta_k} & = - \Divg \vect{J}^{\beta_k}        && \mathrm{in} \  \Omega_0, \\
    \omega^{\beta_k}(\vect{X}, t=0)  & = \omega_0^{\beta_k} && \mathrm{in} \ \Omega_0, \\
    \omega^{\beta_k} & = q^{\beta_k}                        && \mathrm{on} \ \Gamma_{q^{\beta_k}} , \\
    - \vect{J}^{\beta_k} \cdot \vect{N} & = I^{\beta_k}     && \mathrm{on} \ \Gamma_{I^{\beta_k}},
\end{aligned}
\end{equation}
where $\vect{J}^{\beta_k}$ is the referential molar flux of the same ion across the surface $\Gamma_{I^{\beta_k}}$ with an outward normal $\vect{N}$. In numerical implementation, the mass balance equations need to be written for each ionic species present in the PE gel and external solution.

\subsection{Electroneutrality}    
\label{sec:electroneutrality}

The flow of ions in and out of the PE gel results in the accumulation of charges, which is referred to as the referential charge density, $Q$, and can be written as the sum of all charges as,
\begin{equation} \label{eq:charge-density-def}
    Q = F  \left( z^{\mathrm{fix}} C^{\mathrm{fix}} + \sum_{\beta_k} z^{\beta_k} C^{\beta_k}  \right).
\end{equation}
where $F$ is Faraday's constant, $C^{\mathrm{fix}}$ is the referential concentration of polymer fixed charge, $z^{\mathrm{fix}}$ and $z^{\beta_k}$ are the charge numbers of polymer chains and ions, $\beta_k$, respectively. The charge density has a unit of charge per unit volume. The electroneutrality assumption states that the mobile charges in a medium larger than the Debye length scale will migrate in such a way that the net charge remains zero \citep{hongLargeDeformationElectrochemistry2010,narayanCoupledElectrochemomechanicalTheory2022}, which can be written as,
\begin{equation} \label{eq:electroneutrality-def}
    z^{\mathrm{fix}} C^{\mathrm{fix}} + \sum_{\beta_k} z^{\beta_k} C^{\beta_k} = 0.
\end{equation}
At least two ionic species of opposite charges are needed to maintain electroneutrality in the external solution and in the gel. The difference in the ion concentration, due to the presence of the polymer fixed charges, between the gel and the external solution gives rise to an electric potential, $\psi$, within the gel, known as the Donnan potential.

\subsection{Constitutive restrictions}   
\label{sec:constitutive-def}

Based on the thermodynamically-consistent coupled electro-chemo-mechanical theory for PE gels (see Appendix \ref{sec:appendix-theory} for details), we have the following constitutive relations for the Cauchy stress, $\boldsymbol{\upsigma}$, solvent chemical potential, $\mu^w$, and ion electrochemical potential, $\omega^{\beta_k}$,
\begin{equation} \label{eq:constitutive-definitions-final}
\begin{aligned}
    \boldsymbol{\upsigma} & = J^{-1} \left[ \vect{F}^e \left(2 \frac{\partial \Psi}{\partial \vect{C}^e} \right) {\vect{F}^e}^{\top}  \right], \\
    \mu^{w} & = \frac{\partial \Psi}{\partial C^{w}} + p \mathcal{V}^{w}, \\
    \omega^{\beta_k} & =  \frac{\partial \Psi}{\partial C^{\beta_k}} + p \mathcal{V}^{\beta_k} + F \psi z^{\beta_k},
\end{aligned}
\end{equation}
where 
\begin{equation}    \label{eq:mean-pressure-defn}
    p = \frac{-1}{3} J^e \tr(\boldsymbol{\upsigma})
\end{equation}
represents the mean pressure. The standard pull-back operations can be applied on the Cauchy stress, $\boldsymbol{\upsigma}$, to obtain the first Piola-Kirchhoff stress, $\vect{P} = J \boldsymbol{\upsigma} \vect{F}^{-\top}$, and the second Piola-Kirchhoff stress, $ \vect{S} = J \vect{F}^{-1} \vect{\boldsymbol{\upsigma}} \vect{F}^{-\top}$. 

\begin{remark}
    The thermodynamic restrictions in Eq. \eqref{eq:constitutive-definitions-final} are identical to those in \cite{narayanCoupledElectrochemomechanicalTheory2022} except that we do not have any thermodynamic restrictions for the electric field. This is a consequence of our treatment of the PE gel as an electroneutral medium in the global energy balance statement Eq. \eqref{eq:energy-balance-global}.
\end{remark}

To describe the species transport, we can expand the generic form of the flux law we introduced in Eq. \eqref{eq:onsager-principle} to describe the kinetics of diffusion for solvent and ions,
\begin{equation}    \label{eq:generic-transport-law}
\begin{aligned}
    \vect{J}^w & =  \left( \sum_{\gamma \in \{ w, \beta_k \} } L^{w \gamma} \Grad f^{\gamma} \right) \vect{C}^{-1}, \\
    \vect{J}^{\beta_k} &=  \left( \sum_{\gamma \in \{ w, \beta_k \} } L^{\beta_k \gamma} \Grad f^{\gamma} \right) \vect{C}^{-1},
\end{aligned}
\end{equation}
where $L^{w \gamma}$ and $L^{\beta_k \gamma}$ are known as the Onsager matrix of transport coefficients, consisting of phenomenological material parameters that depend on the specific constitutive law for diffusion kinetics. As the thermodynamic force approaches zero, the flow slowly vanishes, and the system reaches equilibrium. The generic thermodynamic force, $\Grad f^{\gamma}$, represents the gradient of solvent chemical potential, $\Grad \mu^w$, or ion electrochemical potential, $\Grad \omega^{\beta_k}$. We will specify the particular forms of $\vect{J}^w$ and $\vect{J}^{\beta_k}$ in Section \ref{sec:constitutive-model}.

\section{Coupled nonlinear finite element method}   
\label{sec:finite-element}

We now present a second Piola-Kirchhoff stress-based Total Lagrangian finite element framework based on our transient coupled nonlinear theory for PE gels. We start with the weak forms of our governing equations, followed by the discretization and constitutive update procedure.

\subsection{Weak form of the governing equations}   
\label{sec:weak-forms}

Let $\vect{u} $ be the trial displacement solution, $\mu^w$ be the trial solution for solvent chemical potential, and $\omega^{\beta_k}$ be the trial solution for electrochemical potential of ionic species that satisfy the essential boundary conditions,  $\vect{u} = \vect{g}$,  $\mu^w = h$, and $\omega^{\beta_k} = q^{\beta_k}$, prescribed on the boundaries $\Gamma_{\vect{g}}$, $\Gamma_{h}$, and $\Gamma_{q^{\beta_k}}$, respectively. We further assume $\vect{W}$, $\Theta$, and $\eta_k$ are the weight functions corresponding to the assumed trial solutions that vanish on the same boundaries. Thus, the weak forms of the corresponding governing equations \eqref{eq:momentum-balance}, \eqref{eq:solvent-mass-balance}, and \eqref{eq:ion-mass-balance}, can be written as,
\begin{equation}    \label{eq:global-weak-forms}
\begin{gathered}
    - \int\limits_{\Omega_0} \vect{FS} : \Grad \vect{W} \ dV 
    + \int\limits_{\Omega_0} \rho_{\R} \vect{B} \cdot \vect{W} \ dV 
    + \int\limits_{\Gamma_{\vect{T}}} \vect{T} \cdot \vect{W} \ dS = 0, \\
    - \int\limits_{\Omega_0} \dot{C}^w \ \Theta \ dV 
    +  \int\limits_{\Omega_0} \vect{J}^w \cdot \Grad \Theta \ dV 
    + \int\limits_{\Gamma_{I^w}} I^w \ \Theta \ dS = 0, \\
    - \int\limits_{\Omega_0} \dot{C}^{\beta_k} \ \eta_k \ dV  
    + \int\limits_{\Omega_0} \vect{J}^{\beta_k} \cdot \Grad \eta_k \ dV 
    + \int\limits_{\Gamma_{I^{\beta_k}}} I^{\beta_k}\ \eta_k \ dS = 0.
\end{gathered}
\end{equation}
These standard weak forms are obtained after multiplying the governing equations by their corresponding weight functions, integrating them over the domain, and applying integration by parts and the divergence theorem.

\subsection{Finite element discretization}  
\label{sec:discretization}

Consistent with the essential boundary conditions prescribed in the strong form Eqs. \eqref {eq:momentum-balance}, \eqref{eq:solvent-mass-balance}, and \eqref{eq:ion-mass-balance}, we chose displacement, $\vect{u}$, solvent chemical potential, $\mu^w$, and electrochemical potential for ions, $\omega^{\beta_k}$, as the nodal degrees of freedom. In what follows, we discretize the domain and its boundaries using finite elements such that $\Omega_0 = \bigcup_{e=1}^{n_{\mathrm{el}}} \ \Omega_0^e$ and $\partial \Omega_0 = \bigcup_{e=1}^{n_{\mathrm{el}}} \ \partial \Omega_0^e$. Thus, by using the standard finite element procedure \citep{zienkiewiczFiniteElementMethod2014,hughesFiniteElementMethod2000}, for each element, $e \in  \{1, 2, 3, \cdots \ \cdots, n_{\mathrm{el}} \}$, the nodal degrees of freedom, $\vect{d}$, can be presented as,
\begin{equation}    \label{eq:nodal-vars}
    \vect{d} = \{\vect{u}_a, \ \mu^w_a, \ \omega^{\beta_k}_a \}^{\T}_{n_{\mathrm{en}} \times 1},
\end{equation}
where $a \in \{1, 2, 3, \cdots \ \cdots, n_{\mathrm{en}} \}$ represents the nodes in an element.
We now employ the standard Galerkin procedure to approximate the degrees of freedom, $\vect{u}$, $\mu^w$, and $\omega^{\beta_k}$, and their corresponding weight functions within each element as follows,
\begin{equation}    \label{eq:galerkin-approx}
\begin{aligned}
    u^i(\vect{X})             &= \sum_a  N^{\vect{u}}_a(\vect{X}) \ u_{a}^i 
    &\text{and} \quad
    W^i(\vect{X})             &= \sum_a  N^{\vect{u}}_a(\vect{X}) \ W_{a}^i, \\
    \mu^w (\vect{X})          &= \sum_a N^{\mu}_a(\vect{X}) \ \mu^w_a 
    &\text{and} \quad
    \Theta(\vect{X})          &= \sum_a N^{\mu}_a(\vect{X}) \ \Theta_a, \\
    \omega^{\beta_k}(\vect{X}) &= \sum_a N^{\omega}_a(\vect{X}) \ \omega^{\beta_k}_a 
    &\text{and} \quad
    \eta_k(\vect{X})          &= \sum_a N^{\omega}_a(\vect{X}) \ \eta_{ka}.
\end{aligned}
\end{equation}
Here, $N^{\vect{u}}_a$,  $N^{\mu}_a$, and $N^{\omega}_a$ are the interpolation functions correspond to the nodal values of $\vect{u}_a$, $\mu^w_a$, and $\omega^{\beta_k}_a$, respectively. The standard Galerkin weighted residual approach uses the same interpolation function for any nodal variable and its weight function pair. However, it does not restrict the use of different interpolation functions for other nodal fields and their weight functions in a multi-field finite element framework.

Furthermore, a first-order backward differencing scheme is applied to discretize the time derivative of referential concentrations at the current time $t + \Delta t$
\begin{equation}    \label{eq:euler-backward}
    \dot{C}^{w} = \frac{C^{w} - C^{w}_{t}}{\Delta t}
    \quad \text{and} \quad
    \dot{C}^{\beta_k} = \frac{C^{\beta_k} - C^{\beta_k}_{t}}{\Delta t},
\end{equation}
where $C^w$ and $C^{\beta_k}$ are the referential concentrations of the corresponding species (as indicated by the superscript) at time $t+ \Delta t$, and  $C^w_t$ and $C^{\beta_k}_t$ are the referential concentrations of the corresponding species (as indicated in the superscript) at the previous time $t$. By using the spatial and the temporal discretizations Eqs. \eqref {eq:galerkin-approx}-\eqref{eq:euler-backward}, we now write the element-level residuals in index notation as follows,
\begin{equation}
\begin{aligned} \label{eq:elem-residuals}
    r^{u_i}_{a}
    & = - \int\limits_{\Omega_0^e} \frac{\partial N^{\vect{u}}_a}{\partial X_I} F_{iJ} S_{JI} \ dV 
    + \int\limits_{\Omega_0^e} \rho_{\R} N^{\vect{u}}_a B_{i} \ dV
    + \int\limits_{\Gamma_{\vect{T}}^e} N^{\vect{u}}_a T_i \ dS, \\
    r^{\mu}_a
    & = - \int\limits_{\Omega_0^e} N^{\mu}_a \left( \frac{C^w - C^w_{t}}{\Delta t} \right) \ dV 
    + \int\limits_{\Omega_0^e} \frac{\partial N^{\mu}_a}{\partial X_I} J^w_I \ dV
    + \int\limits_{\Gamma_{I^w}^e} N^{\mu}_a {I^w}  \ dS, \\
    r^{\omega_k}_a
    & = - \int\limits_{\Omega_0^e} N^{\omega}_a \left( \frac{C^{\beta_k} - C^{\beta_k}_{t}}{\Delta t} \right) \ dV 
    + \int\limits_{\Omega_0^e} \frac{\partial N^{\omega}_a}{\partial X_I} J^{\beta_k}_I \ dV
    + \int\limits_{\Gamma_{I^{\beta_k}}^e} N^{\omega}_a {I^{\beta_k}} \ dS.
\end{aligned}
\end{equation}
The element residual vector, $\vect{r}_e$, can be constructed using its components as follows,
\begin{equation}
    \vect{r}_e = \{ r^{u_i}_{a}, \ r^{\mu}_a, \ r^{\omega_k}_a \}^{\T}_{n_{\mathrm{en}}\times1}.
\end{equation}
In what follows, we will employ the standard Newton-Raphson procedure to linearize the residuals and evaluate all the constitutive quantities and their derivatives at time $t+ \Delta t$ by adopting a backward Euler scheme to ensure the stability of the formulation.

\subsection{Constitutive update procedure}   
\label{sec:constitutive-update}

To solve for the coupled electro-chemo-mechanical problem posed by the system of element residuals Eq. \eqref {eq:elem-residuals}, we must perform constitutive evaluations at the current time $t+\Delta t$. Based on the constitutive definitions Eqs. \eqref {eq:constitutive-definitions-final} and \eqref{eq:generic-transport-law}, we assume the following functional forms for constitutive evaluation,
\begin{equation}    \label{eq:constitutive-functions}
\begin{aligned}
    \boldsymbol{\mathcal{F}} (\Grad \vect{d}; \vect{q})
    = \{ \vect{S} (\Grad \vect{d}; \vect{q}), \ \vect{J}^w(\Grad \vect{d}; \vect{q}), \ \vect{J}^{\beta_k} (\Grad \vect{d}; \vect{q}) \},
\end{aligned}
\end{equation}
where $\Grad \vect{d} = \{ \vect{F}, \ \Grad \mu^w, \ \Grad \omega^{\beta_k} \}$ represents the gradient of nodal degrees of freedom in the material coordinate and their computation procedure is standard, whereas $\vect{q}$ is the set of algorithmic internal variables at time, $t+\Delta t$, given as follows,
\begin{equation}    \label{eq:internal-vars}
    \vect{q} = \{ C^w, \ C^{\beta_k}, \ \psi \}^{\T}.
\end{equation}

Within each element, the constitutive expressions for solvent chemical potential, ion electrochemical potentials, and electroneutrality form a system of nonlinear equations collected in a single residual vector, $\boldsymbol{\mathcal{G}}$, as follows
\begin{equation}    \label{eq:internal-vars-system}
    \boldsymbol{\mathcal{G}}(\vect{d}, \Grad \vect{d};\vect{q}) 
    = \{ \mathcal{G}_1(\vect{d}, \Grad \vect{d};\vect{q}), 
    \ \mathcal{G}_2(\vect{d}, \Grad \vect{d};\vect{q}), 
    \ \cdots \ , 
    \ \mathcal{G}_{n+2}(\vect{d}, \Grad \vect{d};\vect{q}) \}^{\T},
\end{equation}
where $n$ represents the number of ionic species. To solve for the algorithmic internal variables, at each integration point, a local Newton-Raphson procedure is employed as follows until convergence is achieved,
\begin{equation}    \label{eq:local-newton-iteration}
    \vect{q}^{(j+1)} = \vect{q}^{(j)} + \Delta \vect{q}^{(j+1)},
    \quad \mathrm{where}, \
    \Delta \vect{q}^{(j+1)} = - \left( \frac{d \boldsymbol{\mathcal{G}}} {d \vect{q}} \right )^{-1} \boldsymbol{\mathcal{G}}^{(j)},
\end{equation}
where $(j)$ and $(j+1)$ are the consecutive iteration counters for Newton-Raphson procedure. Material-specific constitutive expressions for the system of local nonlinear equations are given in Eq. \eqref {eq:local-residual-system}, and the corresponding Jacobian matrix is given in Appendix \ref{sec:appendix-local-newton}. Since $\psi$ is being computed locally using the electroneutrality assumption as a constraint, it acts similarly to a Lagrange multiplier. Once the solution for $\vect{q}$ is obtained, we can compute the constitutive response of the PE gel from Eq. \eqref {eq:constitutive-functions}.

\subsection{Consistent linearization and element tangent stiffness matrix}  
\label{sec:element-tangents}

By assuming nonlinear constitutive relations for the PE gel Eq. \eqref {eq:constitutive-functions}, the element residual vector Eq. \eqref {eq:elem-residuals} can be linearized at the current time step, $t+ \Delta t$, to obtain the element tangent stiffness matrix, $\mat{K}_{ab}$, as 
\begin{equation}    \label{eq:elem-linearization}
    \mat{K}_{ab} = - \frac{d \vect{r}_a}{d \vect{d}_b},
\end{equation}
where the subscripts, $a,b$, represent the nodes in an element. Computation of the components of the element tangent stiffness matrix Eq. \eqref {eq:elem-linearization} is a standard, however, quite a lengthy procedure. Hence, to keep our presentation concise, we only list the final expressions of the components of the element tangent stiffness matrix below in index notation (see Appendix \ref{sec:appendix-fem} for the matrix-vector form) by omitting all the intermediate steps,
\begingroup
\allowdisplaybreaks
\begin{align*}      \label{eq:elem-tangents}
    & k_{ab}^{u_i u_k}
    && = - \frac{\partial r^{u_i}_{a}} {\partial u_b^k}
    && = \int\limits_{\Omega_0^e} \frac{\partial N^{\vect{u}}_a}{\partial X_J}  S_{JL} \delta_{ik} \frac{\partial N_b^{\vect{u}}}{\partial X_L} \ dV
    + \int\limits_{\Omega_0^e} \frac{\partial N^{\vect{u}}_a}{\partial X_J} F_{iI}  \left( 2 \frac{d S_{IJ}}{d C_{KL}} \right) F_{kK} \frac{\partial N_b^{\vect{u}}}{\partial X_L} \ dV, \\
    & k_{ab}^{u_i \mu}
    && = - \frac{\partial r^{u_i}_{a}} {\partial \mu^w_b}
    && = \int\limits_{\Omega_0^e}  \frac{\partial N^{\vect{u}}_a}{\partial X_I} \left( F_{iJ} \frac{d S_{IJ}}{d \mu^w} \right) N_b^{\mu} \ dV, \\
    & k_{ab}^{u_i \omega_k}
    && = - \frac{\partial r^{u_i}_{a}} {\partial \omega^{\beta_k}_b}
    && = \int\limits_{\Omega_0^e}  \frac{\partial N^{\vect{u}}_a}{\partial X_I} \left( F_{iJ} \frac{d S_{IJ}}{d \omega^{\beta_k}}  \right)  N_b^{\omega} \ dV, \\
    & k_{ab}^{\mu u_k}
    && = - \frac{\partial r^{\mu}_a} {\partial u_b^k}
    && =  \int\limits_{\Omega_0^e} N^{\mu}_a \left( \frac{1}{\Delta t} \frac{d C^w}{d F_{kL}} \right) \frac{\partial N_b^{\vect{u}}}{\partial X_L} \ dV
    -  \int\limits_{\Omega_0^e}  \frac{\partial N^{\mu}_a}{\partial X_I} \left( \frac{d J^w_I}{d F_{kL}} \right) \frac{\partial N_b^{\vect{u}}}{\partial X_L}  \ dV, \\
    & k_{ab}^{\mu \mu}
    && = - \frac{\partial r^{\mu}_a} {\partial \mu^w_b}
    && = \int\limits_{\Omega_0^e} N^{\mu}_a \left( \frac{1}{\Delta t} \frac{d C^w}{d \mu^w} \right)  N_b^{\mu} \ dV 
    - \int\limits_{\Omega_0^e} \frac{\partial N^{\mu}_a}{\partial X_I} \left( \frac{d J^w_I}{d \mu^w} \right) N_b^{\mu} \ dV \\
    & && && \quad -  \int\limits_{\Omega_0^e} \frac{\partial N^{\mu}_a}{\partial X_I} \left( \frac{d J^w_I}{d (\Grad \mu^w)_J} \right) \frac{\partial N_b^{\mu}}{\partial X_J} \ dV,
    \stepcounter{equation}\tag{\theequation} \\
    & k_{ab}^{\mu \omega_k}
    && = - \frac{\partial r^{\mu}_a} {\partial \omega^{\beta_k}_b}
    && = \int\limits_{\Omega_0^e} N^{\mu}_a \left( \frac{1}{\Delta t} \frac{d C^w}{d \omega^{\beta_k}} \right) N_b^{\omega} \ dV 
    - \int\limits_{\Omega_0^e} \frac{\partial N^{\mu}_a}{\partial X_I} \left( \frac{d J^w_I}{d \omega^{\beta_k}} \right) N_b^{\omega}  \ dV \\
    & && && \quad -  \int\limits_{\Omega_0^e} \frac{\partial N^{\mu}_a}{\partial X_I} \left( \frac{d J^w_I}{d (\Grad \omega^{\beta_k})_J} \right) \frac{\partial N_b^{\omega}}{\partial X_J} \ dV, \\
    & k_{ab}^{\omega_i u_k}
    && = - \frac{\partial r^{\omega_i}_a} {\partial u_b^k}
    && = \int\limits_{\Omega_0^e} N^{\omega}_a \left( \frac{1}{\Delta t} \frac{d C^{\beta_i}}{d F_{kL}} \right) \frac{\partial N_b^{\vect{u}}}{\partial X_L} \ dV
    -  \int\limits_{\Omega_0^e}  \frac{\partial N^{\omega}_a}{\partial X_I} 
    \left( \frac{d J^{\beta_i}_I}{d F_{kL}} \right) \frac{\partial N_b^{\vect{u}}}{\partial X_L}  \ dV, \\
    & k_{ab}^{\omega_i \mu}
    && = - \frac{\partial r^{\omega_i}_a} {\partial \mu^w_b}
    && = \int\limits_{\Omega_0^e} N^{\omega}_a \left( \frac{1}{\Delta t} \frac{d C^{\beta_i}}{d \mu^w} \right) N_b^{\mu} \ dV
    - \int\limits_{\Omega_0^e} \frac{\partial N^{\omega}_a}{\partial X_I} \left( \frac{d J^{\beta_i}_I}{d \mu^w} \right) N_b^{\mu} \ dV, \\
    & k_{ab}^{\omega_i \omega_k}
    && = - \frac{\partial r^{\omega_i}_a} {\partial \omega^{\beta_k}_b}
    && = \int\limits_{\Omega_0^e} N^{\omega}_a  \left( \frac{1}{\Delta t} \frac{d C^{\beta_i}}{d \omega^{\beta_k}} \right)  N_b^{\omega} \ dV 
    - \int\limits_{\Omega_0^e} \frac{\partial N^{\omega}_a}{\partial X_I} \left( \frac{d J^{\beta_i}_I}{d \omega^{\beta_k}} \right) N_b^{\omega} \ dV \\
    & && && \quad -  \int\limits_{\Omega_0^e} \frac{\partial N^{\omega}_a}{\partial X_I} \left( \frac{d J^{\beta_i}_I}{d (\Grad \omega^{\beta_k})_J} \right) \frac{\partial N_b^{\omega}}{\partial X_J} \ dV.
\end{align*}
\endgroup

Calculating the components of the element tangent stiffness matrix as given in Eq. \eqref {eq:elem-tangents} requires computing the derivatives of the internal variable with respect to the degrees of freedom or their gradients. By using the implicit function theorem for a multi-variable system, we can obtain the following form,
\begin{equation}    \label{eq:internal-vars-derivatives}
    \frac{d \vect{q}}{d (\bullet)} 
    = - \left( \frac{\partial \boldsymbol{\mathcal{G}}}{\partial \vect{q}} \right)^{-1} \frac{\partial \boldsymbol{\mathcal{G}}}{\partial (\bullet)}
\end{equation}
where $(\bullet)$ represents the degrees of freedom, $\vect{d}$, or their material gradients, $\Grad \vect{d}$. Subsequently, by applying the chain rule, consistent material tangent moduli can be computed as
\begin{equation}    \label{eq:material-tangent-moduli}
    \frac{d \boldsymbol{\mathcal{F}}}{d (\bullet)}
    = \frac{\partial \boldsymbol{\mathcal{F}}}{\partial (\bullet)}
    + \frac{\partial \boldsymbol{\mathcal{F}}}{d \vect{q}} \frac{d \vect{q}}{\partial (\bullet)},
\end{equation}
where $\frac{d \boldsymbol{\mathcal{F}}}{d (\bullet)}$, a second- or higher-order tensor, is the total derivative of the constitutive functions, $\boldsymbol{\mathcal{F}}$, with respect to the degrees of freedom, $\vect{d}$, or their gradients in material coordinates, $\Grad \vect{d}$. 

Once computed, the element tangent stiffness matrices and element residual vectors can be assembled for all the elements in the computational domain to obtain the following,
\begin{equation}    \label{eq:global-system}
    \bigA_{e=1}^{n_\mathrm{el}} \left[ \mat{K}_e \right] \{\Delta \vect{d} \} 
    = \bigA_{e=1}^{n_\mathrm{el}} \{ \vect{r}_e \},
\end{equation}
where $\bigA$ denotes the standard assembly operator over all the elements in the computational domain and $\Delta \vect{d}$ is the incremental nodal degrees of freedom. The global system of equations Eq. \eqref {eq:global-system} can be solved iteratively as,
\begin{equation}    \label{eq:global-iteration}
    \vect{d}^{(j+1)} = \vect{d}^{(j)} + \Delta \vect{d}^{(j+1)},
\end{equation}
where $(j)$ and $(j+1)$ denote the counters for two consecutive iterations for the global Newton-Raphson procedure.

We implemented the proposed finite element procedure for three-dimensional, axisymmetric, and plane strain first-order Lagrangian elements with reduced and full Gaussian integration schemes as an Abaqus/Standard user element subroutine (UEL). Appendix \ref{sec:abaqus-implementation} provides a detailed description of the  Abaqus UEL implementation, and a generic pseudocode of the finite element algorithm is presented in Algorithm \ref{alg:pe-gel-implementation}.

\begin{algorithm}[htbp]     
\caption{Algorithm for calculating the element residual vector and tangent stiffness matrix}
\label{alg:pe-gel-implementation}

\DontPrintSemicolon
\SetAlgoNoLine
\LinesNumbered

\SetKwInOut{Input}{Input}
\SetKwInOut{Output}{Output}
\SetKw{EndFor}{end for}
\SetKw{EndWhile}{end while}
\SetKw{EndIf}{end if}
\SetKw{EndProcedure}{end material procedure}
\SetKw{EndSolver}{end local Newton solver}

\SetKwBlock{GaussForLoop}{for $k = 1$ to ${n_\mathrm{int}}$}{\EndFor}
\SetKwBlock{MaterialSubroutine}{material procedure:}{\EndProcedure}
\SetKwBlock{localNewtonSolver}{local Newton solver:}{\EndSolver}
\SetKwBlock{localWhileLoop}{while $j \le N_{\mathrm{iter}}$}{\EndWhile}
\SetKwBlock{localIfBlock}{if $ \left\| \boldsymbol{\mathcal{G}} \right\|_2 \le \varepsilon_\mathrm{tol} $ or $ \left\| \boldsymbol{\mathcal{G}} \right\|_2 / \left\| \boldsymbol{\mathcal{G}}_0 \right\|_2\le \varepsilon_\mathrm{tol} $}{\EndIf}

\Input{ 
$\vect{X}$, $\Delta t$,  
$\vect{d} = \{ \vect{u}, \ \mu^w, \ \omega^{\beta_k} \}$, 
$\vect{q}_{t} = \{C^w_{t}, \ C^{\beta_k}_{t}, \ \psi_{t} \}$ 
}
\Output{ $\mat{K}_e$, $\vect{r}_e$, $\vect{q} =\{ C^w, \ C^{\beta_k}, \ \psi \}$ }

\BlankLine
Set Gaussian quadrature points and weights: $\boldsymbol{\upxi}$ and $\vect{w}$ \;
\BlankLine

\GaussForLoop{
    Evaluate interpolation functions and their derivatives: 
    $\vect{N}(\boldsymbol{\upxi})$ and
    $\frac{\partial \vect{N}}{\partial \boldsymbol{\upxi}}$     \;

    Compute element Jacobian and map interpolation function derivatives:
    $J_{\boldsymbol{\upxi}} = \det \left[ \frac{\partial \vect{X}}{\partial \boldsymbol{\upxi}} \right]$  
    and
    $\frac{\partial \mat{N}}{\partial \vect{X}} 
    = J_{\boldsymbol{\upxi}}^{-1} \left( \frac{\partial \mat{N}}{\partial \boldsymbol{\upxi}} \right) $    \;

    Form finite element matrix operators: 
    $\mat{N}_{\vect{u}}, \mat{B}_{\vect{u}}, \mat{G}_{\vect{u}}$,  
    $\mat{N}_{\mu}, \mat{B}_{\mu}$,
    $\mat{N}_{\omega}, \mat{B}_{\omega}$

    Compute the gradients of degrees of freedom: 
    $ \Grad \vect{d} = \{ \vect{F}, \ \Grad \mu^w, \ \Grad \omega^{\beta_k} \} $ \;

    \BlankLine
    \MaterialSubroutine{
        \BlankLine
        \localNewtonSolver{
            Compute: 
            $\boldsymbol{\mathcal{G}}_0 \left( \vect{d}, \Grad \vect{d}, \vect{q}_{t} \right)$     \;
            \localWhileLoop{
        
                \localIfBlock{
                    \Return
                    } 
                    
                Compute: $\vect{q}^{(j+1)} = \vect{q}^{(j)} - 
                \left( \frac{d \boldsymbol{\mathcal{G}}} {d \vect{q}} \right )^{-1} 
                \boldsymbol{\mathcal{G}}^{(j)}$ \;
            }
        }
        
        \BlankLine
        
        Update the algorithmic internal variables: 
        $\vect{q} \leftarrow \vect{q}^{(j+1)}$ \;

        Compute constitutive response: 
        $ \boldsymbol{\mathcal{F}}  (\Grad \vect{d}; \vect{q}) 
        = \{ \vect{S} (\Grad \vect{d}; \vect{q}), \ \vect{J}^w (\Grad \vect{d}; \vect{q}), \ \vect{J}^{\beta_k} (\Grad \vect{d}; \vect{q}) \}$ \;

        Compute consistent tangent moduli of constitutive functions:
        $ \frac{d \vect{q}}{d (\bullet)} 
        = - \left( \frac{\partial \boldsymbol{\mathcal{G}}}{\partial \vect{q}} \right)^{-1} 
        \frac{\partial \boldsymbol{\mathcal{G}}}{\partial (\bullet)}$
        and 
        $\frac{d \boldsymbol{\mathcal{F}}}{d (\bullet)}
        = \frac{\partial \boldsymbol{\mathcal{F}}}{\partial (\bullet)}
        + \frac{\partial \boldsymbol{\mathcal{F}}}{\partial \vect{q}} \frac{d \vect{q}}{d (\bullet)}$ 
        \tcp*{$(\bullet)$ represents $\vect{d}$ or $\Grad \vect{d}$}
        
    }
    
    \BlankLine
    
    Gauss integration of the element residual vector:
    $\vect{r}^{\vect{d}} = \vect{r}^{\vect{d}} + w(k) \det(J_{\xi}) \left( \bullet \right)$ \;
    \tcp*{$\left( \bullet \right)$ represents the integrand residual expressions in Eq. \eqref {eq:elem-residuals}}

    Gauss integration of the element tangent matrix: 
    $\vect{K}^{\vect{dd}} = \vect{K}^{\vect{dd}} + w(k) \det(J_{\xi}) \left( \bullet \right)$ \;
    \tcp*{$\left( \bullet \right)$ represents the integrand in the tangent matrix expressions in Eq.  \eqref{eq:elem-tangents}}
}
\BlankLine
Assemble the element residual vector and element tangent stiffness matrix: 
$\vect{r}_e = \bigA_{a=1}^{n_\mathrm{en}} 
\begin{Bmatrix}
    r^{u_i}_{a} \\ r^{\mu}_a \\ \vdots \\ r^{\omega_n}_a
\end{Bmatrix}$
and
$ \mat{K}_e = \bigA_{(a,b)=1}^{n_\mathrm{en}}
\begin{bmatrix}
    k^{\vect{u} \vect{u}}_{ab}        & k^{\vect{u} \mu}_{ab}   & \cdots    & k^{\vect{u} \omega_n}_{ab}         \\
    k^{\mu \vect{u}}_{ab}             & k^{\mu \mu}_{ab}        & \cdots    & k^{\mu \omega_n}_{ab}         \\
    \vdots                            & \vdots                  & \ddots    & \vdots     \\
    k^{\omega_n \vect{u}}_{ab}        & k^{\omega_n \mu}_{ab}   & \cdots    & k^{\omega_n \omega_n}_{ab}  
\end{bmatrix}
$ \;

\Return $\mat{K}_e$, $\vect{r}_e$, $\vect{q}$

\end{algorithm}

\section{Constitutive model for PE gel in dilute ionic solution}       
\label{sec:constitutive-model}

In the previous section, a finite element framework was presented for the general coupled electro-chemo-mechanical theory, which did not consider specific constitutive models for the stress and flux of ions and solvent. However, to study the multiphysical behaviors of PE gels quantitatively, we now need to prescribe specific constitutive laws. For a standard polyelectrolyte gel, the total Helmholtz free energy density, $\Psi$, can be additively decomposed as follows,
\begin{equation} \label{eq:free-energy-decomposition}
    \Psi= \Psi^w + \sum_{\beta_k} \Psi^{\beta_k} + \Psi^{\mathrm{mech}} + \Psi^{\mathrm{mix,pol}} + \Psi^{\mathrm{mix,ion}},
\end{equation}
where $\Psi^w$ and $\Psi^{\beta_k} $ represent the free energy density of unmixed pure solvent, $w$, and any pure ionic species, $ \beta_k $, respectively. The entropic elasticity of the polymer network is described by $\Psi^{\text{mech}}$, whereas $\Psi^{\text{mix,pol}}$ and $\Psi^{\text{mix,ion}}$ represent the polymer-solvent mixing energy and the mixing energy between the ions and solvent, respectively. Assuming the ionic mixture is dilute, we neglected the interaction between the polymer and ions in this work. We also omit the contribution of electrostatic energy, which is not needed because of the electroneutrality constraint we impose in our framework. Considering these assumptions, the following is specified for the total free energy density for the PE gel in the hydrated reference state.
\begin{equation} \label{eq:total_psi}
\begin{gathered}
    \Psi =
    \mu^{0,w} C^w
    + \sum_{\beta_k} \omega^{0,\beta_k} C^{\beta_k}
    + \frac{ G}{2} \left[ I_1 - 3 - 2 (\phi^p_0)^{2/3} \ln \left( J \right) \right] 
    + \frac{\phi^p_0 \kappa}{2} J^s (\ln J^e)^2 \\
    + \frac{ \phi^p_0 R \theta} {\mathcal{V}^{w}} \left[ \left( \frac{1}{\phi^p}-1 \right) \ln \left( 1-\phi^{p} \right) + \chi (1-\phi^{p})  \right]
    + R \theta  \sum_{\beta_k}  C^{\beta_k} \left[ \ln \left( \frac{C^{\beta_k}}{C^{w}} \right) -1 \right],
\end{gathered}
\end{equation}
where 
\begin{equation} \label{eq:polymer-vol-frac-def}
    \phi^p 
    = \frac{\phi^p_0}{C^p \mathcal{V}^p + C^w \mathcal{V}^w} 
    =  \frac{\phi^p_0}{\phi^p_0 + C^w \mathcal{V}^w} 
    = \frac{1}{J^s}
\end{equation}
is defined as the polymer volume fraction, $R$ is the universal gas constant, and $\theta$ is the absolute temperature. The standard reference chemical potential for the solvent, $w$, and ion, $\beta_k$, are represented by $\mu^{0,w}$ and $\omega^{0, {\beta_k}}$, respectively. We chose the Neo-Hookean potential with a volumetric constraint to represent the near-incompressible entropic elasticity of the polymer network. The shear and bulk moduli of the gel in its hydrated reference state are represented by $G$ and $\kappa$, respectively. In our numerical simulations, we set $\kappa \gg G$ to enforce quasi-incompressibility. Finally, $\chi$ is the Flory-Huggins interaction parameter, representing the enthalpic contribution of a binary polymer-solvent mixture \citep{floryThermodynamicsHighPolymer1942,doiIntroductionPolymerPhysics1996}. The last term in the prescribed free energy density in Eq. \eqref{eq:total_psi} is attributed to the dilute mixture formed by the solvent and the ions within the gel \citep{narayanCoupledElectrochemomechanicalTheory2022}. We should note that, in our numerical implementation, by invoking the dilute mixture assumption, we ignored the volume occupied by the ions to define the polymer volume fraction, $\phi^p$, and swelling volume change, $J^s$, as seen in Eq. \eqref {eq:polymer-vol-frac-def}.

\begin{remark}
    Since we prescribed the free energy density of PE gel in its pre-swollen hydrated reference state, the specific forms of $\Psi^{\text{mech}}$ and $\Psi^{\text{mix,pol}}$ given in Eq. \eqref {eq:total_psi} may appear somewhat non-standard. These potentials can be obtained either from statistical mechanics as given in Appendix \ref{sec:appendix-free-energy-derivation} or by using a push forward operation on the free energy density at the dry state to the pre-swollen state as $\Psi = \frac{\Psi_{\mathrm{dry}}}{J^d} = \phi^p_0 \Psi_{\mathrm{dry}}$.
\end{remark}

Using the expression for the total free energy density of PE gel in Eq. \eqref {eq:total_psi} and the constitutive definitions in Eq. \eqref {eq:constitutive-definitions-final}$_1$, the Cauchy stress tensor,  $\boldsymbol{\upsigma}$, can be written as,
\begin{equation} \label{eq:cauchy-stress-def}
    \boldsymbol{\upsigma} 
    = J^{-1} \left[ G \vect{b}
    - G (\phi^p_0)^{2/3} \mathds{1} 
    + \kappa \phi^p_0 J^s (\ln J^e) \mathds{1}\right].
\end{equation}
The mean pressure, $p$, can be obtained using Eq. \eqref {eq:mean-pressure-defn} on the Cauchy stress expression in Eq.  \eqref{eq:cauchy-stress-def} as follows,
\begin{equation} \label{eq:mean-pressure-def-2}
    p = \frac{-G}{3\phi^p_0 J^s} \left( I_1 
    - 3 (\phi^p_0)^{2/3} \right) - \kappa (\ln J^e).
\end{equation}
The first and second Piola-Kirchhoff stress tensors are determined from the Piola transformation and a push forward of $\boldsymbol{\upsigma}$, respectively,
\begin{equation} \label{eq:piola-stress-def}
\begin{gathered}
    \vect{P} 
    = J \boldsymbol{\upsigma} \vect{F}^{-\top} =
    G \vect{F} - \left[ G (\phi^p_0)^{2/3} - \kappa \phi^p_0 J^s (\ln J^e) \right] \vect{F}^{-\T}, \\
    \vect{S} 
    = J \vect{F}^{-1} \boldsymbol{\upsigma} \vect{F}^{-\top}
    = G \mathds{1} - \left[ G (\phi^p_0)^{2/3} - \kappa \phi^p_0 J^s (\ln J^e) \right] \vect{C}^{-1}.
\end{gathered}
\end{equation}

Using Eq. \eqref {eq:constitutive-definitions-final}$_2$ on the total free energy density in Eq. \eqref{eq:total_psi} gives  the expression for solvent chemical potential, $\mu^w$, as below
\begin{equation} \label{eq:chem-pot-def}
    \mu^w  
    = \mu^{0,w} 
    + R \theta \left[ \phi^{p} + \ln(1-\phi^{p}) + \chi (\phi^{p})^2 \right]
    + \mathcal{P} \mathcal{V}^{w}
    - R \theta \sum_{\beta_k} \frac{C^{\beta_k}}{C^{w}},
\end{equation}
where
\begin{equation}    \label{eq:lagrange-mult-def}
    \mathcal{P} = \frac{\kappa }{2} \left( \ln J^e \right)^2 - \kappa \ln J^e,
\end{equation}
acts as a penalty function for volume change, which allows us to model quasi-incompressibility by specifying a large bulk modulus.

Finally, for any ion, $\beta_k$, the following expression for the ion electrochemical potential, $\omega^{\beta_k}$, can be obtained by applying Eq. \eqref {eq:constitutive-definitions-final}$_3$ to the total free energy density in Eq. \eqref{eq:total_psi}.
\begin{equation}  \label{eq:echem-pot-def}
    \omega^{\beta_k} 
    = \omega^{0,\beta_k}
    + R \theta  \ln \left( \frac{C^{\beta_k}}{C^{w}}\right) 
    + F \psi z^{\beta_k}
    + p \mathcal{V}^{\beta_k}.
\end{equation}

In our theory, we assumed that the non-equilibrium and irreversible thermodynamic phenomena are attributed to the diffusion of the solvent and ions. Based on the thermodynamic restrictions on the solvent transport Eq. \eqref {eq:generic-transport-law}$_1$, we, therefore, prescribe the following kinetic law for the diffusion of the solvent,
\begin{equation} \label{eq:flux-law-solvent}
    \vect{J}^w = - \vect{M}^{w} \Grad \mu^w, 
    \qquad \mathrm{where} \quad 
    \vect{M}^{w} = \frac{D^{w} C^{w}}{R \theta} \vect{C}^{-1}.
\end{equation}
Here, the diffusion coefficient of the solvent in the PE gel is represented by $D^w$, and $\vect{M}^w$ is known as the mobility tensor in the reference configuration and adopted from \cite{chesterCoupledTheoryFluid2010} and \cite{narayanCoupledElectrochemomechanicalTheory2022}.

Similar to the case for the solvent transport, we assumed a diffusion-based referential molar flux relation for the ions as,
\begin{equation} \label{eq:flux-law-ion}
    \vect{J}^{\beta_k} = - \vect{M}^{\beta_k} \Grad \omega^{\beta_k},
    \qquad \mathrm{where} \quad 
    \vect{M}^{\beta_k} = \frac{D^{\beta_k} C^{\beta_k}}{R \theta} \vect{C}^{-1}.
\end{equation}
where $D^{\beta_k}$ is the diffusion coefficient and $\vect{M}^{\beta_k}$ is referred to as the referential mobility tensor of the ion, $\beta_k$.

\begin{remark}
    To simplify our implementation, we chose the simplest forms of constitutive equations for referential molar fluxes for solvent in Eq. \eqref{eq:flux-law-solvent} and ions in Eq.  \eqref{eq:flux-law-ion} that satisfy the positive dissipation criterion in Eq. \eqref{eq:transport-inequality}. It is a straightforward extension to include the effect of multi-species cross-diffusion by adopting the flux laws prescribed by \cite{zhangKineticsPolyelectrolyteGels2020} and \cite{narayanCoupledElectrochemomechanicalTheory2022}.
\end{remark}

At the current time, $t+ \Delta t$, the local residual vector in Eq. \eqref{eq:internal-vars-system} can be formed by collecting the nonlinear constitutive equations for solvent chemical potential in Eq. \eqref{eq:chem-pot-def}, ion electrochemical potential in Eq. \eqref{eq:echem-pot-def}, and the electroneutrality assumption in Eq. \eqref{eq:electroneutrality-def} as follows
\begin{equation} \label{eq:local-residual-system}
\begin{aligned}
    \mathcal{G}_1 & = 
    \mu^{0,w} 
    + R \theta \left[ \phi^{p} + \ln(1-\phi^{p}) + \chi (\phi^{p})^2 \right]
    + \mathcal{P} \mathcal{V}^{w}
    - R \theta \sum_{\beta_k} \frac{C^{\beta_k}}{C^{w}} - \mu^w, \\
    \mathcal{G}_2 & = 
    \omega^{0,{\beta_1}} + R \theta  \ln \left( \frac{C^{\beta_1}}{C^{w}}\right) 
    + F \psi z^{\beta_1} + p \mathcal{V}^{\beta_1} - \omega^{\beta_1}, \\
    \mathcal{G}_3 & = 
    \omega^{0,{\beta_2}} + R \theta  \ln \left( \frac{C^{\beta_2}}{C^{w}}\right) 
    + F \psi z^{\beta_2} + p \mathcal{V}^{\beta_2} - \omega^{\beta_2}, \\
    \vdots \\
    \mathcal{G}_{n+2} & =
    C^w  \sum_{\beta_k} z^{\beta_k} \exp \left( \frac{\omega^{\beta_k} 
    - F \psi z^{\beta_k} - p \mathcal{V}^{\beta_k} - \omega^{0,{\beta_k}} } {R \theta} \right)  + C^{\mathrm{fix}} z^{\mathrm{fix}}.
\end{aligned}
\end{equation}
The Jacobian of the local residuals is given in Appendix \ref{sec:appendix-local-newton}, which is used in the local iterative solution procedure and in computing consistent material tangent moduli.

\section{Numerical validation and examples}          
\label{sec:numerical-example}

To demonstrate the applicability of the finite element framework, we present three different numerical examples in this section. In the first example, we simulated the cyclic swelling and deswelling experiments of cationic PE gels presented in \cite{sunMultiresponsiveToughHydrogels2015} to validate the model. We then studied the consolidation of PE gels in a salt solution under confined compression, as well as the anomalous transient bending behavior of a PE gel-elastomer bilayer. By utilizing symmetry of the geometries and boundary conditions, the numerical examples were modeled using axisymmetric and plane strain elements and two different ions, which also helped reduce the computational cost.

\subsection{Transient swelling and deswelling of polyelectrolyte gel in salt bath}
\label{sec:validation-sun}

\cite{sunMultiresponsiveToughHydrogels2015} reported two sets of experiments: (1) transient swelling of as-prepared cationic PE gels in salt solution with different ionic strengths and (2)  cyclic deswelling and swelling of the same gel in aqueous solvent baths of different ionic strengths and pH, which were numerically reproduced by \cite{zhangKineticsPolyelectrolyteGels2020} and \cite{narayanCoupledElectrochemomechanicalTheory2022}. In the current work, the first set of experiments was used to calibrate the diffusion coefficient of water in the PE gel, $D^w$, and the Flory-Huggins interaction parameter, $\chi$. The remaining parameters were obtained from the literature as described below. These, along with the calibrated values for $D^w$ and $\chi$, are listed in Table \ref{tab:sun_pe_gel_props}. The same set of model parameters was used to simulate a second set of experiments by \cite{sunMultiresponsiveToughHydrogels2015} to validate the finite element model.  

\cite{sunMultiresponsiveToughHydrogels2015} synthesized the hydrogels using acrylamide (Am), quaternized N,N-dimethylamino ethylacrylate (DMAEA), and cross-linking agent pluronic F127 diacrylate micelles. In a saline aqueous solvent (NaCl with water), DMAEA releases Cl$^-$ ion in the solvent; thus, the remaining --N$^+$(CH$_3$)$_3$ on the polymer chain gives rise to a cationic charge with $z^{\mathrm{fix}} = +1$. The concentrations of H$^+$ and OH$^-$ ions are significantly smaller than Na$^+$ and Cl$^-$ ions within the gel, and the polymer was fully ionized as the external solution bath was maintained at a constant pH = 7.0. Hence, we only included Na$^+$ and Cl$^-$ ions in our finite element model. For these ions, the ionic strength of the solvent, $I$, can be approximated as,
\begin{equation}    \label{eq:ionic-strength}
    I \approx \frac{1}{2} \left( C^{{\mathrm{Na}}^+}_{\mathrm{sol}} + C^{{\mathrm{Cl}}^-}_{\mathrm{sol}} \right),
\end{equation}
where $C^{{\mathrm{Na}}^+}_{\mathrm{sol}} =  C^{{\mathrm{Cl}}^-}_{\mathrm{sol}}$ because of the electroneutral behavior of the external solution.

\cite{sunMultiresponsiveToughHydrogels2015} performed the oscillatory swelling-deswelling experiments by changing the ionic strength of the aqueous solvent from $I = 0.05$ M to $I = 0.2$ M every 24 hours while keeping a constant pH = 7. Based on the definition of ionic strength, we calculated the concentration of the ions within the solvent bath as follows:
\begin{equation}        \label{eq:sun-concentration-values}
\begin{aligned}
    & I = 0.05 \ \mathrm{M} 
    \quad && \Rightarrow \quad 
    C^{{\mathrm{Na}}^+}_{\mathrm{sol}} = C^{{\mathrm{Cl}}^-}_{\mathrm{sol}} \approx 50 \ \mathrm{mol/m}^3, \\
    & I = 0.1 \ \mathrm{M} 
    \quad && \Rightarrow \quad 
    C^{{\mathrm{Na}}^+}_{\mathrm{sol}} = C^{{\mathrm{Cl}}^-}_{\mathrm{sol}} \approx 100 \ \mathrm{mol/m}^3, \\
    & I = 0.15 \ \mathrm{M} 
    \quad && \Rightarrow \quad 
    C^{{\mathrm{Na}}^+}_{\mathrm{sol}} = C^{{\mathrm{Cl}}^-}_{\mathrm{sol}} \approx 150 \ \mathrm{mol/m}^3, \\
    & I = 0.20 \ \mathrm{M} 
    \quad && \Rightarrow \quad 
    C^{{\mathrm{Na}}^+}_{\mathrm{sol}} = C^{{\mathrm{Cl}}^-}_{\mathrm{sol}} \approx 200 \ \mathrm{mol/m}^3.
\end{aligned}
\end{equation}
For any external ionic solution, we can assume the polymer volume fraction, $\phi^p = 0$, the mean pressure, $p = 0$, and the electric potential, $\psi = 0$. By using these assumptions in Eq. \eqref {eq:chem-pot-def} and Eq. \eqref {eq:echem-pot-def}, the expressions for solvent chemical potential, $\mu^w_{\mathrm{sol}}$, and ion electrochemical potentials, $\omega^{\mathrm{Na}^+}_{\mathrm{sol}}$ and $\omega^{\mathrm{Cl}^-}_{\mathrm{sol}}$, within the external solution bath, can be obtained as follows,
\begin{equation}
\begin{aligned}     \label{eq:echem-sol}
    \mu^w_{\mathrm{sol}} & = - \frac{R \theta}{C^w_{\mathrm{sol}}} \left( C^{{\mathrm{Na}}^+}_{\mathrm{sol}} + C^{{\mathrm{Cl}}^-}_{\mathrm{sol}} \right), \\
    \omega^{\mathrm{Na}^+}_{\mathrm{sol}} & = R \theta \log \left( \frac{C^{{\mathrm{Na}}^+}_{\mathrm{sol}}}{C^w_{\mathrm{sol}}} \right), \\
    \omega^{\mathrm{Cl}^-}_{\mathrm{sol}} & = R \theta \log \left( \frac{C^{{\mathrm{Cl}}^-}_{\mathrm{sol}}}{C^w_{\mathrm{sol}}} \right),
\end{aligned}
\end{equation}
By using the density of water, $\rho^w_{\mathrm{sol}} = 1000 $ kg/m$^3$ and molar mass of water, $\mathcal{M}^w_{\mathrm{sol}} = 18$ g/mol, we approximated $C^w_{\mathrm{sol}} = \frac{\rho_{\mathrm{sol}}}{\mathcal{M}_{\mathrm{sol}}} \approx  55000$ mol/m$^3$ to represent the concentration of water in dilute ionic solutions.

\begin{figure}[htp]
\begin{center}
    \begin{subfigure}{0.8\textwidth}
    \includegraphics[width=\textwidth, trim={4.25cm 5.5cm 4.5cm 4.5cm}, clip] {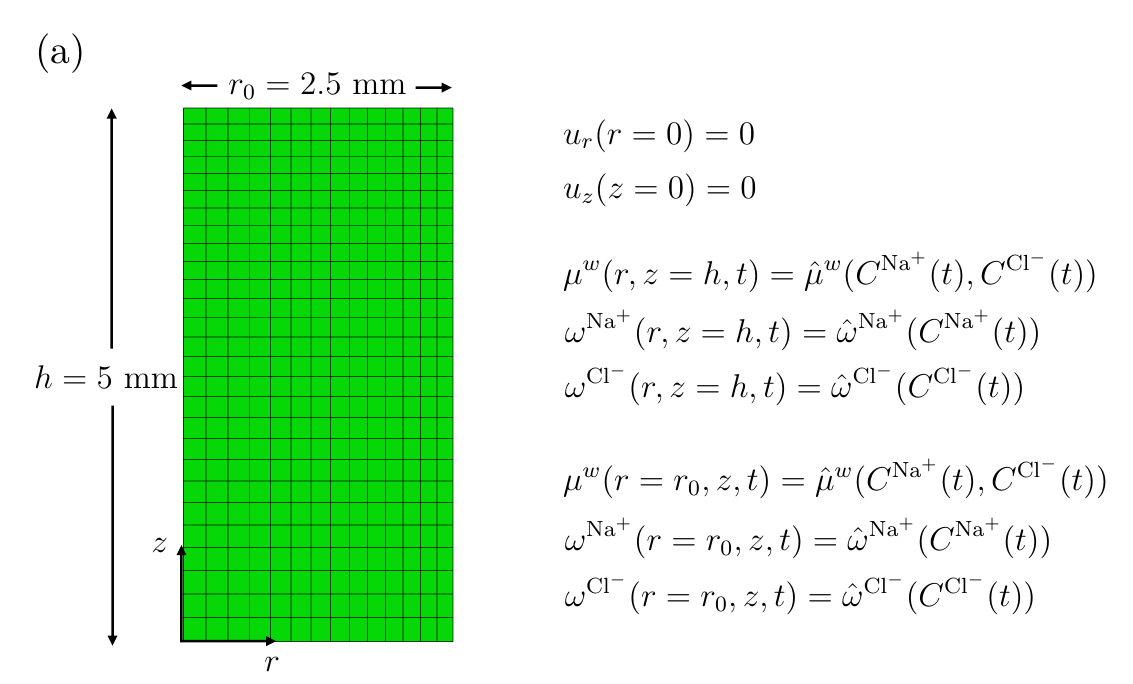}
    \end{subfigure}
    
    \smallskip
    \begin{subfigure}{\textwidth}
    \includegraphics[width=\textwidth, trim={0.5cm 6.5cm 0.5cm 6cm}, clip] {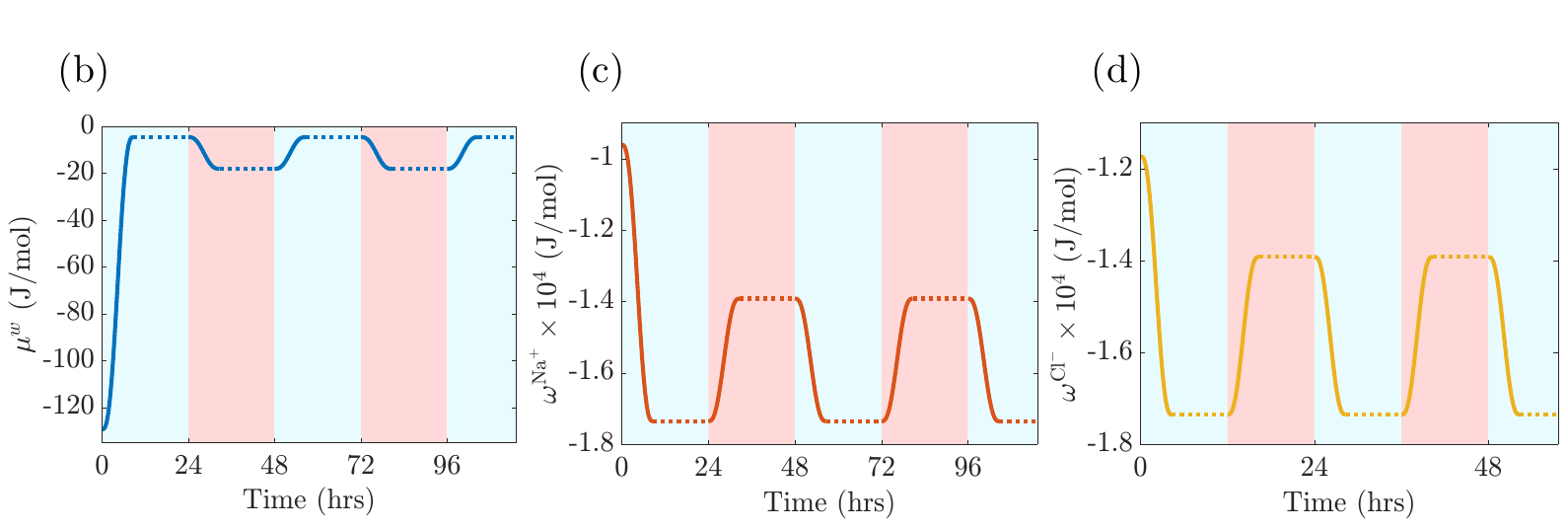}
    \end{subfigure}
    
    \caption{Boundary conditions and time-varying amplitude profiles of solvent chemical potential and ion electrochemical potentials ($I = 0.05$ M) for modeling the swelling-deswelling experiments performed by \cite{sunMultiresponsiveToughHydrogels2015}: (a) Axisymmetric finite element model with mesh and prescribed boundary conditions for cationic polyelectrolyte gel cylinder used in swelling-deswelling experiments, (b) amplitude profile of chemical potential of the solvent, $\mu^w$, (c) amplitude profile of electrochemical potential of Na$^+$ ion, $\omega^{\mathrm{Na}^+}$, and (d) amplitude electrochemical potential of Cl$^-$ ion, $\omega^{\mathrm{Cl}^-}$. Note that the amplitude profiles were prescribed over 180 s at the beginning of each load step in Abaqus, and the dashed lines connecting the profiles represent the continuation of the curve. The variations in amplitude curves were calculated by substituting the ion concentration values from Eq. \eqref {eq:sun-concentration-values} in Eq. \eqref {eq:echem-sol}. We only used the first 24 hours of the amplitude profiles for material parameter calibration purposes, whereas to perform a cyclic deswelling-swelling simulation, we used the entire profile over the 120 hours.}
    \label{fig:sun-cylinder-model}
\end{center}
\end{figure}

As-prepared cylindrical PE gel specimens used in the experiments had a diameter of 5 mm and a height of 10 mm with an initial polymer volume fraction, $\phi^p_0 = 0.312$. To model the free swelling and deswelling of the hydrogel specimen, we utilized the axisymmetric condition and modeled only one quarter of the geometry, as shown in Figure 
\ref{fig:sun-cylinder-model}. We discretized the domain using axisymmetric 4-node quadrilateral elements, with the smallest element being 0.15 mm $\times$ 0.15 mm and the largest element being 0.25 mm $\times$ 0.25 mm. We adopted the initial concentration of the fixed charge, $C_0^{\mathrm{fix}}$, shear modulus, $G$, and diffusion coefficient of the ions from \cite{narayanCoupledElectrochemomechanicalTheory2022}. However, since our model accommodates specifying an initial polymer volume fraction, we adjusted the shear modulus, $G = G_{\mathrm{dry}} 
(\phi^p_0)^{1/3}$ = 48 kPa, by using the scaling law for shear modulus in Eq. \eqref{eq:free-energy-neo-hookean-stat-mech}, and similarly the initial concentration of the polymer fixed charge, $C_0^{\mathrm{fix}} = C_{\mathrm{dry}}^{\mathrm{fix}}/ J^d = \phi^p_0 C_{\mathrm{dry}}^{\mathrm{fix}}$ = 460 mol/m$^3$, for the prescribed hydrated reference state $(\phi^p_0 = 0.312)$. These adjustments are consistent with our kinematic prescription of the hydrated reference state.

\textbf{Mechanical boundary conditions:} As shown in Figure \ref{fig:sun-cylinder-model}, by using symmetry conditions, we fixed the radial displacement at the left edge, \emph{i.e.}, $u_r(r=0,z) = 0$ and axial displacement at the bottom edge, \emph{i.e.}, $u_z (r,z=0) =0$. These displacement boundary conditions were maintained throughout the simulation.

\textbf{Computation of initial chemical conditions:} Since the initial concentrations for the coions and counterions are unknown, we assumed $C_0^{\mathrm{Na}^+}$ = 340 mol/m$^3$ and $C_0^{\mathrm{Cl}^-}$ = 800  mol/m$^3$ such that the electroneutrality condition is satisfied, \emph{i.e.}, $C_0^{\mathrm{fix}} + C_0^{\mathrm{Na}^+} - C_0^{\mathrm{Cl}^-} = 0$. We should note that the results were unaltered when we repeated the simulations with different values for $C_0^{\mathrm{Na}^+}$ and $C_0^{\mathrm{Cl}^-}$. We further assumed zero electric potential, $\psi = 0$, throughout the gel in its initial state. By substituting the initial polymer volume fraction, $\phi^p_0 = 0.312$, assumed ion concentrations, $C^{\mathrm{Na}^+}$ and $C^{\mathrm{Cl}^-}$, and the assumed electric potential, $\psi$, in Eqs. \eqref{eq:chem-pot-def} and \eqref{eq:echem-pot-def}, we computed the initial conditions for the solvent chemical potential, $\mu^w_0(r,z,t=0)$, and ion electrochemical potentials, $\omega^{\mathrm{Na}^+}_0(r,z,t=0)$ and $\omega^{\mathrm{Cl}^-}_0(r,z,t=0)$, which are the starting values of the time-varying profiles shown in Figure \ref{fig:sun-cylinder-model}(b)--(d). We then prescribed these uniform initial values for the solvent chemical potential, $\mu^w(r,z) = \mu^w_0$, and electrochemical potentials of the ions, $\omega^{\mathrm{Na}^+}(r,z) = \omega^{\mathrm{Na}^+}_0$, and $\omega^{\mathrm{Cl}^-}(r,z) = \omega^{\mathrm{Cl}^-}_0$, throughout the computational domain as field values to compute the electric potential, $\psi$.

\textbf{Prescribed chemical boundary conditions:} To simulate the transient swelling of cationic PE gel immersed in an external solution of a specific ionic strength, at the right outer egde and the top edge, we applied boundary conditions for solvent chemical potential, \emph{i.e.}, $\mu^w(r=r_0,z,t) = \mu^w(r,z=h,t)$, and ion electrochemical potentials, $\omega^{\mathrm{Na}^+}(r=r_0,z,t) = \omega^{\mathrm{Na}^+}(r,z=h,t)$ and $\omega^{\mathrm{Cl}^-}(r=r_0,z,t) = \omega^{\mathrm{Cl}^-}(r,z=h,t)$. We computed the solvent chemical potential, $\mu^w$, and ion electrochemical potentials, $ \omega^{\mathrm{Na}^+}$ and $\omega^{\mathrm{Cl}^-}$, by substituting the ion concentrations, $C^{{\mathrm{Na}}^+}_{\mathrm{sol}}$ and $C^{{\mathrm{Cl}}^-}_{\mathrm{sol}}$, corresponding to the ionic strength in the external solution from Eq. \eqref{eq:sun-concentration-values} in Eq. \eqref{eq:echem-sol}. For computational tractability, we changed the solvent chemical potential and ion electrochemical potentials boundary conditions from the initial values to the computed values over $t_{\mathrm{ramp}} = 180$ s using the smooth step amplitude feature available in Abaqus and subsequently held the boundary conditions for a simulation time of 24 hours. The first 24 hours of the time-varying profiles of the solvent chemical potential and ion electrochemical potentials in Figure \ref{fig:sun-cylinder-model}(b)-(d) represent the application of time-varying chemical boundary conditions for the case of $I = 0.05$ M. In our simulation, we specified the initial time step, $\Delta t_{\mathrm{init}} = 10^{-3}$ s, the minimum time step, $\Delta t_{\mathrm{min}} = 10^{-15}$ s, and maximum time step, $\Delta t_{\mathrm{max}} = 300$ s. In our subsequent simulations, we maintained the same time-stepping parameters\footnote{Further details of the calibration simulation can be found in the Abaqus input file, \texttt{0\_15x0\_25\_mm\_bias\_mesh\_378elm\_Dw9e7\_salt\_50mM.inp}, under the \texttt{test/sun\_calibration\_validation} directory of the GitHub repository. The details of the multi-cycle deswelling-swelling simulation can be found in the \texttt{0\_15x0\_25\_mm\_bias\_mesh\_378elm\_Dw9e7\_50\_200mM\_cycle.inp} in the same directory.}.

\begin{figure}[ht]
\begin{center}
    \includegraphics[width=0.6\textwidth, trim={7cm 5.5cm 7cm 5.5cm},clip] {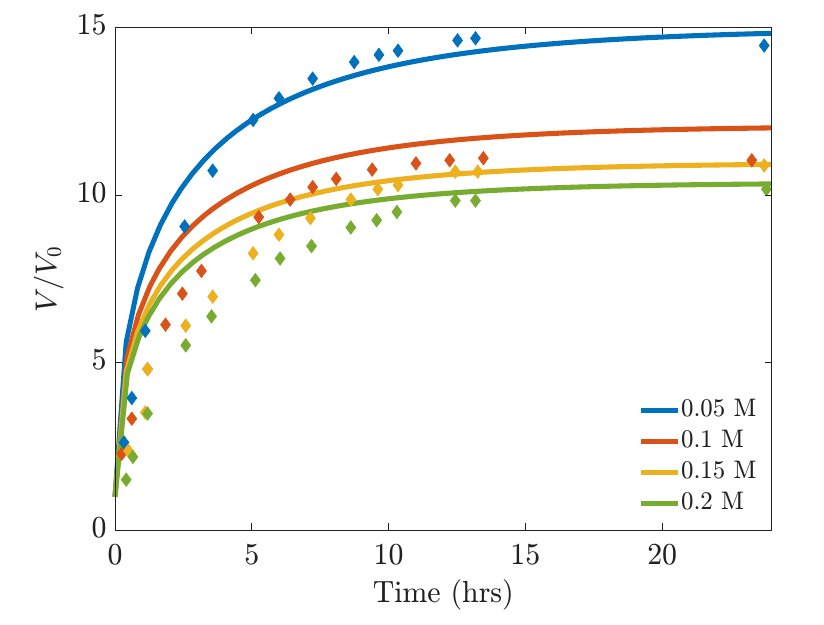}
    \caption{Calibration of the diffusion coefficient of the solvent $D^w$ and Flory-Huggins parameter, $\chi$, by comparing transient volumetric swelling ratio from \cite{sunMultiresponsiveToughHydrogels2015} in NaCl salt solution with different ionic strengths. The solid line represents finite element results, and the symbols of the same color represent corresponding experimental results.}
    \label{fig:sun-calibration}
\end{center}
\end{figure}

\begin{figure}[ht]
\begin{center}
    \includegraphics[width=\textwidth, trim={3cm 6cm 3cm 6cm}, clip] {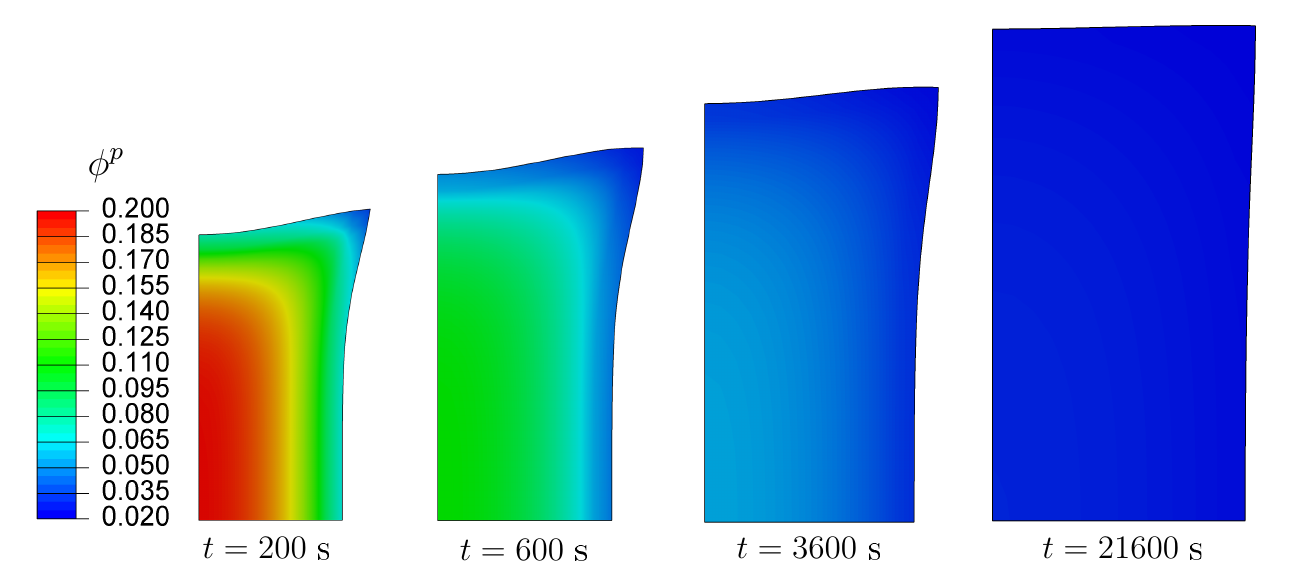}
    \caption{Time series snapshots of polymer volume fraction contour after the cylindrical gel specimen was immersed in an ionic solution bath of $I=0.05$ M.}
    \label{fig:sun-init-swell-phi}
\end{center}
\end{figure}

The transient volumetric swelling response of the cationic PE gel in ionic solutions of different strength levels is presented in Figure \ref{fig:sun-calibration}. Recall that the simulations were used to calibrate the diffusion constant for water in PE gel $D^w$ and the Flory-Huggins interaction parameter $\chi$. For comparison with experiments, the volumetric swelling ratio of the PE gel plug with respect to its hydrated reference state was calculated from the change in height at the center of the cylinder. For $I=0.05$ M, we can observe reasonable agreement between finite element and experimental results for $D^w = 9\times10^{-7}$ m$^2$/s and $\chi = 0.40$. Figure \ref{fig:sun-init-swell-phi} shows the time evolution of polymer volume fraction in $I=0.05$ M salt solution as well as its deformed shape. Surfaces exposed to the external solution deform at a higher rate because of the transient diffusion process. While the total immersion time was 24 hours, we can observe that the gel reached a near-equilibrium state within the first 6 hours.  The same experimental study was used by the two other pioneering works on PE gels \citep{zhangKineticsPolyelectrolyteGels2020,narayanCoupledElectrochemomechanicalTheory2022} for theoretical validation. While the diffusion coefficient and Flory-Huggins parameter listed by us match closely with the values reported in \cite{narayanCoupledElectrochemomechanicalTheory2022}, they differ vastly from those reported in \cite{zhangKineticsPolyelectrolyteGels2020}. Especially, we should note that the diffusion coefficient of the solvent is significantly higher than the self-diffusion coefficient of water, $D^w_{\mathrm{self}} \approx 2\times 10^{-9}$ m$^2$/s \citep{eastealDiaphragmCellHightemperature1989}. For highly swellable PE gel, the diffusion coefficient may not only depend on the polymer volume fraction, but also on the concentration of ions and polymer fixed charge. In our proposed model, we can consider the solvent diffusion coefficient, $D^w$, to be an effective diffusion coefficient encompassing all other complex phenomena that are not explicitly included. We then repeated the simulations with the parameters listed in Table \ref{tab:sun_pe_gel_props} to observe the transient swelling response of the gel in other ionic solutions and compared with experimental results in Figure \ref{fig:sun-calibration}. While our model predicts the equilibrium swelling ratio and near-equilibrium transient response fairly well, it overpredicts the transient response in the early stage. We attribute this difference to our simple constitutive model for the diffusion coefficient and Flory-Huggins parameter, which is independent of the swelling ratio, concentration of ions, and concentration of fixed charge. In comparison,  \citep{narayanCoupledElectrochemomechanicalTheory2022} assumed that $\chi$ varied with $C^{\mathrm{OH}^{-}}$. However, even with these simplifying constitutive modeling choices and the electroneutrality assumption, the model reproduced the experimental trends.

\begin{table}[ht]
\centering
\begin{tabular}{c c}
    \toprule \toprule
    \textbf{Properties}                             & \textbf{Value} \\
    \toprule  \toprule
    Universal gas constant, $R$                     & 8.314 J/mol-K \\
    Faraday's constant, $F$                         & 96485 C/mol \\
    Absolute temperature, $\theta$                  & 298 K \\
    Initial polymer volume fraction, $\phi^p_0$     & 0.312 \\
    Shear modulus at hydrated reference state, $G$              & 48 kPa \\
    Bulk modulus at hydrated reference state, $\kappa$          & 50 $G$ \\
    Charge number of polymer, $z^{\mathrm{fix}}$                & +1 \\
    Initial fixed charge of polymer, $C_0^{\mathrm{fix}}$       & 460 mol/m$^3$ \\
    Reference chemical potential of solvent, $\mu^{0,w}$        & 0 J/mol \\
    Molar volume of the solvent, $\mathcal{V}^w$                & $1.8 \times 10^{-5}$ m$^3$/mol \\ 
    Flory-Huggins interaction parameter, $\chi$                 &  0.40 \\
    Diffusion coefficient of the solvent, $D^w$                 & $9 \times 10^{-7}$ m$^2$/s \\
    Reference electrochemical potentials of pure ions, 
    $\omega^{0,{\mathrm{Na}^+}}$        & 0 J/mol\\
    $\omega^{0,{\mathrm{Cl}^-}}$        & 0 J/mol\\
    Molar volumes of ions,  
    $\mathcal{V}^{\mathrm{Na}^+}$       & $2.38 \times 10^{-5}$ m$^3$/mol   \\
    $\mathcal{V}^{\mathrm{Cl}^-}$       & $2.24 \times 10^{-5}$ m$^3$/mol   \\
    Charge numbers of ions,  
    $z^{\mathrm{Na}^+}$   & $+1$ \\
    $z^{\mathrm{Cl}^-}$   & $-1$ \\
    Diffusion coefficients of ions, 
    $D^{\mathrm{Na}^+}$   & $4 \times 10^{-8}$ m$^2$/s    \\
    $D^{\mathrm{Cl}^-}$   & $4 \times 10^{-8}$ m$^2$/s    \\
    \bottomrule
\end{tabular}
\caption{List of properties used in finite element simulation of swelling and deswelling of polyelectrolyte gel reported by \cite{sunMultiresponsiveToughHydrogels2015}.}
\label{tab:sun_pe_gel_props}
\end{table}

We next applied the parameters listed in Table \ref{tab:sun_pe_gel_props} to simulate the cyclic deswelling and swelling experiments of the cationic PE gel to numerically validate our proposed framework. In the experimental studies, an as-prepared PE gel is first swollen in a low ionic strength solution ($I=0.05 $ M) for 24 hours, followed by two complete cycles of deswelling and swelling which was achieved by alternatively placing the PE gel in the external solution of ionic strengths of $I = 0.2$ M and $I = 0.05$ M. A single deswelling-swelling cycle comprises 48 hours with the deswelling phase in a high ionic strength solvent ($I = 0.2$ M) spanning 24 hours and the swelling phase in a low ionic strength solvent ($I = 0.05$ M) spanning the next 24 hours. Similar to the experiments, we first let the gel swell and then performed two complete deswelling-swelling cycles, resulting in a total of 120 hours of simulation time. For mechanical boundary conditions, once again, we fixed the radial displacement at the left edge ($u_r(r=0,z) = 0$) and axial displacement at the bottom edge ($u_z (r,z=0) =0$), which were maintained throughout the simulation time of 120 hours. We prescribed the same initial conditions for the solvent chemical potential, $\mu^w_0(r,z,t=0)$, and ion electrochemical potentials, $\omega^{\mathrm{Na}^+}_0(r,z,t=0)$ and $\omega^{\mathrm{Cl}^-}_0(r,z,t=0)$, as the previous calibration simulation. To perform swelling and deswelling simulation of the cationic PE gel, we prescribed time-varying boundary conditions for the solvent chemical potential at the right edge and the top edge, \emph{i.e.}, $\mu^w(r=r_0,z,t) = \mu^w(r,z=h,t)$, and similarly for the ion electrochemical potentials, $\omega^{\mathrm{Na}^+}(r=r_0,z,t) = \omega^{\mathrm{Na}^+}(r,z=h,t)$ and $\omega^{\mathrm{Cl}^-}(r=r_0,z,t) = \omega^{\mathrm{Cl}^-}(r,z=h,t)$, as shown in Figure \ref{fig:sun-cylinder-model}(b)--(d). The detailed description of the time-varying applied boundary conditions is given below:

\begin{itemize} [leftmargin=0.5cm, topsep=0pt, itemsep=-3pt]

    \item \textbf{0--24 hours (pre-swelling):} To simulate the swelling of the as-prepared PE gel in a low ionic strength solution ($I = 0.05$ M), the boundary conditions for solvent chemical potential, $\mu^w$, and ion electrochemical potentials, $\omega^{\mathrm{Na}^+}$ and $\omega^{\mathrm{Cl}^-}$, were changed from their initial values to the values corresponding to $I = 0.05$ M over a short time period of $t_{\mathrm{ramp}} = 180$ s using the Abaqus smooth amplitude feature and held there until 24 hours. Prescribed boundary values for the solvent chemical potential, $\mu^w$, and electrochemical potentials of the ions, $\omega^{\mathrm{Na}^+}$, and $\omega^{\mathrm{Cl}^-}$, were calculated by substituting the ion concentrations corresponding to $I = 0.05$ M from \eqref{eq:sun-concentration-values} in \eqref{eq:echem-sol}.

    \item \textbf{24--48 hours (deswelling):} In this time period, the boundary values for $\mu^w$, $\omega^{\mathrm{Na}^+}$, and $\omega^{\mathrm{Cl}^-}$ were computed using the ion concentration values corresponding to the high ionic strength, $I = 0.2$ M, from \eqref{eq:sun-concentration-values} and subsequently by substituting them in \eqref{eq:echem-sol}. The variation was again applied over $t_{\mathrm{ramp}} = 180$ and held there for 24 hours, which resulted in swelling of the PE gel.

    \item \textbf{48--72 hours (swelling):} To simulate deswelling, in this time period, we prescribed and held the boundary conditions for $\mu^w$, $\omega^{\mathrm{Na}^+}$, and $\omega^{\mathrm{Cl}^-}$ computed by substituting the ion concentrations corresponding to $I = 0.05$ M, from \eqref{eq:sun-concentration-values} in \eqref{eq:echem-sol}. Similar to the previous application of the boundary conditions, we utilized the smooth amplitude feature in Abaqus and applied the boundary conditions over $t_{\mathrm{ramp}} = 180$ and held there for 24 hours, which resulted in swelling of the cationic PE gel.

    \item \textbf{72--120 hours (deswelling and swelling):} We then repeated the same procedure for prescribing boundary conditions from the 24--48 hours time period to simulate the second deswelling phase, followed by the same procedure as the time period 48--72 hours to simulate the second swelling phase.

\end{itemize}

\begin{figure}[ht]
\begin{center}
    \includegraphics[width=0.6\textwidth, trim={7cm 5.25cm 7cm 5.25cm}, clip] {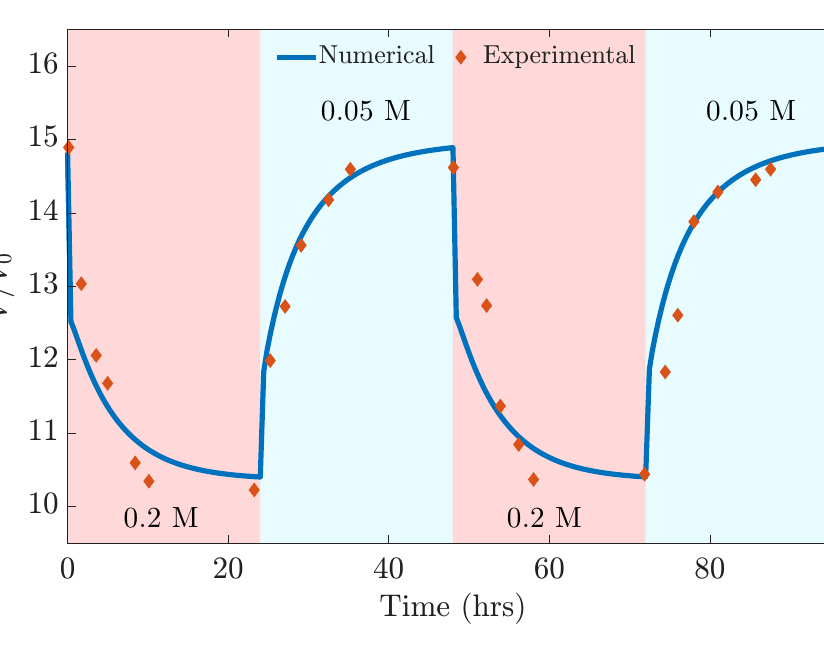}
    \caption{Comparison of volumetric swelling ratio between finite element analyses and experimental study of immersing cationic polyelectrolyte gel alternatively in high ($I=0.2$ M) and low ($I=0.05$ M) ionic strength solutions. For a direct comparison with experimental results, we omitted the transient swelling response of the gel in the first 24 hours in the plot.}
    \label{fig:sun-cylic-comparison}
\end{center}
\end{figure}

The volumetric swelling ratio from simulations and experiments is plotted in Figure \ref{fig:sun-cylic-comparison}. The model predictions are in reasonable agreement with the experimental results reported by the authors. At the beginning, as the gel was immersed in the salt solution with high ionic strength ($I=0.2$ M) from its previously swollen state, the ionic pressure, which is also known as the Donnan osmotic pressure, dropped, causing the gel to release water and subsequently deswell. Subsequently, as the gel was transferred back into the salt solution with low ionic strength ($I=0.05$ M), an increase in ionic pressure caused the gel to absorb water from the bath and swell accordingly. Throughout our simulation, we assumed the concentration of the polymer fixed charge to be constant; however, in reality, the presence of the OH$^-$ ion alters the degree of ionization of the fixed charge. Neglecting this effect contributed to the discrepancy between the experimental measurement and the simulation results.

\begin{figure}[ht]
\begin{center}
    \includegraphics[width=0.6\textwidth, trim={7cm 5.25cm 7cm 5cm}, clip] {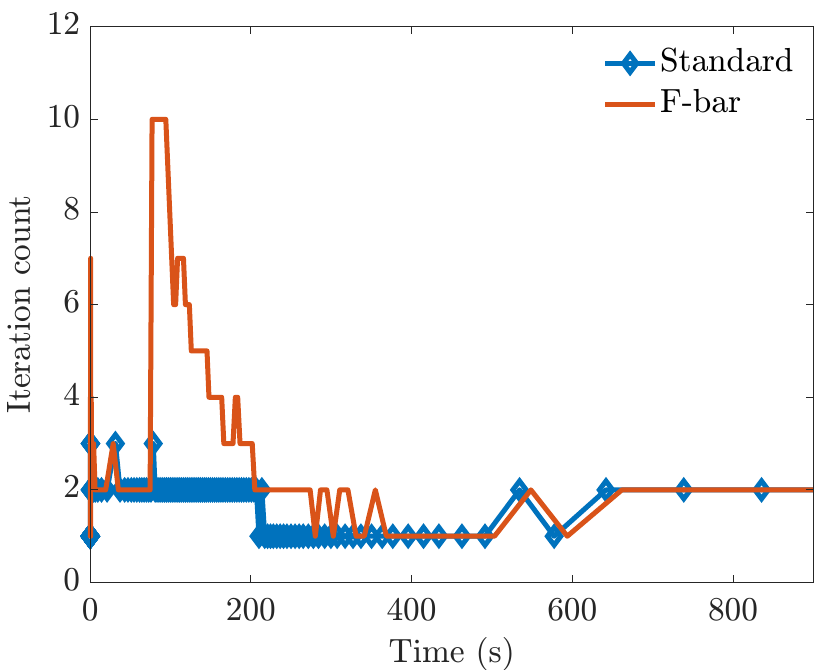}
    \caption{Number of global Newton iterations required to obtain the equilibrium solution for the standard and F-bar PE hydrogel element in the first 900 s of the swelling phase of the first cycle.}
    \label{fig:iteration-counts}
\end{center}
\end{figure}

We performed the cyclic deswelling-swelling simulation using both the standard and F-bar element formulations and obtained identical solutions. Figure \ref{fig:iteration-counts} shows the number of iterations required to obtain the equilibrium solution for the first 900 s of the swelling phase of the first cycle. We can observe, except for a few time step increments, that the standard element formulation reached equilibrium within one or two iterations. On the other hand, the F-bar element required a larger number of iterations in the first 180 s when the boundary conditions were gradually applied to the gel. This trend is persistent across the phases of the second cycle; hence, we did not include them here. For the F-bar element, we only modified the mechanical residual and associated tangent stiffness matrices with F-bar, keeping other residuals and tangent stiffness matrices the same as the standard element formulation. This approximation of the residual and tangent stiffness matrices resulted in a poorer convergence rate for the F-bar element formulation. To understand the convergence of the standard PE gel element, we examined the absolute maximum residual force reported by Abaqus at different time points in the early stage of the swelling phase of the first cycle for the standard element formulation in Table \ref{tab:residual-iteration}. At these reported time points, Abaqus increased the time step size, $\Delta t$, and hence required more than two iterations to achieve convergence. After the first 3600 s, the standard element formulation required a single iteration to obtain convergence until the boundary conditions were changed for deswelling. This reflects the robustness of the proposed finite element formulation.

\begin{table}[ht]
\centering
\begin{tabular}{c c c c c c}
    \toprule \toprule
    \textbf{Iterations}     &   $t=0.35$ s      &  $t=0.60$ s     &  $t=31.53$ s  & $t=125.7$ s \\
    \toprule  \toprule
    1      & $1.186724 \times 10^{-8}$  & $6.604551 \times 10^{-9}$     & $2.382792 \times 10^{-3}$ & $2.870839 \times 10^{-5}$   \\
    2      & $2.679409 \times 10^{-9}$  & $3.483787 \times 10^{-9}$     & $5.124843 \times 10^{-6}$ & $6.544267 \times 10^{-8}$   \\
    3      & $1.541962 \times 10^{-14}$ & $1.275113 \times 10^{-14}$    & $3.742290 \times 10^{-10}$ & --     \\
    \bottomrule
\end{tabular}
\caption{Absolute maximum force residual reported by Abaqus/Standard solver at different time points in the swelling phase of the first cycle for standard PE gel element formulation.}
\label{tab:residual-iteration}
\end{table}

\subsection{Transient bending of PE gel-elastomer bilayer beam}
\label{sec:bilayer-bending}

\begin{figure}[ht]
\begin{center}
    \includegraphics[width=\textwidth, trim={0.5cm 7.75cm 0.5cm 7.75cm}, clip] {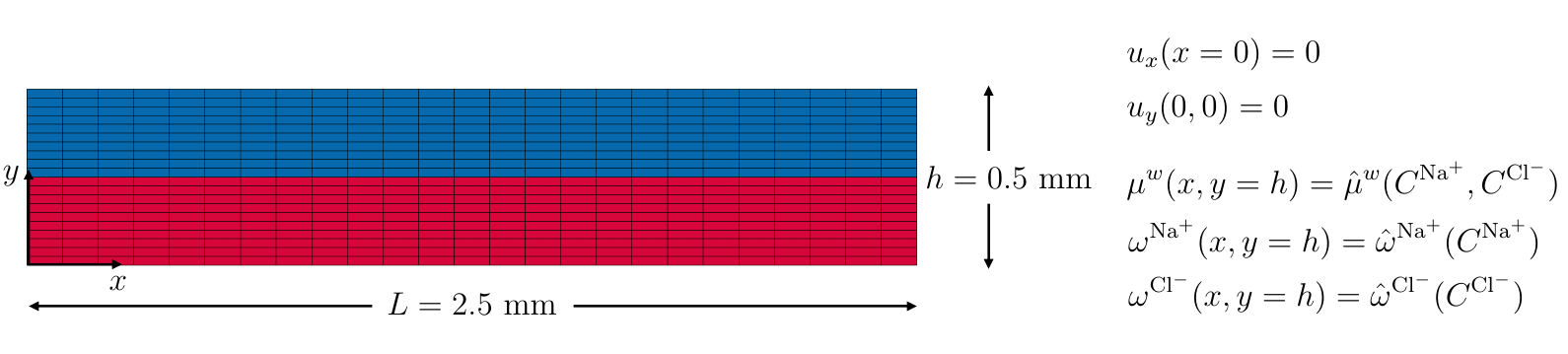}
    \caption{Plane strain finite element model of bilayer actuator. The top half of the domain (blue) represents the polyelectrolyte gel, and the bottom half of the domain (red) represents the elastomeric rubber substrate.}
    \label{fig:bilayer-fem-model}
\end{center}
\end{figure}

One of the fascinating applications of PE gels is in the field of soft robotics and actuators. The most common building block of such actuators is gel-gel or gel-elastomer bilayers \citep{duanBilayerHydrogelActuators2017,xiaoSaltresponsiveBilayerHydrogels2017,heOnepotOnestepFabrication2019}. In this example, we studied the transient bending characteristics of PE gel-rubber bilayer structures. Figure \ref{fig:bilayer-fem-model} shows the geometry, mesh, and boundary conditions applied to the baseline case of the bilayer structure. The bilayer has a length of $L=2.5$ mm and a height of $h=0.5$ mm. The bottom substrate (red) is an elastomeric rubber and was modeled using the standard Neo-Hookean material in Abaqus/Standard with CPE4H elements \citep{dassaultsystemesSIMULIAUserAssistance2023a}. The top half of the bilayer structure (blue) is the as-prepared cationic PE gel, which was modeled using the plane-strain elements we developed for PE gels in this work. We discretized the whole domain using uniform 4-node quadrilateral elements having dimensions 0.1 mm $\times$ 0.025 mm.

We considered the PE gel to be pre-swollen with initial polymer volume fraction, $\phi^p_0 = 0.1$, at its as-prepared state, which is reasonable considering hydrogels are often synthesized at 10 wt\% polymer formulation. We kept most of the properties the same as listed in Table \ref{tab:sun_pe_gel_props}. A list of properties that are different from our previous analyses is given in Table \ref{tab:bilayer_props}. Since the initial polymer volume fraction, $\phi^p_0$, is different in this case, we adjusted the shear modulus of the PE gel following the scaling law $G \sim (\phi^p_0)^{1/3}$ from \eqref{eq:free-energy-neo-hookean-stat-mech} as well as the initial concentration of fixed charge within the PE gel by scaling it with the hydrated reference state volume corresponding to $\phi^p_0 = 0.1$. For the elastomeric Neo-Hookean substrate, we chose the shear moduli, $G_{\mathrm{sub}}$, to be twice that of the PE gel, which is approximately the same as if the PE gel were dry. Additionally, to ensure quasi-incompressible behavior of the rubber substrate, we set the bulk modulus to be 100 times larger than the shear modulus. For the baseline case, we immersed the bilayer in a NaCl solution with ionic strength $I=0.7$ M, which is similar to the ionic strength of seawater. Additionally, immersing a highly pre-swollen gel with $\phi^p_0 = 0.1$ in a moderately saline solution ($I=0.7$ M) causes sufficient bi-directional bending of the bilayer structure for us to understand the effect of the material and geometric parameters. If the same PE gel were immersed in a low-salinity solution, such as $I=0.05$ M, it would have caused excessive swelling and bending for us to develop a thorough understanding.

To perform the swelling-induced bending simulations, we fixed the horizontal displacement along the left edge by setting $u_x(x=0,y)$. We additionally fixed the geometric origin of the computational model along the vertical direction as well, \emph{i.e.}, $u_y(0,0) = 0$, to avoid rigid body motion. These displacement boundary conditions were maintained throughout the simulation. For the baseline case ($C_0^{\mathrm{fix}} = 150$ mol/m$^3$), we computed the initial values for ion electrochemical potential using Eqs. \eqref {eq:chem-pot-def} and \eqref {eq:echem-pot-def} by assuming the initial concentration of the Na$^+$ coion, $C_0^{\mathrm{Na}^+} = 50$ mol/m$^3$, and Cl$^-$ counterion, $C_0^{\mathrm{Cl}^-} = 200$ mol/m$^3$ to ensure electroneutrality. We further assumed the initial electric potential, $\psi = 0$, and unlike the previous study, we did not recompute it before our simulation with the so-called equilibrium step. Since the equilibration step introduces a small amount of curvature, which depends on the geometry and parameters of the bilayer structure, we intentionally decided to start from a no-curvature state by eliminating the equilibrium step to make a clear comparison in the parametric studies.

We computed solvent chemical potential, $\mu^w$, and ion electrochemical potentials, $\omega^{\mathrm{Na}^+}$ and $\omega^{\mathrm{Cl}^-}$, by substituting the ion concentrations corresponding to the ionic strength, $I = 0.7$ M in Eq. \eqref {eq:echem-sol}, and then subsequently we ramped the solvent chemical potential, $\mu^w(x,y=h)$, and the ion electrochemical potentials, $\omega^{\mathrm{Na}^+}(x,y=h)$ and $\omega^{\mathrm{Cl}^-}(x,y=h)$, along the top edge from their initial values to the computed values over $t_{\mathrm{ramp}} = 180$ s. Because of the aspect ratio of the bilayer beam, the vertical direction is the shortest path for diffusion; thus, we chose to apply the chemical boundary conditions only along the top edge. For this study, we set the initial time step, $t_{\mathrm{init}} = 10^{-6}$ s, minimum time step, $\Delta t_{\mathrm{min}} = 10^{-15}$ s, maximum time step, $\Delta t_{\mathrm{max}} = 300$ s, and the total simulation time, $t_{\mathrm{total}} = 6$ hours\footnote{The Abaqus input file for the baseline case, \texttt{Bilayer2D\_5x0\_5mm\_mesh\_0\_1x0\_025mm\_700mM.inp}. under the \texttt{test/bilayer\_bending} directory contains the details required to perform the simulation.}.

\begin{table}[ht]
\centering
\begin{tabular}{c c}
    \toprule \toprule
    \textbf{Properties}                                         & \textbf{Value} \\
    \toprule  \toprule
    Initial polymer volume fraction, $\phi^p_0$                 & 0.10          \\
    Shear modulus at reference state, $G$                       & 33 kPa        \\
    Initial fixed charge of polymer, $C_0^{\mathrm{fix}}$       & 150 mol/m$^3$ \\
    Flory-Huggins interaction parameter, $\chi$                 &  0.495        \\
    \midrule
    Shear modulus of rubber substrate, $G_{\mathrm{sub}}$       & 66 kPa        \\
    Bulk modulus of rubber substrate, $\kappa_{\mathrm{sub}}$   & 100 $G_{\mathrm{sub}}$ \\
    \bottomrule
\end{tabular}
\caption{List of properties that are different from the previous studies for the baseline case. The rest of the required properties are used as listed in Table \ref{tab:sun_pe_gel_props}.}
\label{tab:bilayer_props}
\end{table}

\begin{figure}[ht]
\begin{center}
    \includegraphics[width=\textwidth, trim={2cm 7.5cm 1cm 5.5cm}, clip] {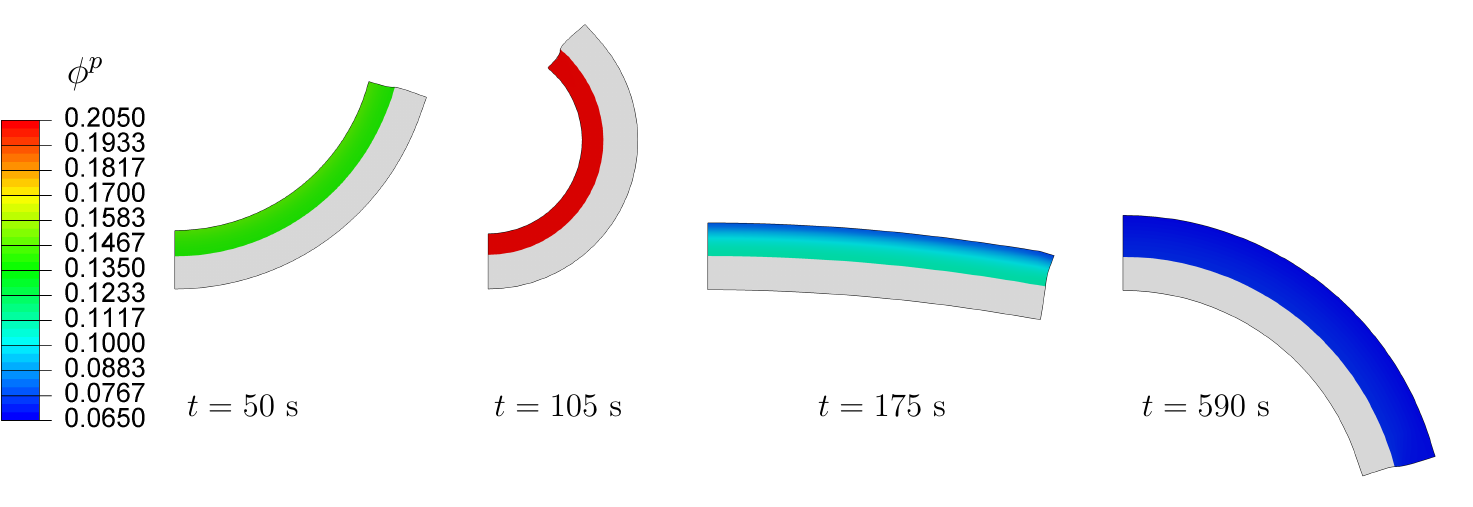}
    \caption{Time series snapshots of the evolution of polymer volume fraction in the PE gel layer. The gray color of the substrate represents zero water permeation in the substrate.}
    \label{fig:bilayer-phi-timeshot}
\end{center}
\end{figure}

Figure \ref{fig:bilayer-phi-timeshot} shows the evolution of the polymer volume fraction contour in the first 10 minutes of the bilayer being immersed in the solution with NaCl concentration of 0.7 M. Contact of the bilayer with the salt solution caused water to diffuse out since the chemical potential of the as-prepared gel was higher than the external solution. This led the gel to deswell to a higher polymer volume fraction than its initial state and bend upward because of the strain mismatch. After a few seconds, the ions started diffusing into the gel from the solution as the electrochemical potentials of the ions were higher in the solution than in the gel. The flow of ions dragged water into the gel, causing the gel to swell and bend downward. This transient bending behavior of the bilayer is primarily attributed to the larger diffusion coefficient of water compared to the ions \citep{zhangKineticsPolyelectrolyteGels2020}.

Design of a bilayer actuator based on PE gel will require careful consideration of this transient bending behavior. We hypothesized that the ionic strength of the solution bath, fixed charge, and Flory-Huggins parameter can affect the behavior of a PE gel-based bilayer structure. Additionally, the extension of the Stoney formula for bilayer structures suggests that the bending curvature depends on the moduli ratio and the thickness ratio of the structure \citep{freundExtensionsStoneyFormula1999}. Hence, to gain further insight into the transient bending of the bilayer structure, we performed the following studies by varying our baseline case:
\begin{enumerate} [label=(\alph*), leftmargin=0.9cm, topsep=0pt, itemsep=-6pt]
    \item ionic strength of the external solution was varied as 0.05 M, 0.2 M, and 0.5 M,
    \item initial polymer fixed charge was varied as 0.3 M, 0.5 M, and 0.7 M\footnote{For this study, we assumed $C_0^{\mathrm{Na}^{+}} = 100, 150, \ \text{and} \ 200$ mol/m$^3$, and $C_0^{\mathrm{Cl}^{-}} = 400, 650, \ \text{and} \ 1000$ mol/m$^3$, for the listed cases, respectively to compute the initial ion electrochemical potentials. These values for the coion and counterion concentrations were chosen to maintain electroneutrality.},
    \item Flory-Huggins interaction parameter was varied as 0.3, 0.4, and 0.6,
    \item gel to substrate moduli ratio, $G_{\mathrm{gel}}/G_\mathrm{sub}$, was varied as 1, 0.25, and 0.125,
    \item gel to substrate thickness ratio, $h_{\mathrm{gel}}/h_\mathrm{sub}$, was varied  as 2, 0.5, and 0.25.
\end{enumerate}

\begin{figure}[htp]
\begin{center}
    \begin{subfigure}{0.96\textwidth}
    \includegraphics[width=\textwidth, trim={1cm 5.25cm 1cm 5.5cm}, clip] {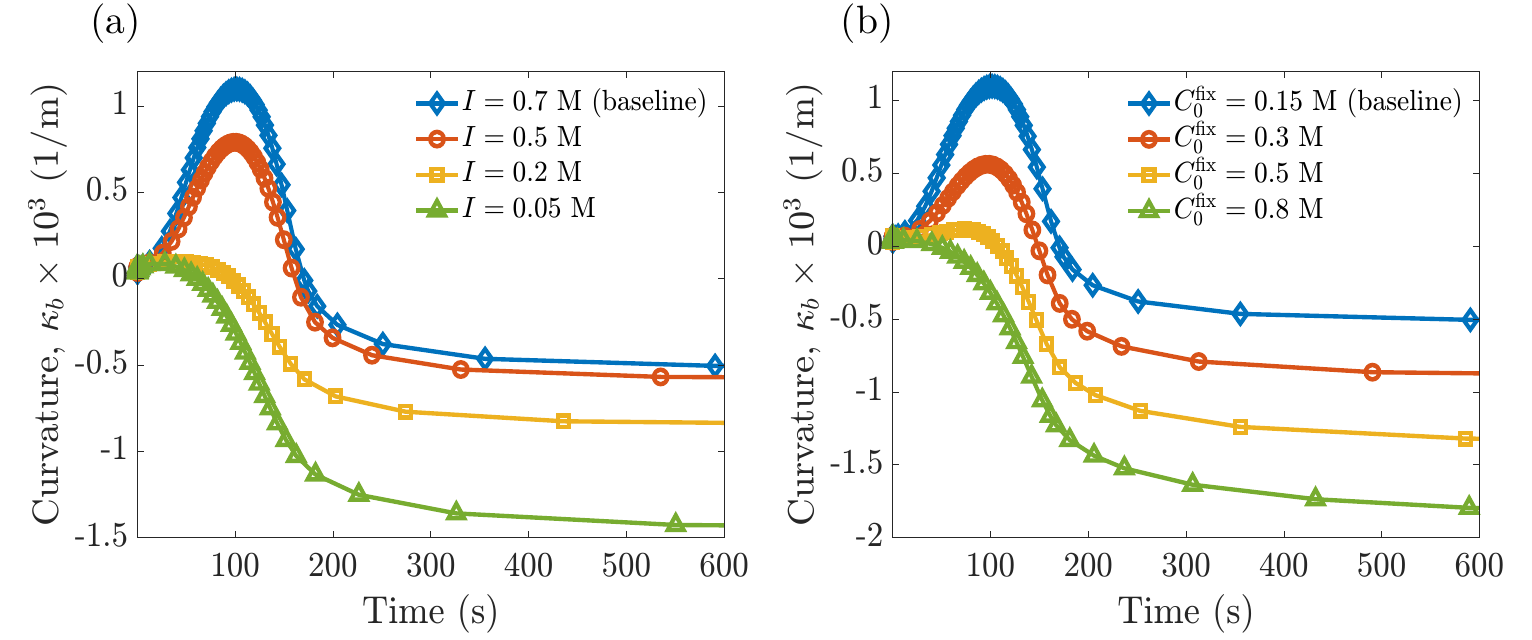}
    \end{subfigure}
    
    \smallskip
    \begin{subfigure}{0.96\textwidth}
    \includegraphics[width=\textwidth, trim={1cm 5.25cm 1cm 5.5cm}, clip] {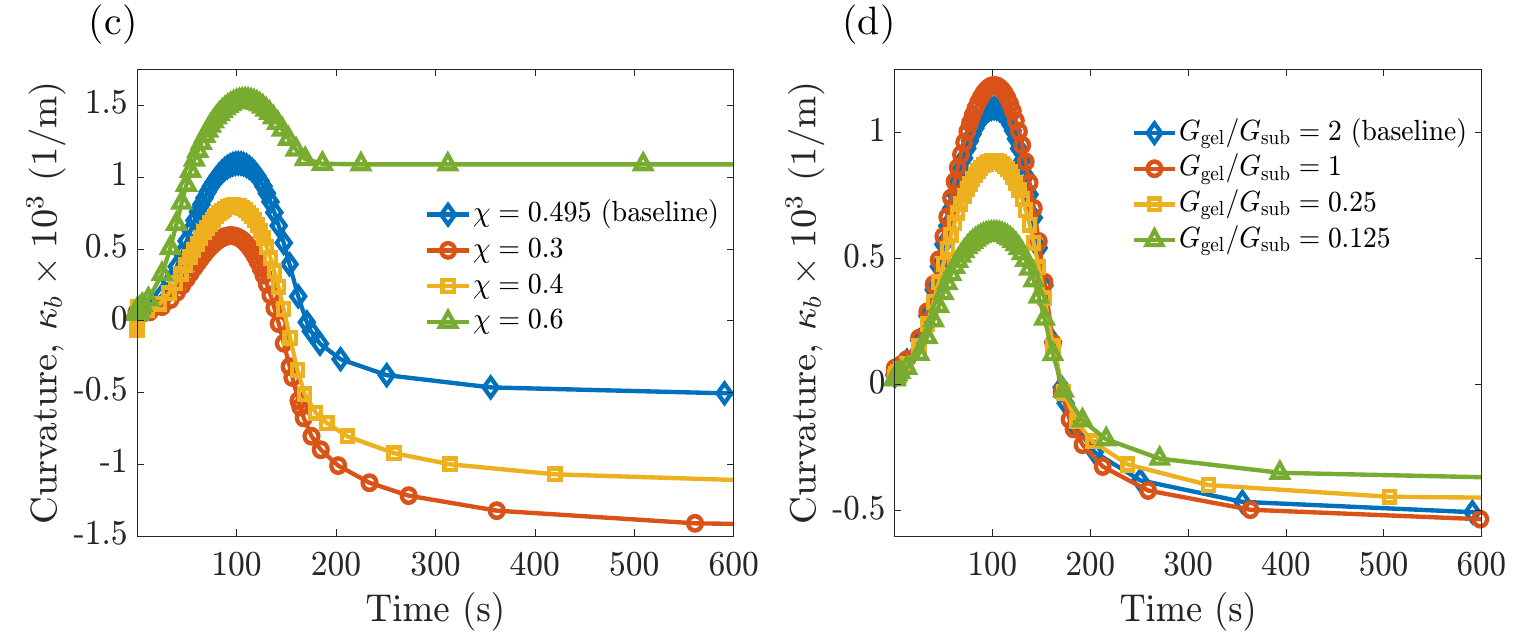}
    \end{subfigure}

    \smallskip
    \begin{subfigure}{0.48\textwidth}
    \includegraphics[width=\textwidth, trim={7.25cm 5.25cm 7.75cm 5.5cm}, clip] {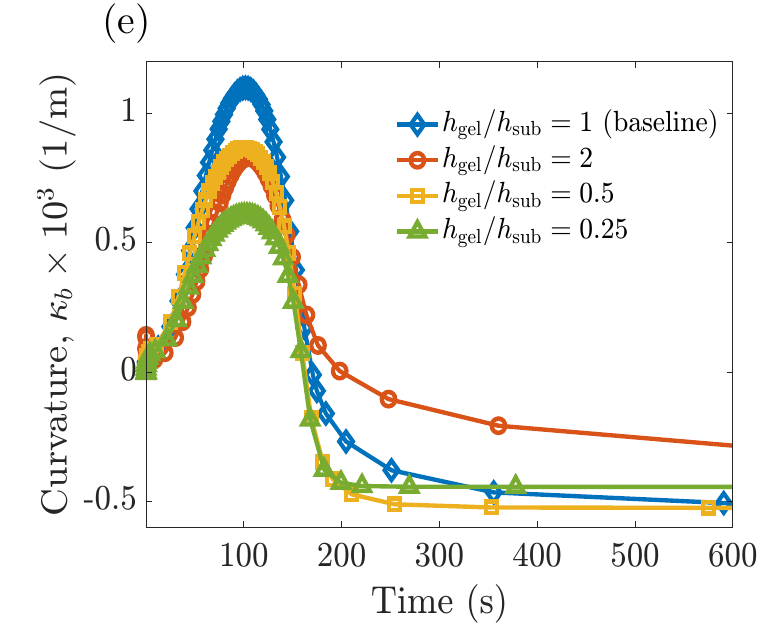}
    \end{subfigure}
    
    \caption{Temporal evolution of the bending curvature of the gel-elastomer bilayer under parametric variations: (a) external solution concentration, (b) polymer fixed charge density, (c) Flory–Huggins interaction parameter, (d) modulus ratio between gel and substrate, and (e) thickness ratio between gel and substrate.}
    \label{fig:bilayer-parmetric study}
\end{center}
\end{figure}

The result of these parametric studies is presented in Figure \ref{fig:bilayer-parmetric study} where each subfigure panel corresponds to the cases listed above. We plotted the bending curvature of the bilayer over the first 600 s. The curvature was calculated at the bottom surface $(y=0)$ of the elastomeric substrate using a second-order accurate ($\mathcal{O} (h^2)$) forward finite difference approach. From Figure \ref{fig:bilayer-parmetric study}, it is evident that, except for the case with $\chi=0.6$ in Figure \ref{fig:bilayer-parmetric study}(c), we observed a positive-to-negative curvature transition. A larger Flory-Huggins interaction parameter, $\chi$, represents an unfavorable mixture between the polymer and the solvent. Hence, the gel deswelled from its initial state by diffusing water out and bent in the upward direction without any transition afterwards. For most cases, the positive-to-negative curvature transition time was between 100 and 200 s, except for the case with lower ionic strength in Figure \ref{fig:bilayer-parmetric study}(a) and high fixed charges in Figure \ref{fig:bilayer-parmetric study}(b). For these cases, the differences in electrochemical potentials between the ions inside the gel and the external solvent were sufficiently small, resulting in almost instantaneous balancing out and curvature transition. A quick observation of Figure \ref{fig:bilayer-parmetric study} reveals that a lower ionic strength of the external solution bath, a higher concentration of fixed charge, and lower Flory-Huggins interaction parameters have a prominent effect on the equilibrium bending curvature, whereas the gel-to-substrate thickness ratio and gel-to-substrate moduli ratio have a small effect on the equilibrium bending curvature in the range of parameters we studied.

\subsection{Consolidation during confined compression}   
\label{sec:gel-consolidation-study}

\begin{figure}[ht]
\begin{center}
    \includegraphics[width=\textwidth, trim={2.5cm 6cm 2.5cm 6cm}, clip] {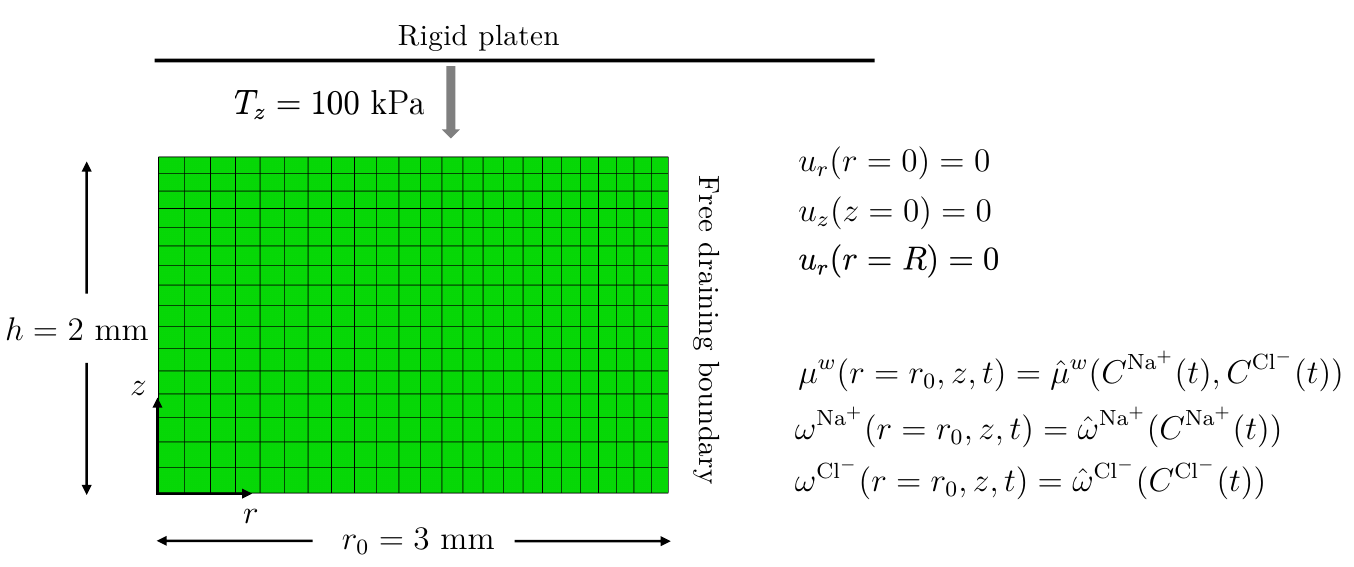}
    \caption{Axisymmetric finite element model of the polyelectrolyte gel disk showing mesh and boundary conditions for the consolidation study under confinement. Boundary conditions for the solvent chemical potential and ion electrochemical potentials were computed using Eq. \eqref {eq:echem-sol}. Confinement boundary condition, $u_r(r=R) = 0$, was applied at the swollen radial position, $R$, of the PE gel.}
    \label{fig:consolidation-disk-model}
\end{center}
\end{figure}

There exists a significant amount of interest among researchers to use PE gels as a replacement for diseased and damaged tissue, such as articular cartilage \citep{yangSyntheticHydrogelComposite2020,hashemi-afzalAdvancementsHydrogelDesign2025,azumaBiodegradationHightoughnessDouble2007}. In those applications, hydrogels function as a load-bearing matrix and exude fluid during compression. In this example, we considered the same PE gel from our previous case, \emph{i.e.}, the cationic PE gel with the material properties listed in Table \ref{tab:sun_pe_gel_props}. By exploiting the symmetry condition, we modeled one half of an as-prepared cylindrical disk of 2 mm height and 6 mm diameter with axisymmetric elements as shown in Figure \ref{fig:consolidation-disk-model}. This aspect ratio prevents buckling under compressive loading. The smallest element size was 0.1 mm $\times$ 0.1 mm, whereas the largest element was 0.15 mm $\times$ 0.15 mm in our model.

The simulations proceeded as follows:
\begin{itemize} [leftmargin=0.5cm, topsep=0pt, itemsep=-3pt]

    \item \textbf{Step 1:} Similar to the first example, the first step was performed to compute the algorithmic internal variables within the gel to ensure equilibrium has been established with its prescribed initial conditions. As shown in Figure \ref{fig:consolidation-disk-model}, we fixed the radial displacement at the left edge, $u_r(r=0,z) = 0$, and axial displacement at the bottom edge, $u_z(r,z=0) = 0$, as the displacement boundary conditions. Since the initial concentration of the polymer fixed charge was taken to be $C_0^{\mathrm{fix}} = 460 $ mol/m$^3$, we assumed the coion concentration, $C^{\mathrm{Na}^+} = 340$ mol/m$^3$, counterion concentration, $C^{\mathrm{Cl}^-} = 800$ mol/m$^3$, initial polymer volume fraction, $\phi^p_0 = 0.312$, and electric potential, $\psi = 0$, to ensure electroneutrality. These values were substituted in the expressions for solvent chemical potential in Eq. \eqref {eq:chem-pot-def} and ion electrochemical potentials in Eq. \eqref{eq:echem-pot-def} to obtain the initial values for $\mu^w(r,z)$, $\omega^{\mathrm{Na}^+}(r,z)$, and $\omega^{\mathrm{Cl}^-}(r,z)$ which was then prescribed these initial values to as uniform fields throughout the computational domain to compute the electric potential, $\psi$. Performing this step ensures the unknown internal variables are initialized accurately to the swollen reference state and eliminates the need for smaller time steps to ensure convergence.

    \item \textbf{Step 2:} Once the PE gel was equilibrated to its prescribed initial conditions, we performed equilibrium free-swelling analyses by immersing the PE gel in external solution baths of different ionic strengths. We chose external solution baths of three different ionic strengths: $I=0.05$ M, $I=0.15$ M, and $I=0.25$ M. Based on these chosen ionic strengths, we first computed the concentrations of the coion and counterion in the external solution bath using Eq. \eqref {eq:ionic-strength}. We then substituted these ionic concentrations in Eq. \eqref {eq:echem-sol} to compute solvent chemical potential, $\mu^w(r,z)$, and ion electrochemical potentials, $\omega^{\mathrm{Na}^+}(r,z)$ and $\omega^{\mathrm{Cl}^-}(r,z)$. These values were then applied as uniform fields throughout the computational domain of the PE gel. At the end of this step, the PE gel swelled to different extents depending on the ionic strength of the external solution bath.

    \item \textbf{Step 3:} In this step, we radially confined the PE gel and applied compressive traction on the top surface of the swollen PE gel while allowing free drainage of the solvent and ions through the boundary at $r=r_0$. To ensure the drainage of the solvent and the ions, we maintained the same numerical values at the swollen outer edge from the previous step for the solvent chemical potential, $\mu^w(r=R,z,t)$, and ion electrochemical potentials, $\omega^{\mathrm{Na}^+}(r=R,z,t)$ and $\omega^{\mathrm{Cl}^-}(r=R,z,t)$. To realize the confinement, we fixed the radial displacement at the swollen outer edge, \emph{i.e.}, we prescribed $u_r(r=R) = 0$ as the displacement boundary condition. We placed a rigid platen in contact with the top surface of the swollen PE gel to perform compression. Using the rigid platen, we applied an engineering traction, $T_{z} = 100$ kPa, with respect to the swollen configuration, on the top surface ($z=h$) over 10 s and maintained the traction load for a total of 12 hours. We set the initial time step, $\Delta t_{\mathrm{init}} = 10^{-6}$ s, minimum time step, $\Delta t_{\mathrm{min}} = 10^{-15}$ s, and the maximum time step $\Delta t_{\mathrm{max}} = 300 $ s. We chose a smaller initial time step and maximum time step so that we can capture the compression characteristic of the gel at a finer timescale, as well as to tackle the potential convergence issues related to contact.

\end{itemize}

\begin{figure}[t!]

    \centering
    \begin{subfigure}[t]{0.49\textwidth}
        \centering
        \includegraphics[width=\textwidth, trim={8.25cm 3.5cm 8cm 3.5cm}, clip] {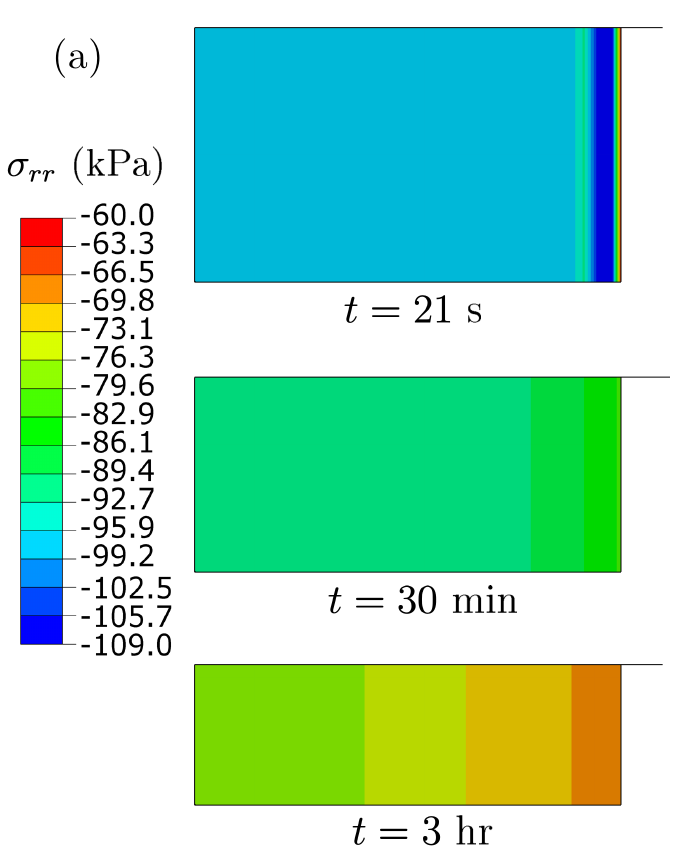}
    \end{subfigure}%
    ~ 
    \begin{subfigure}[t]{0.49\textwidth}
        \centering
        \includegraphics[width=\textwidth, trim={8.25cm 3.5cm 8cm 3.5cm},clip] {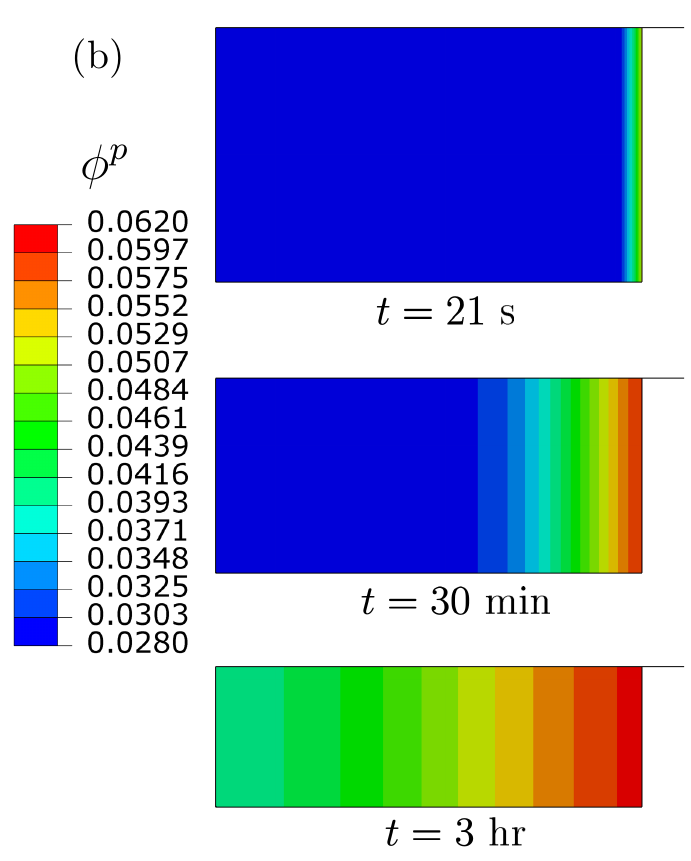}
    \end{subfigure}
    
    \caption{Time series snapshots of polyelectrolyte gel under compression loading: (a) contour plot of distribution of radial Cauchy stress, $\sigma_{rr}$, and (b) contour plot of polymer volume fraction. In the first 10 s, the gel was compressed axially, and then the load was held constant, causing the gel to shrink from the loss of solvent and ions.}
    \label{fig:consolidation-contour-plot}
    
\end{figure}

Using the above procedure, first, we performed a finite element analysis of our baseline case ($I=0.15$ M). Figure \ref{fig:consolidation-contour-plot}(a) shows the radial Cauchy stress contour, and Figure \ref{fig:consolidation-contour-plot}(b) shows the time evolution of polymer volume fraction at different stages during the confined compression. In the first 10 s, as the traction is being ramped to $T_{z}$ = 100 kPa, the gel exhibits rapid axial shrinkage. At $t=10$ s, when the compressive load has settled, we can observe the development of high compressive stress near the free-draining boundary (see Figure \ref{fig:consolidation-contour-plot}(a)). To release the compressive stress, the gel started exuding solvent and ions immediately, resulting in a higher polymer volume fraction near the free-drainage boundary compared to the rest of the PE gel (see the snapshot at $t=21$ s in Figure \ref{fig:consolidation-contour-plot}(b)). As the load remains steady on the gel for 12 hours, to minimize the compressive stress, it continues to exude more solvent and ions over time through its free-drainage boundary. We can observe from the contour plots presented in Figure \ref{fig:consolidation-contour-plot}(a) that the compressive stress is slowly being relaxed at $t=30$ min and $t=3$ hr. This exudation-based stress relaxation results in an increase of polymer volume fraction and a volume shrinkage of the gel as the gel slowly moves toward equilibrium, which can be seen in the snapshots at $t=30$ min and $t=3$ hr in Figure \ref{fig:consolidation-contour-plot}(b). Since the process of solvent and ion leakage is diffusion-based, the rate of deformation in the consolidation stage is slower compared to the early-stage compression. Similar phenomena have been observed in the consolidation of soil \citep{muradMicromechanicalComputationalModeling2001} and non-ionic hydrogels \citep{klugeConsolidationBehaviorSilk2010}.

\begin{figure}[ht]
\begin{center}
    \includegraphics[width=\textwidth, trim={0.5cm 5.5cm 1cm 4.75cm}, clip] {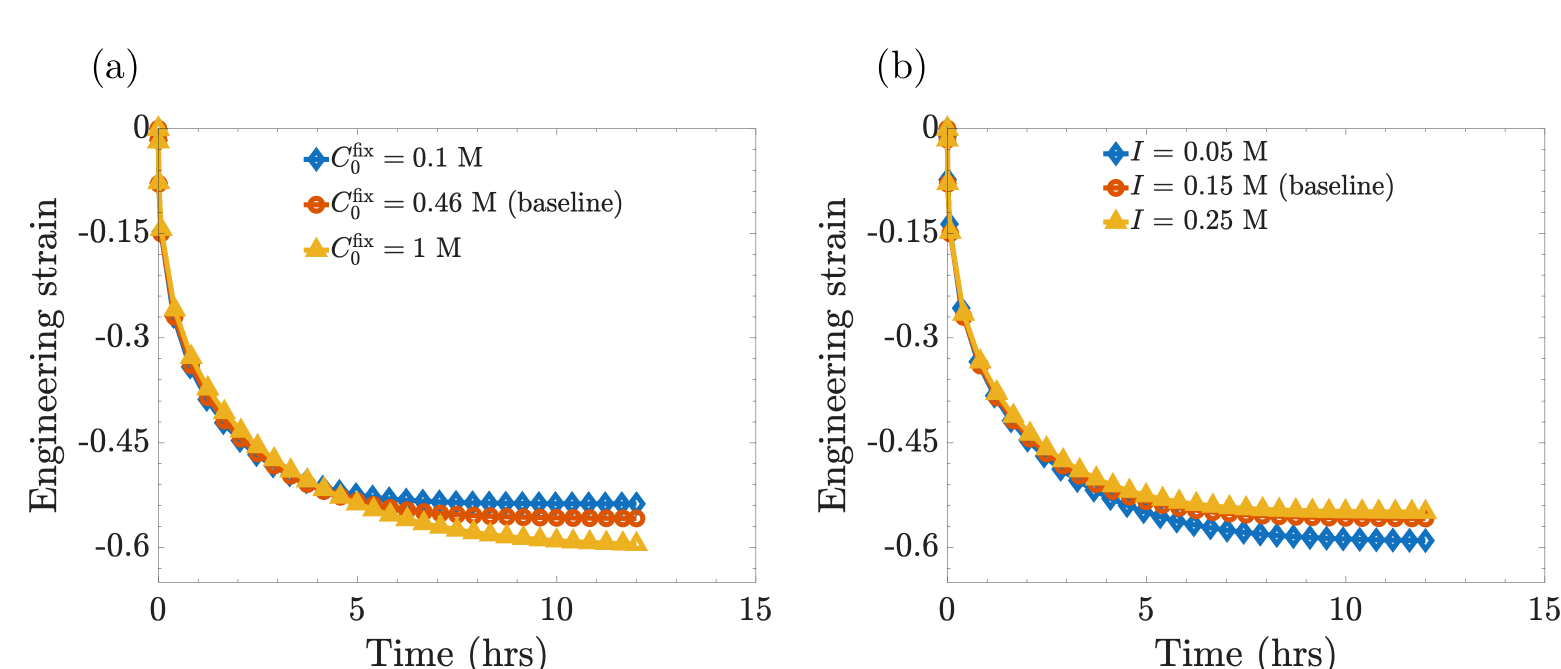}
    \caption{Consolidation curve of polyelectrolyte gel under a constant engineering traction, $T_{z} = 100$ kPa (a) in NaCl solution with different ionic strengths with fixed charge $C_0^{\mathrm{fix}}=460$ mol/m$^3$ and (b) for different initial polymer fixed charge concentrations with ionic strength of the external solution, $I=0.15$ M.}
    \label{fig:consolidation-curve}
\end{center}
\end{figure}

Next, to understand the effect of the ionic strength of the external solution bath and initial polymer fixed charge, we repeated the simulations for (a) two different ionic strengths (0.05 M, 0.25 M) while keeping the fixed charge concentration 460 mol/m$^3$ and (b) two different initial fixed charge concentrations (100 mol/m$^3$, 1000 mol/m$^3$) while maintaining the ionic strength of NaCl solution same as the baseline case, \emph{i.e.}, $I=0.15$ M. The consolidation curves for the PE gel with different initial fixed charges and immersed in external solutions of different ionic strengths are plotted in Figure \ref{fig:consolidation-curve}(a) and (b), respectively. Under the prescribed loading condition, compressive engineering strain (with respect to the swollen height) is in the range of 54\%--60\%. For the highest initial fixed charge concentration (Figure \ref{fig:consolidation-curve}(a)) and the external solution with the lowest ionic strength (Figure \ref{fig:consolidation-curve}(b)), we can observe the maximum equilibrium strain at the end of the consolidation stage in the respective studies. This can be attributed to the fact that those were the two most swollen gels in their respective categories and had the lowest shear modulus, which scales following $G \sim (\phi^p)^{1/3}$. For both extreme cases, the gel had the highest amount of swelling before compression, resulting in softer behavior, leading to a higher amount of strain. On the opposite end, the gels that were less swollen and, therefore, stiffer before compression, \emph{i.e.}, PE gel with the lowest initial fixed charge concentration and immersed in the highest ionic strength solution, exhibit the smallest equilibrium strain in their groups. 

As mentioned previously, for all of these cases, ions diffuse out of the gel across the free-draining boundary alongside the solvent. However, with appropriate boundary conditions, our framework can be used to model leakage-free PE gels, in which ions are contained and redistribute themselves within the gel when any pressure is applied, for sensing applications. However, these findings are in qualitative agreement with those reported for cartilage \citep{ateshianMultiphasicFiniteElement2013}.

\section{Concluding remarks}
\label{sec:conclusions}

In this work, we presented an efficient and robust finite element framework for a transient coupled electro-chemo-mechanical theory for PE gel. As opposed to the currently available theories and computational models, our treatment of PE gel as an electroneutral medium and proposed kinematics and free energy density functions at the pre-swollen state resulted in an efficient and robust framework. In particular, we validated the proposed finite element framework for the free swelling-deswelling behaviors of cationic PE gel based on the experiments performed by \cite{sunMultiresponsiveToughHydrogels2015}. While the focus of this article was not on developing constitutive models for PE gel, we should admit that the simplified flux laws we utilized in this work, as well as the cross-diffusion-based flux laws proposed by \cite{zhangKineticsPolyelectrolyteGels2020} and \cite{narayanCoupledElectrochemomechanicalTheory2022}, required overestimating the solvent diffusion coefficient to obtain reasonable agreement with free swelling-deswelling experiments. Our numerical simulation highlights the importance of performing thorough experimental studies of the solvent and solute kinetics within a PE gel by varying the initial polymer volume fraction, initial fixed charge concentration, and ionic strength of the external solution. These experimental findings will aid in developing a physically-based constitutive model for solvent diffusion in a polyelectrolyte gel system. Our future work will address this in detail.

Despite the limitations, we demonstrated the compressive behavior of PE gels in dilute salt solution and the application of PE gel-based structures as bending actuators in the salt baths. For both cases, we studied the effect of different parameters to gain a better understanding of the physical mechanisms by leveraging our proposed numerical framework, which would have been difficult to perform analytically and experimentally. Although a broader class of PE gels is pH-responsive, we did not demonstrate any examples of this category in this work. Modeling pH-responsive PE gel requires modifications of the constitutive model, such as introducing the degree of ionization to compute the variation of polymer fixed charge, and our future work will include studies of pH-responsive PE gels along with the implementation procedure. Taking the intricacy of the theory and finite element procedure into account, we explicitly listed all the steps with substantial details on how to perform multiphysics analyses of PE gels. We also made the source code of our implementation available to interested readers so that they can either directly adopt and extend the work with new constitutive models to describe rather complicated phenomena for PE gels and other materials of the same category. The supplementary information will guide the readers on how to set up simulations using our framework and extend the implementation. We hope this work helps close the gap between experiments and modeling of PE gels and provides a tool to study important phenomena such as surface instability, electro-mechanical actuation, and electroosmosis, with potential engineering applications in various industries and academic research.

\section*{CRediT authorship contribution statement}

\textbf{Bibekananda Datta:} Conceptualization, Methodology, Software, Data curation, Visualization, Formal analysis,  Validation, Investigation, Writing – original draft.
\textbf{Brandon K. Zimmerman:} Validation, Writing - Review \& Editing.
\textbf{Thao D. Nguyen:} Methodology, Resources, Supervision, Project administration, Funding acquisition, Writing - Review \& Editing.

\section*{Declaration of competing interest}

The authors declare that they have no known competing financial interests or personal relationships that could have appeared to influence the work reported in this paper.

\section*{Data availability}

The Abaqus UEL source code and input files used in producing the results in this article, along with the instructions to run the simulations, are available for download from \url{https://github.com/bibekanandadatta/Abaqus-UEL-Polyelectrolyte_Gel} or \url{https://github.com/NguyenLabJHU/Abaqus-UEL-Polyelectrolyte_Gel}.

\section*{Acknowledgments}

The authors would like to acknowledge the funding support from the National Science Foundation (EFMA-1830893) for part of the research reported in this article. Additionally, B.D. likes to thank Dr. Tingting Xu and Yuefeng Jiang for numerous discussions on deriving the analytical tangents for the finite element model.

\bibliographystyle{elsarticle-harv}     
\bibliography{polyelectrolyte-gel.bib}         

\begin{appendices}

\renewcommand{\thetable}{\Alph{section}\arabic{table}}
\numberwithin{equation}{section}
\counterwithin{remark}{section}

\section{Detailed derivation of the coupled nonlinear theory}     
\label{sec:appendix-theory}
\setcounter{table}{0}

In section \ref{sec:theory-summary}, we omitted detailed calculations essential to establishing our thermodynamically-consistent coupled electro-chemo-mechanical theory, which we describe in detail here for the interested readers.

\subsection{Additional kinematic relations}

Based on the hydrated reference state and current deformed state, we further define velocity, $\vect{v}$, and velocity gradient, $\vect{L}$, as
\begin{equation}    \label{eq:velocity-and-vel-gradient}
    \vect{v} = \dot{\vect{x}}, 
    \quad \text{and} \quad
    \vect{L} = \grad \vect{v} = \dot{\vect{F}} {\vect{F}}^{-1}
\end{equation}

\begin{remark}
In developing our theory, we employ the standard notations of modern continuum mechanics. Specifically, $\grad(\bullet)$, $\divg(\bullet)$, and $\curl(\bullet)$ denote the gradient, divergence, and curl in the spatial coordinate, $\vect{x}$. $\Grad(\bullet)$, $\Divg(\bullet)$, and $\Curl(\bullet)$ denote the gradient, divergence, and the curl of quantity in the material coordinate, $\vect{X}$. Additionally, we should note that $\frac{D}{Dt} (\bullet) = \dot{\overline{(\bullet)}}$ represents the material time derivative whereas $\frac{\partial }{\partial t}(\bullet)$ represents the spatial time derivative. Finally, for tensor algebra, we use $\sym(\bullet)$ and $\skw(\bullet)$ to denote the symmetric and skew-symmetric parts of a tensor, respectively.
\end{remark}

We now use the prescribed multiplicative decomposition of deformation gradient in Eq. \eqref{eq:def-grad} to obtain the following deformation tensors. Here,
\begin{equation} \label{eq:cauchy-green}
    \vect{b} = \vect{F}\vect{F}^{\top}, 
    \quad 
    \vect{C} = \vect{F}^{\top} \vect{F}, 
    \quad \text{and}  \quad 
    \vect{b}^e = \vect{F}^e {\vect{F}^e}^{\top}, 
    \quad 
    \vect{C}^e = {\vect{F}^e}^{\top} \vect{F}^e,
\end{equation}
denote the total left and right Cauchy-Green deformation tensors and the elastic left and right Cauchy-Green deformation tensors, respectively. Using the definitions of deformation tensors in Eq. \eqref{eq:cauchy-green} and our prescribed kinematic descriptions of the deformation gradient for a pre-swollen hydrogel  in Eq. \eqref{eq:def-grad}, we obtain,
\begin{equation} \label{eq:cauchy-green-2}
\begin{aligned}
    \vect{b} = (\phi^p_0 J^s)^{2/3} \vect{b}^e, 
    \quad \text{and} \quad
    \vect{C} = (\phi^p_0 J^s)^{2/3} \vect{C}^e.
\end{aligned}
\end{equation}
For later use, we express the invariants $(I_1, I_2, I_3)$ of the total left and right Cauchy-Green tensor $(\vect{b} \ \mathrm{ or } \ \vect{C} )$ in terms of the invariants $(I_1^e, I_2^e, I_3^e)$ of elastic Cauchy-Green tensor $(\vect{b}^e \ \mathrm{ or } \ \vect{C}^e)$ as follows,
\begin{equation}    \label{eq:cauchy-green-invariants}
\begin{aligned}
    I_1 &= \tr(\vect{C}) =  (\phi^p_0 J^s)^{2/3} I_1^e, \\
    I_2 & = \frac{1}{2} [(\tr(\vect{C}))^2 - \tr(\vect{C}^2)] = \frac{(\phi^p_0 J^s)^{4/3}} {2} [(I_1^e)^2 - \tr({I_1^e}^2)], \\
    I_3 & = \det(\vect{C}) = (\phi^p_0 J^s)^{2} I_3^e.
\end{aligned}
\end{equation}
We now use the kinematic description in Eq. \eqref{eq:def-grad} to further decompose the velocity gradient in Eq. \eqref{eq:velocity-and-vel-gradient}$_2$ as,
\begin{equation}    \label{eq:L-decomposition}
\begin{aligned}
    \vect{L} & = \dot{\vect{F}} \vect{F}^{-1}, \\
    & = (\phi^p_0)^{1/3} \left( \dot{\vect{F}}^e \vect{F}^s + \vect{F}^e \dot{\vect{F}}^s \right) \vect{F}^{-1}, \\
    & = (\phi^p_0)^{1/3} \left( \dot{\vect{F}}^e \vect{F}^s \vect{F}^{-1} + \vect{F}^e \dot{\vect{F}}^s \vect{F}^{-1} \right), \\
    & = (\phi^p_0)^{1/3} \left( \dot{\vect{F}}^e \vect{F}^s \left[ (\phi^p_0)^{-1/3} (\vect{F}^s)^{-1} {\vect{F}^e}^{-1} \right] + \vect{F}^e \dot{\vect{F}}^s \left[ (\phi^p_0)^{-1/3} (\vect{F}^s)^{-1} {\vect{F}^e}^{-1} \right] \right), \\
    & =  \dot{\vect{F}}^e \underbrace{\vect{F}^s  (\vect{F}^s)^{-1}}_{\mathds{1}} {\vect{F}^e}^{-1} + \vect{F}^e \underbrace{\dot{\vect{F}}^s  (\vect{F}^s)^{-1}}_{\vect{L}^s} {\vect{F}^e}^{-1}, \\
    & =  \dot{\vect{F}}^e {\vect{F}^e}^{-1} + \vect{F}^e \vect{L}^s {\vect{F}^e}^{-1}, \\
    & =  \vect{L}^e  + \vect{F}^e \vect{L}^s {\vect{F}^e}^{-1}.
\end{aligned}
\end{equation}
where $\vect{L}^e$ and $\vect{L}^s$ are the elastic and swelling velocity gradients, respectively. As is standard, we can decompose the elastic and swelling velocity gradients into their respective stretch and spin tensors,
\begin{equation} \label{eq:velocity-gradient}
\begin{gathered}
    \vect{L}^e = \vect{D}^e + \vect{W}^e
    \quad \text{and} \quad
    \vect{L}^s = \vect{D}^s + \vect{W}^s, \\
    \text{where} \quad
    \vect{D}^e = \sym(\vect{L}^e), 
    \quad
    \vect{W}^e = \skw(\vect{L}^e),
    \quad
    \vect{D}^s = \sym(\vect{L}^s), 
    \quad \text{and} \quad
    \vect{W}^s = \skw(\vect{L}^s).
\end{gathered}
\end{equation}
Here, $\vect{D}^e$ and $\vect{D}^s$ are elastic and swelling stretch tensors, and $\vect{W}^e$ and $\vect{W}^s$ are the elastic and swelling spin tensors, respectively. Using the assumption of isotropic swelling deformation, \emph{i.e.}, $\vect{F}^s = \lambda^s \mathds{1}$, and Eq. \eqref{eq:velocity-gradient}, we obtain,
\begin{equation} \label{eq:swell-strecht-tensor}
    \vect{D}^s = \vect{L}^s = \dot{\vect{F}}^s  (\vect{F}^s)^{-1} = \frac{\dot{\lambda^s}}{\lambda^s} \mathds{1}, \quad \text{and} \quad \vect{W}^s = \vect{0}.
\end{equation}
The material time derivative of the volume change due to swelling can be written as,
\begin{equation}
    \dot{J}^s = J^s \tr(\vect{L}^s) = J^s \tr(\vect{D}^s) \quad \Rightarrow \tr(\vect{D}^s) = \dot{J}^s (J^s)^{-1}.
\end{equation}
Since we assumed the swelling deformation to be isotropic, we may write,
\begin{equation}
    \vect{D}^s = \frac{1}{3} \frac{\dot{J}^s} {J^s} \mathds{1}.
\end{equation}

\subsection{Global form of momentum and mass balance laws}

For an arbitrary subregion in the reference configuration, $ \mathcal{B}_0 \in \Omega_{0}$, the global form of linear momentum balance for the PE gel can be written as,
\begin{equation} \label{eq:momentum-balance-global}
\begin{gathered}
    \frac{D}{Dt} \int\limits_{\mathcal{B}_0} \rho_{\R} \vect{v} \ dV 
    =
    \int\limits_{\mathcal{\partial B}_0} \vect{P} \cdot \vect{N} \ dS 
    + \int\limits_{\mathcal{B}_0}  \rho_{\R} \vect{B} \ dV
\end{gathered}
\end{equation}

For the same arbitrary subregion, $\mathcal{B}_0$, the global mass balance law for the solvent is given by,
\begin{equation} \label{eq:solvent-mass-balance-global}
    \frac{D}{Dt} \int\limits_{\mathcal{B}_0} {C^{w} \ dV} 
    = - \int\limits_{\partial \mathcal{B}_0}{\vect{J}^{w} \cdot \vect{N} \ dS}.
\end{equation}

Similarly, for any ion, $\beta_k$, the global form of mass balance is as follows,
\begin{equation} \label{eq:ion-mass-balance-global}
    \frac{D}{Dt} \int\limits_{\mathcal{B}_0} {C^{\beta_k} \ dV} 
    = - \int\limits_{\partial \mathcal{B}_0}{\vect{J}^{\beta_k} \cdot \vect{N} \ dS} 
\end{equation}
As is standard, the local forms of the governing equations  \eqref{eq:momentum-balance} for linear momentum balance, \eqref{eq:solvent-mass-balance} for the solvent mass balance, and Eq. \eqref{eq:ion-mass-balance} for the ion mass balance were obtained by applying the divergence theorem to Eqs. \eqref{eq:momentum-balance-global}, \eqref{eq:solvent-mass-balance-global}, and \eqref{eq:ion-mass-balance-global}, respectively.

\subsection{Energy balance and entropy inequality}

A PE gel occupying the subregion $\mathcal{B}_0 \in \Omega_0 $, which is under mechanical equilibrium and charged species are being transported in and out of $\mathcal{B}_0$ across its boundary, $ \partial \mathcal{B}_0 $, with an outward unit normal, $\vect{N}$, the global energy balance can be written as,
\begin{equation} \label{eq:energy-balance-global}
\begin{aligned}
    \frac{D}{Dt} \int \limits_{\mathcal{B}_0} \left( \varepsilon + \psi Q  \right) \ dV = & 
    \underbrace{\int \limits_{\partial \mathcal{B}_0} \vect{P N} \cdot \vect{v} \ dS  
    + \int \limits_{\mathcal{B}_0} \rho_{\R} \vect{B} \cdot \vect{v} \ dV}_{\text{mechanical power}} 
    \underbrace{ - \int \limits_{\partial \mathcal{B}_0} \mu^{w} \vect{J}^{w} \cdot \vect{N} \ dS}_{\text{solvent transport power}} \\
    & \underbrace{ - \sum_{\beta_k} \int \limits_{\partial \mathcal{B}_0} \mu^{\beta_k} \vect{J}^{\beta_k} \cdot \vect{N} \ dS 
    - \int \limits_{\partial \mathcal{B}_0} \psi \vect{I_e} \cdot \vect{N} \ dS}_{\text{ionic species transport power}},
\end{aligned}
\end{equation}
where $\varepsilon$ is the internal energy density of the gel in the hydrated reference state encompassing total potential and thermal vibration of all possible microstates. The electric potential, $\psi$, is the energy required to move a unit charge from an infinite distance to an electric field of that electric potential. Thus, the second term on the left side of the equation represents the rate of change of electric energy density accumulating within the PE gel. In existing coupled theories for PE gels \citep{hongLargeDeformationElectrochemistry2010,zhangKineticsPolyelectrolyteGels2020,narayanCoupledElectrochemomechanicalTheory2022}, the authors considered the electrical energy to be a part of internal energy, assuming the gel to be a dielectric material. However, we decided to include the electric energy term explicitly since the rise of the electric potential happens because of external ion transport. The first two terms on the right side of the equation represent the conventional mechanical power expended by the first Piola-Kirchhoff traction, $\vect{T} = \vect{P} \cdot \vect{N}$, acting over the surface $\partial \mathcal{B}_0$, and body force per unit mass, $\vect{B}$, and the subregion $\mathcal{B}_0$, respectively. The third term represents the chemical power associated with the transport of solvent. The solvent chemical potential, $\mu^{w}$, represents the energy required for the absorption or release of a unit mole of the solvent. The chemical potential of an ion, $\mu^{\beta_k}$, possesses the same physical meaning as the solvent chemical potential, $\mu^w$, thus representing the chemical power expended by the flow of ions. The fifth term represents the electrical energy expended by the flow of ionic species with a current density, $\vect{I_e}$. The referential current density has a unit of charge per unit area, per unit time, and is defined as
\begin{equation} \label{eq:current-density}
    \vect{I_e} = F \sum_{\beta_k}  z^{\beta_k} \vect{J}^{\beta_k}
\end{equation}
We can reorganize the global energy balance statement in Eq. \eqref{eq:energy-balance-global} as follows,
\begin{equation} \label{eq:energy-balance-global-2}
\begin{aligned}
    \frac{D}{Dt} \int \limits_{\mathcal{B}_0} \varepsilon \ dV = & 
    \int \limits_{\partial \mathcal{B}_0} \vect{P N} \cdot \vect{v} \ dS  
    + \int \limits_{\mathcal{B}_0} \rho_{\R} \vect{B} \cdot \vect{v} \ dV
    - \int \limits_{\partial \mathcal{B}_0} \mu^{w} \vect{J}^{w} \cdot \vect{N} \ dS \\
    & - \sum_{\beta_k} \int \limits_{\partial \mathcal{B}_0} \mu^{\beta_k} \vect{J}^{\beta_k} \cdot \vect{N} \ dS 
    - \int \limits_{\partial \mathcal{B}_0} \psi \vect{I_e} \cdot \vect{N} \ dS
    - \int \limits_{\mathcal{B}_0}  \psi \dot{Q} \ dV.
\end{aligned}
\end{equation}

Using the divergence theorem and the governing equation for the balance of linear momentum in Eq. \eqref{eq:momentum-balance} on the first two terms of the right side of the global energy balance in Eq.  \eqref{eq:energy-balance-global-2}, we get,
\begin{equation} \label{eq:mechanical-power}
\begin{aligned}
    \int \limits_{\partial \mathcal{B}_0} \vect{P N} \cdot \vect{v} \ dS 
    + \int \limits_{\mathcal{B}_0} \rho_{\R} \vect{B} \cdot \vect{v} \ dV  
    & = \int \limits_{ \mathcal{B}_0} \vect{P}:\dot{\vect{F}} \ dV.
\end{aligned}
\end{equation}
Application of the divergence theorem and the governing equation for the solvent mass balance in Eq.  \eqref{eq:solvent-mass-balance} gives us the following for the third term on the right-hand side of the global energy balance in Eq. \eqref{eq:energy-balance-global-2}.
\begin{equation} \label{eq:chemical-power}
\begin{aligned}
    - \int \limits_{\partial \mathcal{B}_0} \mu^{w} \vect{J}^{w} \cdot \vect{N} \ dS
    & = \int \limits_{\partial \mathcal{B}_0} \left( \mu^w \dot{C}^w - \Grad \mu^w \cdot \vect{J}^w \right) \ dV.
\end{aligned}
\end{equation}
Similarly, by using the definition of the referential current density in Eq. \eqref{eq:current-density} and ion electrochemical potential in Eq. \eqref{eq:echem-pot-def}, and subsequently, by applying the divergence theorem and the governing equation for the mass balance for ionic species in Eq.  \eqref{eq:ion-mass-balance}, the fourth and fifth terms on the right-hand side of the global energy balance  in Eq. \eqref{eq:energy-balance-global-2} can be written as follows,
\begingroup
\allowdisplaybreaks
\begin{align*} \label{eq:electrochemical-power}
    - \sum_{\beta_k} \int \limits_{\partial \mathcal{B}_0} \mu^{\beta_k} \vect{J}^{\beta_k} \cdot \vect{N} \ dS 
    - \int \limits_{\partial \mathcal{B}_0} \psi \vect{I_e} \cdot \vect{N} \ dS
    & = - \sum_{\beta_k} \int \limits_{\partial \mathcal{B}_0} \mu^{\beta_k} \vect{J}^{\beta_k} \cdot \vect{N} \ dS 
    - \sum_{\beta_k} \int \limits_{\partial \mathcal{B}_0} \psi  F  z^{\beta_k} \vect{J}^{\beta_k} \cdot \vect{N} \ dS, \\
    & = - \sum_{\beta_k}  \int \limits_{\partial \mathcal{B}_0} \omega^{\beta_k} \vect{J}^{\beta_k} \cdot \vect{N} \ dS,     
    \stepcounter{equation}\tag{\theequation} \\
    & = \sum_{\beta_k} \int \limits_{ \mathcal{B}_0} \left( \omega^{\beta_k} \dot{C}^{\beta_k}  - \Grad \omega^{\beta_k} \cdot \vect{J}^{\beta_k} \right) \ dV,
\end{align*}
\endgroup
where
\begin{equation} \label{eq:echem-pot}
    \omega^{\beta_k} = \mu^{\beta_k} + F \psi z^{\beta_k},
\end{equation}
is defined as the electrochemical potential of the ion, $\beta_k$.

Using the definition of referential charge density in Eq. \eqref{eq:charge-density-def} on the last term appearing on the right-hand side of the global energy balance in Eq. \eqref{eq:energy-balance-global-2}, we obtain,
\begin{equation} \label{eq:electric-power}
\begin{aligned}
    - \frac{D}{Dt} \int \limits_{\mathcal{B}_0} \psi Q \ dV
    = - \int \limits_{\mathcal{B}_0} F \psi  \sum_{\beta_k} z^{\beta_k} \dot{C}^{\beta_k}  \ dV.
\end{aligned}
\end{equation}

By substituting Eqs. \eqref{eq:mechanical-power}, \eqref{eq:chemical-power}, \eqref{eq:electrochemical-power}, and \eqref{eq:electric-power} into the global energy balance equation \eqref{eq:energy-balance-global-2}, we have,
\allowdisplaybreaks
\begingroup
\begin{align*}
    \frac{D}{Dt} \int \limits_{\mathcal{B}_0}  \varepsilon \ dV 
    & =  
    \int \limits_{ \mathcal{B}_0} \vect{P}:\dot{\vect{F}} \ dV 
    + \int \limits_{\partial \mathcal{B}_0} \left( \mu^w \dot{C}^w - \Grad \mu^w \cdot \vect{J}^w \right) \ dV \\
    & \quad + \sum_{\beta_k} \int \limits_{ \mathcal{B}_0} \left( \omega^{\beta_k} \dot{C}^{\beta_k}  - \Grad \omega^{\beta_k} \cdot \vect{J}^{\beta_k} \right) \ dV
    - \sum_{\beta_k} \int \limits_{\mathcal{B}_0} F \psi  \sum_{\beta_k} z^{\beta_k} \dot{C}^{\beta_k}  \ dV, 
    \stepcounter{equation}\tag{\theequation} \\
    & =  \int \limits_{ \mathcal{B}_0} \left( \vect{P}:\dot{\vect{F}} 
    +  \mu^w \dot{C}^w - \Grad \mu^w \cdot \vect{J}^w 
    + \sum_{\beta_k} \left( \mu^{\beta_k} \dot{C}^{\beta_k} - \Grad \mu^{\beta_k} \cdot \vect{J}^{\beta_k} \right) \right) \ dV.
\end{align*}
\endgroup

For an arbitrary volume, $\mathcal{B}_0$, within the PE gel, the energy balance equation can be localized as,
\begin{equation} \label{eq:energy-balance-local}
\begin{aligned}
    \dot{\varepsilon} = \vect{P}:\dot{\vect{F}} 
    + \mu^{w} \dot{C}^{w}  
    - \Grad \mu^{w} \cdot \vect{J}^{w} 
    + \sum_{\beta_k} \left( \mu^{\beta_k} \dot{C}^{\beta_k}
    - \Grad \omega^{\beta_k} \cdot \vect{J}^{\beta_k} \right).
\end{aligned}
\end{equation}
This local form of energy balance is the same as what is presented in \cite{narayanCoupledElectrochemomechanicalTheory2022} except for the electrical power term, since we considered the gel to be electroneutral. However, this local form is different from what we presented in our previous work \citep{zimmermanReactiveElectrochemomechanicalTheory2024}, as the concentration was defined per fluid volume and the electroneutrality constraint was enforced later using the Lagrange multiplier approach.

From the second law of thermodynamics, the positive dissipation criterion requires the difference between the internal power, $\dot{\varepsilon}$, and the temporal rate of Helmholtz free energy density, $\dot{\Psi}$, to be positive. The abstract form of the positive dissipation criterion can be written as,
\begin{equation} \label{eq:entropy-inequality}
\begin{aligned}
    \dot{H} \geq 0 
    \qquad \Rightarrow 
    \mathcal{D}_{\mathrm{int}} = \dot{\varepsilon} - \dot{\Psi} \geq 0,
\end{aligned}
\end{equation}
where $\dot{H}$ is the rate of entropy generation, $\mathcal{D}_{\mathrm{int}}$ is the rate of internal dissipation, and $\dot{\Psi}$ is the rate change of Helmholtz free energy density. Substituting the local energy balance Eq.\eqref{eq:energy-balance-local} in the positive dissipation criterion in Eq. \eqref{eq:entropy-inequality} leads to the following,
\begin{equation} \label{eq:dissipation-criterion}
    \vect{P}:\dot{\vect{F}} 
    + \mu^{w} \dot{C}^{w}  
    - \Grad \mu^{w} \cdot \vect{J}^{w} 
    + \sum_{\beta_k} \left( \mu^{\beta_k} \dot{C}^{\beta_k}
    -  \Grad \omega^{\beta_k} \cdot \vect{J}^{\beta_k} \right)
    - \dot{\Psi} \geq 0.
\end{equation}
Application of the symmetry of the Cauchy stress tensor, $\boldsymbol{\upsigma} = \boldsymbol{\upsigma}^{\top}$, and the kinematic constraint in Eq. \eqref{eq:L-decomposition}, let us write the stress power term in the positive dissipation criterion in Eq. \eqref{eq:dissipation-criterion} as,
\begin{equation} \label{eq:work-conjugate-1}
\begin{aligned} 
    \vect{P}:\dot{\vect{F}} & = J \boldsymbol{\upsigma}:\vect{D}, \\
    & = J \boldsymbol{\upsigma}:\vect{L}, \\
    & =  J \boldsymbol{\upsigma}: \left( \vect{L}^e  + \vect{F}^e \vect{L}^s {\vect{F}^e}^{-1} \right), \\
    & =  J \boldsymbol{\upsigma}:\vect{L}^e +  J \boldsymbol{\upsigma}:\left( \vect{F}^e \vect{L}^s {\vect{F}^e}^{-1} \right), \\
    & = \frac{1}{2} \underbrace{J {\vect{F}^e}^{-1}\boldsymbol{\upsigma} {\vect{F}^e}^{-\top}}_{\vect{S}^e}:\dot{\vect{C}}^e + \underbrace{J {\vect{F}^e}^{\top} \boldsymbol{\upsigma} {\vect{F}^e}^{-\top}}_{\vect{M}^e} : \vect{L}^s, \\
    & = \frac{1}{2} \vect{S}^e : \dot{\vect{C}}^e + \vect{M}^e : \vect{L}^s,
\end{aligned}    
\end{equation}
where 
\begin{equation} \label{eq:elastic-stress-def}
\begin{gathered}
    \vect{S}^e = J {\vect{F}^e}^{-1}\boldsymbol{\upsigma} {\vect{F}^e}^{-\top},
    \quad \text{and} \quad
    \vect{M}^e = J {\vect{F}^e}^{\top} \boldsymbol{\upsigma} {\vect{F}^e}^{-\top},
\end{gathered}
\end{equation}
denote the elastic second Piola-Kirchhoff stress and elastic Mandel stress, respectively. We can further decompose the second term in in Eq. \eqref{eq:work-conjugate-1} and use Eq. \eqref{eq:swell-strecht-tensor} and Eq. \eqref{eq:elastic-stress-def}$_2$ to obtain the following,
\begin{equation} \label{eq:work-conjugate-2}
\begin{aligned} 
    \vect{M}^e : \vect{L}^s & =  \tr(\vect{M}^e {\vect{D}^s}^{\top}), \\
    & = \frac{1}{3} (J^s)^{-1} \tr(\vect{M}^e)  \dot{J}^s, \\
    & = \frac{1}{3}  \tr \left( \phi^p_0 J^e \ \boldsymbol{\upsigma} {\vect{F}^e}^{\top} {\vect{F}^e}^{-\top}  \right) \dot{J}^s, \\
    & = \phi^p_0 \frac{1}{3} J^e \tr(\boldsymbol{\upsigma}) \dot{J}^s, \\
    & = - \phi^p_0 \ p \dot{J}^s,
\end{aligned}    
\end{equation}
where $\dot{J}^s$ is the rate change of swelling volume ratio and $p$ is the mean pressure  as defined as follows
\begin{equation} \label{eq:mean-pressure-def}
    p = \frac{-1}{3} J^e \tr(\boldsymbol{\upsigma})
\end{equation}
Substituting Eq. \eqref{eq:work-conjugate-1} and subsequently Eq. \eqref{eq:work-conjugate-2} in the positive dissipation criterion in Eq.\eqref{eq:dissipation-criterion} furnishes the following,
\begin{equation} \label{eq:dissipation-criterion-2}
    \frac{1}{2} \vect{S}^e : \dot{\vect{C}}^e - \phi^p_0 \ p \dot{J}^s
    +
    \mu^{w} \dot{C}^{w} - \Grad \mu^{w} \cdot \vect{J}^{w} 
    + \sum_{\beta_k} \left( \mu^{\beta_k} \dot{C}^{\beta_k} - \Grad \omega^{\beta_k} \cdot \vect{J}^{\beta_k} \right)
    - \dot{\Psi} \geq 0.
\end{equation}
Substituting $p$ and $J^s$ in \eqref{eq:dissipation-criterion-2} yields the following,
\begin{equation} \label{eq:dissipation-criterion-final}
    \frac{1}{2} \vect{S}^e : \dot{\vect{C}}^e
    + \left( \mu^{w} - p \mathcal{V}^w \right) \dot{C}^{w} 
    - \Grad \mu^{w} \cdot \vect{J}^{w} 
    + \sum_{\beta_k} \left[ \left( \mu^{\beta_k} - p \mathcal{V}^{\beta_k} \right) \dot{C}^{\beta_k} 
    - \Grad \omega^{\beta_k} \cdot \vect{J}^{\beta_k} \right]
    - \dot{\Psi} \geq 0.
\end{equation}
Similar to the localized energy balance statement, our localized dissipation inequality is the same as \cite{narayanCoupledElectrochemomechanicalTheory2022}, except for the electrical energy contribution. In contrast, the dissipation inequality in our previous work on dynamic polymerization of copolymerized DNA hydrogel \citep{zimmermanReactiveElectrochemomechanicalTheory2024} is different since we split the Cauchy stress into elastic and pressure terms and had additional contributions from reactive species.

\subsection{Coleman-Noll procedure}

Guided by the entropy inequality in Eq. \eqref{eq:dissipation-criterion-final}, we assume the following set of state variables to describe the total free energy density of the PE gel
\begin{equation} \label{eq:psi_state_vars}
    \Psi = \Psi \left( \vect{C}^e, C^w, C^{\beta_1},  C^{\beta_2}, \cdots \cdots , C^{\beta_n} \right).
\end{equation}
Hence, the material time derivative of the free energy density, $\dot{\Psi}$, can be written as,
\begin{equation} \label{eq:psi_time_derivative}
    \dot{\Psi} = \frac{\partial \Psi}{\partial \vect{C}^e}: \dot{\vect{C}}^e  
    + \frac{\partial \Psi}{\partial C^{w}} \dot{C}^{w} 
    + \sum_{\beta_k} \frac{\partial \Psi}{\partial C^{\beta_k}} \dot{C}^{\beta_k} .
\end{equation}
Substituting Eq. \eqref{eq:psi_time_derivative} into the positive dissipation criterion Eq. \eqref{eq:dissipation-criterion-final} furnishes the following,
\begin{equation}    \label{eq:coleman-noll-1}
\begin{gathered}
    \left(  \frac{1}{2} \vect{S}^e - \frac{\partial \Psi}{\partial \vect{C}^e}  \right) : \dot{\vect{C}}^e 
    + \left( \mu^{w} - p \mathcal{V}^w - \frac{\partial \Psi}{\partial C^{w}}  \right) \dot{C}^{w}
    + \sum_{\beta_k} \left( \mu^{\beta_k} - p \mathcal{V}^{\beta_k} - \frac{\partial \Psi}{\partial C^{\beta_k}}  \right) \dot{C}^{\beta_k}  \\
    - \Grad \mu^{w} \cdot \vect{J}^{w}
    - \sum_{\beta_k} \Grad \omega^{\beta_k} \cdot \vect{J}^{\beta_k} 
    \geq 0.
\end{gathered}
\end{equation}
The first three terms of the dissipation criterion Eq. \eqref{eq:coleman-noll-1} are reversible thermodynamic processes, whereas the last two terms represent irreversible kinetic processes. According to the Coleman-Noll argument \citep{colemanThermodynamicsElasticMaterials1963}, for an arbitrary approach to equilibrium, the time derivatives of the extrinsic state variables should disappear. Thus, by applying this principle and then subsequently applying the push-forward operation on the elastic second Piola-Kirchhoff stress, $\vect{S}^e$, and the definition of electrochemical potential, $\omega^{\beta_k}$, we can obtain the following thermodynamic restrictions on the state variables.
\begin{equation} \label{eq:constitutive-definitions-primary}
\begin{aligned}
    \vect{S}^e &=  2 \frac{\partial \Psi}{\partial \vect{C}^e}, \\
    \mu^w & = \frac{\partial \Psi}{\partial C^{w}} + p \mathcal{V}^{w}, \\
    \mu^{\beta_k} & = \frac{\partial \Psi}{\partial C^{\beta_k}} + p \mathcal{V}^{\beta_k},
\end{aligned}
\end{equation}

The remaining terms in Eq. \eqref{eq:coleman-noll-1} in the residual dissipation, representing the non-equilibrium thermodynamic process, should also satisfy the positive dissipation criterion as follows
\begin{equation}    \label{eq:transport-inequality}
    - \vect{J}^w \cdot \Grad \mu^w
    - \sum_{\beta_k} \vect{J}^{\beta_k} \cdot \Grad \omega^{\beta_k} \geq 0,
\end{equation}
where, $\Grad \mu^w$ and $\Grad \omega^{\beta_k}$ represent the thermodynamic forces, and $\vect{J}^w$ and $\vect{J}^{\beta_k}$ represent the flows corresponding to the forces. To describe the transport of any generic species, $\alpha$, whether solvent $(w)$ or ion $(\beta_k)$, we can assume the following constitutive equation for referential molar flux,
\begin{equation}    \label{eq:onsager-principle}
    \vect{J}^{\alpha} =  \left( \sum_{\gamma \in \{w, \beta_k\} } L^{\alpha \gamma} \Grad f^{\gamma} \right) \vect{C}^{-1},
\end{equation}
where, $L^{\alpha \gamma}$ is known as the Onsager matrix of transport coefficients and $\Grad f^{\gamma}$ is the generic thermodynamic force representing $\Grad \mu^w$ and $\Grad \omega^{\beta_k}$ and responsible for the mass flow with flux $\vect{J}^{\alpha}$. To satisfy the dissipation criterion, $L^{\alpha \gamma}$ needs to be positive semi-definite. Additionally, \citep{onsagerReciprocalRelationsIrreversible1931,onsagerReciprocalRelationsIrreversible1931a} showed that $L^{\alpha \gamma}$ is symmetric following the time invariance principle. Specific forms for the molar flux of the solvent, $J^w$, and molar flux for ionic species, $J^{\beta_k}$, were prescribed in Eq. \eqref{eq:flux-law-solvent} and Eq. \eqref{eq:flux-law-ion}.

\section{Matrix-vector form and F-bar element formulation}
\label{sec:appendix-fem}
\setcounter{table}{0}
\setcounter{figure}{0}

Using index notation is often necessary for developing a nonlinear finite element formulation; however, it is convenient to represent the finite element model in matrix-vector form to ease the effort in programming. Hence, we present a more compact matrix-vector form of our finite element formulation in this section.

\subsection{Standard  element formulation}

In matrix-vector form, the Galerkin approximation for the degrees of freedom Eq. \eqref{eq:galerkin-approx} within an element can be written as follows,
\begin{equation} \label{galerkin-approx-matrix}
\begin{gathered}
    \vect{u} = \mat{N}_{\vect{u}} \vect{u}, 
    \quad
    \mu^w = \mat{N}_{\mu} \mu^w, 
    \quad \text{and} \quad
    \omega^{\beta_k} = \mat{N}_{\omega} \omega^{\beta_k}.
\end{gathered}
\end{equation}
Here, $\mat{N}_{\vect{u}}$, $\mat{N}_{\mu}$, and $\mat{N}_{\omega}$ are the interpolation function matrix for the element nodal displacement vector, $\vect{u}$, nodal solvent chemical potential, $\mu^w$, and nodal electrochemical potential for ions, $\omega^{\beta_k}$, respectively. These matrices are represented as follows,
\begin{equation} \label{eq:interp-func-matrix}
\begin{gathered}
    \mat{N}_{\vect{u}}=
    \begin{bmatrix}
        N^{\vect{u}}_1 &  0  &  0  & N^{\vect{u}}_2 &   0  &  0   & \cdots & \cdots & N^{\vect{u}}_{n_{\mathrm{en}}} & 0    & 0\\
        0  & N^{\vect{u}}_1 &  0  &  0  &  N^{\vect{u}}_2 &  0   & \cdots & \cdots &   0  &  N^{\vect{u}}_{n_{\mathrm{en}}}  & 0 \\
        0  &  0  & N^{\vect{u}}_1 &  0  &   0  & N^{\vect{u}}_2  & \cdots & \cdots &   0  &  0  & N^{\vect{u}}_{n_{\mathrm{en}}}
    \end{bmatrix}, \\[6pt]
    \mat{N}_{\mu} =
    \begin{bmatrix}
        N^{\mu}_1 & N^{\mu}_2 &  N^{\mu}_3  & \cdots & \cdots & N^{\mu}_{n_{\mathrm{en}}}
    \end{bmatrix}, \\[6pt]
    \text{and} \quad
    \mat{N}_{\omega} =
    \begin{bmatrix}
        N^{\omega}_1 & N^{\omega}_2 &  N^{\omega}_3  & \cdots & \cdots & N^{\omega}_{n_{\mathrm{en}}}
    \end{bmatrix}.
\end{gathered}
\end{equation}
For two-dimensional cases,  the third row and every third column of $\mat{N}_{\vect{u}}$ are eliminated.

As is standard in finite elements, the symmetric strain-displacement matrix, $\mat{B}_{\vect{u}}$, is then given by,
\begin{equation} \label{eq:B-u-matrix}
    \mat{B}_{\vect{u}} =
    \begin{bmatrix}
        \mat{B}_{\vect{u}}^1 & \mat{B}_{\vect{u}}^2 & \mat{B}_{\vect{u}}^3 & \cdots & \cdots & \mat{B}_{\vect{u}}^{n_{\mathrm{en}}}
    \end{bmatrix},
\end{equation}
where $\mat{B}_{u}^a$ is the sub-matrix of the symmetric gradient of the interpolation function corresponding to each node within the element. For two-dimensional plane strain, axisymmetric, and general three-dimensional cases, $\mat{B}_{u}^a$ can be represented as,
\begin{equation} \label{eq:Ba-u-matrix}
    \mat{B}_{\vect{u}}^a 
    =
    \begin{bmatrix}
        {N^{\vect{u}}_a}_{,1}      &    0          \\
           0            & {N^{\vect{u}}_a}_{,2}    \\
        {N^{\vect{u}}_a}_{,2}      & {N^{\vect{u}}_a}_{,1} 
    \end{bmatrix},
    \quad 
    \begin{bmatrix}
        {N^{\vect{u}}_a}_{,1}      &    0                       \\
           0            & {N^{\vect{u}}_a}_{,2}                 \\
        {N^{\vect{u}}_a}_{,2}      & {N^{\vect{u}}_a}_{,1}      \\
        \frac{{N^{\vect{u}}_a}}{r} &    0                       \\
    \end{bmatrix},
    \quad \text{and} \quad
    \mat{B}_{\vect{u}}^a =
    \begin{bmatrix}
        {N^{\vect{u}}_a}_{,1}      &    0          &   0           \\
           0            & {N^{\vect{u}}_a}_{,2}    &   0           \\
           0            &    0          & {N^{\vect{u}}_a}_{,3}    \\
           0            & {N^{\vect{u}}_a}_{,3}    & {N^{\vect{u}}_a}_{,2}    \\
        {N^{\vect{u}}_a}_{,3}      &    0          & {N^{\vect{u}}_a}_{,1}    \\
        {N^{\vect{u}}_a}_{,2}      & {N^{\vect{u}}_a}_{,1}    &   0           \\
    \end{bmatrix}.
\end{equation}

The non-symmetric gradient matrix of the displacement interpolation functions, $\mat{G}_{\vect{u}}$, appears as,
\begin{equation} \label{eq:G-u-matrix}
    \mat{G}_{\vect{u}} =
    \begin{bmatrix}
        \mat{G}_{\vect{u}}^1 & \mat{G}_{\vect{u}}^2 & \mat{G}_{\vect{u}}^3 & \cdots & \cdots & \mat{G}_{\vect{u}}^{n_{\mathrm{en}}}
    \end{bmatrix}.
\end{equation}
For two-dimensional plane strain, axisymmetric, and general three-dimensional cases, the nodal sub-matrix, $\mat{G}_{\vect{u}}^a$, is given by,
\begin{equation} \label{eq:Ga-u-matrix}
    \mat{G}_{\vect{u}}^a =
    \begin{bmatrix}
        {N^{\vect{u}}_a}_{,1}  &    0          \\
           0        & {N^{\vect{u}}_a}_{,1}    \\
        {N^{\vect{u}}_a}_{,2}  &    0          \\
           0        & {N^{\vect{u}}_a}_{,2} 
    \end{bmatrix},
    \quad
    \mat{G}_{\vect{u}}^a =
    \begin{bmatrix}
        {N^{\vect{u}}_a}_{,1}  &    0           \\
           0        & {N^{\vect{u}}_a}_{,1}     \\
        {N^{\vect{u}}_a}_{,2}  &    0           \\
           0        & {N^{\vect{u}}_a}_{,2}     \\
        \frac{N^{\vect{u}}_a}{r} & 0            \\
    \end{bmatrix},
    \quad \text{and} \quad
    \mat{G}_{\vect{u}}^a =
    \begin{bmatrix}
        {N^{\vect{u}}_a}_{,1} &    0    &    0    \\
           0    & {N^{\vect{u}}_a}_{,1} &    0    \\
           0    &    0    & {N^{\vect{u}}_a}_{,1} \\
        {N^{\vect{u}}_a}_{,2} &    0    &    0    \\
           0    & {N^{\vect{u}}_a}_{,2} &    0    \\
           0    &    0    & {N^{\vect{u}}_a}_{,2} \\
        {N^{\vect{u}}_a}_{,3} &    0    &    0    \\
           0    & {N^{\vect{u}}_a}_{,3} &    0    \\
           0    &    0    & {N^{\vect{u}}_a}_{,3} \\
    \end{bmatrix}.
\end{equation}

For the scalar nodal degrees of freedom, $\mu^w$, and $\omega^{\beta_k}$, the gradient matrix for the interpolation functions, $\mat{B}_{\mu}$, and $\mat{B}_{\omega}$ are represented as below for three-dimensional cases, 
\begin{equation} \label{eq:B-mu-omega-matrix}
\begin{gathered}
    \mat{B}_{\mu} = 
    \begin{bmatrix}
        N^{\mu}_{1,1} & N^{\mu}_{2,1} & \cdots & \cdots & N^{\mu}_{n_{\mathrm{en}},1} \\
        N^{\mu}_{1,2} & N^{\mu}_{2,2} & \cdots & \cdots & N^{\mu}_{n_{\mathrm{en}},2} \\
        N^{\mu}_{1,3} & N^{\mu}_{2,3} & \cdots & \cdots & N^{\mu}_{n_{\mathrm{en}},3} \\
    \end{bmatrix}, 
    \quad \text{and} \quad
    \mat{B}_{\omega} = 
    \begin{bmatrix}
        N^{\omega}_{1,1} & N^{\omega}_{2,1} & \cdots & \cdots & N^{\omega}_{n_{\mathrm{en}},1} \\
        N^{\omega}_{1,2} & N^{\omega}_{2,2} & \cdots & \cdots & N^{\omega}_{n_{\mathrm{en}},2} \\
        N^{\omega}_{1,3} & N^{\omega}_{2,3} & \cdots & \cdots & N^{\omega}_{n_{\mathrm{en}},3} \\
    \end{bmatrix}
\end{gathered}
\end{equation}
For two-dimensional diffusion cases, the third rows of $\mat{B}_{\mu}$ and $\mat{B}_{\omega}$ need to be eliminated.

Using the matrix form of the interpolation functions for different nodal degrees of freedom and their gradient matrices in Eqs. \eqref{eq:interp-func-matrix}-\eqref{eq:B-mu-omega-matrix}, the matrix forms of the element residual vector in Eq. \eqref{eq:elem-residuals} can be expressed as,
\begin{equation} \label{eq:elem-residuals-matrix}
\begin{aligned}
    & \left[ \vect{r}_e^{\vect{u}} \right]_{n_{\mathrm{en}}*n_{\mathrm{dim}}\times1} 
    && = - \int\limits_{\Omega_0^e} (\mat{B}_{\vect{u}}  \Sigma_{\vect{F}} )^{\T} \widetilde{\vect{S}} \ dV 
    + \int\limits_{\Omega_0^e} \rho_{\R} \mat{N}_{\vect{u}}^{\T} \vect{B} \ dV
    + \int\limits_{\Gamma_{\vect{T}}^e} \mat{N}_{\vect{u}}^{\T}  \vect{T} \ dS, \\
    & \left[ \vect{r}_e^{\mu} \right]_{n_{\mathrm{en}}\times1} 
    && = - \int\limits_{\Omega_0^e} \mat{N}_{\mu}^{\T} \left( \frac{C^w - C^w_{t}}{\Delta t} \right) \ dV  
    + \int\limits_{\Omega_0^e} \mat{B}_{\mu}^{\T} \vect{J}^w  \ dV 
    + \int\limits_{\Gamma_{I^w}^e} \mat{N}_{\mu}^{\T} {I^w} \ dS, \\
    & \left[ \vect{r}_e^{\omega_k} \right]_{n_{\mathrm{en}}\times1} 
    && = - \int\limits_{\Omega_0^e} \mat{N}_{\omega}^{\T} \left( \frac{C^{\beta_k} - C^{\beta_k}_{t}}{\Delta t} \right) \ dV  
    + \int\limits_{\Omega_0^e} \mat{B}_{\omega}^{\T} \vect{J}^{\beta_k}  \ dV 
    + \int\limits_{\Gamma_{I^{\beta_k}}^e} \mat{N}_{\omega}^{\T} {I^{\beta_k}} \ dS.
\end{aligned}
\end{equation}

Here, $\widetilde{\vect{S}}$ is the vector form of the second Piola-Kirchhoff stress, which can be represented as
\begin{equation}
\begin{gathered}
    \widetilde{\vect{S}} = 
    \begin{bmatrix}
        S_{11}  &   S_{22}  &  S_{12}
    \end{bmatrix}^{\T}, 
    \quad
    \widetilde{\vect{S}} = 
    \begin{bmatrix}
        S_{11}  &    S_{22}     &    S_{12}     &   S_{33}
    \end{bmatrix}^{\T}, \\
    \text{and} \
    \widetilde{\vect{S}} = 
    \begin{bmatrix}
        S_{11}  &    S_{22}     &    S_{33}     &
        S_{23}  &    S_{13}     &    S_{12}
    \end{bmatrix}^{\T},
\end{gathered}
\end{equation}
for two-dimensional plane strain, axisymmetry, and three-dimensional cases, respectively. 

$[\Sigma_{\vect{F}}]$ is a square block-diagonal type matrix, and for general three-dimensional cases, it appears as,
\begin{equation}
    \Sigma_{\vect{F}} = 
    \begin{bmatrix}
        F_{11} & F_{12} & F_{13} & 0 & 0 & 0 & \cdots & 0 & 0 & 0 \\
        F_{21} & F_{22} & F_{23} & 0 & 0 & 0 & \cdots & 0 & 0 & 0 \\
        F_{31} & F_{32} & F_{33} & 0 & 0 & 0 & \cdots & 0 & 0 & 0 \\
        & & & & \cdots & & & & & \\
        & & & & \cdots & & & & & \\
        & & & & \cdots & & & & & \\
        & & & & \cdots & & & & & \\
        & & & & \cdots & & & & & \\
        & & & & \cdots & & & & & \\
        0 & 0 & 0 & 0 & 0 & 0 & \cdots & F_{11} & F_{12} & F_{13} \\
        0 & 0 & 0 & 0 & 0 & 0 & \cdots & F_{21} & F_{22} & F_{23} \\
        0 & 0 & 0 & 0 & 0 & 0 & \cdots & F_{31} & F_{32} & F_{33} \\
    \end{bmatrix}_{n_{\mathrm{en}}*n_{\mathrm{dim}}\times n_{\mathrm{en}}*n_{\mathrm{dim}}}.
\end{equation}
For two-dimensional plane strain cases, every third row and column can be eliminated to reduce the dimension of $\Sigma_{\vect{F}}$.

Now, by using the matrix form of the interpolation functions and their gradient matrices in Eqs. \eqref{eq:interp-func-matrix}-\eqref{eq:B-mu-omega-matrix}, the element tangent stiffness matrix in Eq.  \eqref{eq:elem-tangents} along with their dimensions can be represented as follows,
\begingroup
\allowdisplaybreaks
\begin{align*}      \label{eq:elem-tangent-matrix}
    & \left[ \mat{K}_e^{\vect{u}\vect{u}} \right]_{n_{\mathrm{en}}*n_{\mathrm{dim}} \times n_{\mathrm{en}}*n_{\mathrm{dim}}}
    && =  \int\limits_{\Omega_0^e} \left( \mat{G}_{\vect{u}}^{\T} \Sigma_{\vect{S}} \mat{G}_{\vect{u}} 
    + (\mat{B}_{\vect{u}} \Sigma_{\vect{F}})^{\T} \vect{D}_{\mathds{C}} (\mat{B}_{\vect{u}} \Sigma_{\vect{F}}) \right) dV, \\
    & \left[ \mat{K}_e^{\vect{u} \mu} \right]_{n_{\mathrm{en}}*n_{\mathrm{dim}} \times n_{\mathrm{en}} }
    && = \int\limits_{\Omega_0^e} \mat{G}_{\vect{u}}^{\T} \mat{a}_{\vect{u}\mu} \mat{N}_{\mu} \ dV , \\
    & \left[ \mat{K}_e^{\vect{u}\omega_k} \right]_{n_{\mathrm{en}}*n_{\mathrm{dim}} \times n_{\mathrm{en}}}
    && =  \int\limits_{\Omega_0^e} \mat{G}_{\vect{u}}^{\T} \mat{a}_{\vect{u} \omega_k} \mat{N}_{\omega} \ dV , \\
    & \left[ \mat{K}_e^{\mu \vect{u}}  \right]_{n_{\mathrm{en}} \times n_{\mathrm{en}}*n_{\mathrm{dim}} }
    && = \int\limits_{\Omega_0^e} \left( \mat{N}_{\mu}^{\T} \left( \frac{1}{\Delta t} \mat{c}_{\mu \vect{u}} \right) \mat{G}_{\vect{u}}
    - \mat{B}_{\mu}^{\T} \mat{D}_{\mu\vect{u}} \mat{G}_{\vect{u}} \right) dV, \\
    & \left[ \mat{K}_e^{\mu\mu} \right]_{n_{\mathrm{en}} \times n_{\mathrm{en}}}
    && =  \int\limits_{\Omega_0^e} \left( \mat{N}_{\mu}^{\T} \left( \frac{1}{\Delta t}  \frac{d C^w}{d \mu^w} \right) \mat{N}_{\mu}
    - \mat{B}_{\mu}^{\T} \vect{m}_{\mu \mu} \mat{N}_{\mu}
    + \mat{B}_{\mu}^{\T} \mat{M}^{w} \mat{B}_{\mu} \right) dV, \\
    & \left[ \mat{K}_e^{\mu \omega_k} \right]_{n_{\mathrm{en}} \times n_{\mathrm{en}}}
    && =  \int\limits_{\Omega_0^e} \left( \mat{N}_{\mu}^{\T} \left( \frac{1}{\Delta t} \frac{d C^w}{d \omega^{\beta_k}} \right) \mat{N}_{\omega}
    - \mat{B}_{\mu}^{\T} \vect{m}_{\mu \omega_k} \mat{N}_{\omega}
    + \mat{B}_{\mu}^{\T} \mat{M}^{w \omega_k} \mat{B}_{\omega} \right) dV, \\
    & \left[ \mat{K}_e^{\omega_k \vect{u}}  \right]_{n_{\mathrm{en}} \times n_{\mathrm{en}}*n_{\mathrm{dim}} }
    && = \int\limits_{\Omega_0^e} \left( \mat{N}_{\omega}^{\T} \left( \frac{1}{\Delta t}\mat{c}_{\omega_1 \vect{u}} \right)  \mat{G}_{\vect{u}}
    - \mat{B}_{\omega}^{\T} \mat{D}_{\omega_1\vect{u}} \mat{G}_{\vect{u}} \right) dV, 
    \stepcounter{equation}\tag{\theequation} \\
    & \left[ \mat{K}_e^{\omega_k \mu} \right]_{n_{\mathrm{en}} \times n_{\mathrm{en}}}
    && = \int\limits_{\Omega_0^e} \left( \mat{N}_{\omega}^{\T} \left( \frac{1}{\Delta t}  \frac{d C^{\beta_k}}{d \mu^w} \right) \mat{N}_{\mu}
    - \mat{B}_{\omega}^{\T} \mat{m}_{\omega_k \mu} \mat{N}_{\mu} \right) dV, \\
    & \left[ \mat{K}_e^{\omega_i \omega_k} \right]_{n_{\mathrm{en}} \times n_{\mathrm{en}}}
    && = \int\limits_{\Omega_0^e} \left( \mat{N}_{\omega}^{\T}  \left( \frac{1}{\Delta t} \frac{d C^{\beta_i}}{d \omega_k} \right) \mat{N}_{\omega}
    - \mat{B}_{\omega}^{\T} \vect{m}_{\omega_i \omega_k} \mat{N}_{\omega}
    + \mat{B}_{\omega}^{\T} \mat{M}^{\beta_i \beta_k} \mat{B}_{\omega} \right) dV.
\end{align*}
\endgroup

$[\Sigma_{\vect{S}}]$ is the expanded sparse matrix form of the fourth-order tensor $S_{JL} \delta_{ik}$. For two-dimensional plane strain and three-dimensional cases, it is given by,
\begin{equation}
    \Sigma_{\vect{S}} =
    \begin{bmatrix}
        \vect{S}_{11} &     0           & \vect{S}_{12} &       0       \\
             0        & \vect{S}_{11}   &       0       & \vect{S}_{12} \\
        \vect{S}_{12} &     0           & \vect{S}_{22} &       0       \\
             0        & \vect{S}_{12}   &       0       & \vect{S}_{22}
    \end{bmatrix},
    \quad \text{and} \quad
    \Sigma_{\vect{S}} =
    \begin{bmatrix}
        \vect{S}_{11} & 0 & 0  & \vect{S}_{12} & 0 & 0 & \vect{S}_{13} & 0 & 0 \\
        0 & \vect{S}_{11} & 0  & 0 & \vect{S}_{12} & 0 & 0 & \vect{S}_{13} & 0 \\
        0 & 0 & \vect{S}_{11}  & 0 & 0 & \vect{S}_{12} & 0 & 0 & \vect{S}_{13} \\
        \vect{S}_{12} & 0 & 0  & \vect{S}_{22} & 0 & 0 & \vect{S}_{23} & 0 & 0 \\
        0 & \vect{S}_{12} & 0  & 0 & \vect{S}_{22} & 0 & 0 & \vect{S}_{23} & 0 \\
        0 & 0 & \vect{S}_{12}  & 0 & 0 & \vect{S}_{22} & 0 & 0 & \vect{S}_{23} \\
        \vect{S}_{13} & 0 & 0  & \vect{S}_{23} & 0 & 0 & \vect{S}_{33} & 0 & 0 \\
        0 & \vect{S}_{13} & 0  & 0 & \vect{S}_{23} & 0 & 0 & \vect{S}_{33} & 0 \\
        0 & 0 & \vect{S}_{13}  & 0 & 0 & \vect{S}_{23} & 0 & 0 & \vect{S}_{33} \\
    \end{bmatrix}.
\end{equation}

\begin{remark}
    As is standard, for axisymmetric elements, the residual vectors and tangent matrix are multiplied by the factor $ 2\pi R$ during the numerical integration.
\end{remark}

\subsection{F-bar modification for volumetric locking} 
\label{sec:appendix-f-bar-element}

To alleviate volumetric locking in quasi-incompressible materials in the finite deformation regime, \cite{netoDesignSimpleLow1996} developed an ad-hoc approach called the F-bar method for fully-integrated first-order quadrilateral and hexahedral elements\footnote{In a subsequent article, \cite{netoFbarbasedLinearTriangles2005} proposed an element patch-based F-bar element formulation for first-order simplex elements; however, the implementation is not straightforward. Hence, we did not adopt and implement this formulation in this work for simplex elements.}. Here, we adopt and extend the three-dimensional F-bar element formulation for PE gels. The central idea of the F-bar method is to construct a so-called modified deformation gradient, $\overline{\vect{F}}$, in which the volumetric part is evaluated at the centroid and the deviatoric part is evaluated at the integration point as usual. This leads to the following definition of $\vect{\overline{F}}$ for a fully-integrated hexahedral element.
\begin{equation} \label{eq:f-bar-def}
    \overline{\vect{F}} = 
    \vect{F}^{\mathrm{dev}} \vect{F}_0^{\mathrm{vol}} 
    = \left( \frac{\det \vect{F}_0}{\det \vect{F}} \right)^{1/3} \vect{F}.
\end{equation}
where $\vect{F}$ is the deformation gradient at the integration point and $\vect{F}_0$ is the deformation gradient at the centroid of the element. Once the modified deformation gradient, $\overline{\vect{F}}$, is calculated, the second Piola-Kirchhoff stress, $\vect{S}$, can be obtained using,
\begin{equation} \label{eq:PK2-stress-f-bar-3D}
    \vect{S}
    = J \vect{F}^{-1} \boldsymbol{\upsigma} \vect{F}^{-\top} 
    = \left( \frac{\det \vect{F}_0}{\det \vect{F}} \right)^{-1/3} \vect{S} \rvert_{\vect{F} = \overline{\vect{F}}}.
\end{equation}
By substituting the expression for Eq. \eqref{eq:PK2-stress-f-bar-3D} in the mechanical residual for an element in Eq. \eqref{eq:elem-residuals}$_1$, we now obtain the modified element mechanical residual as follows,
\begin{equation} \label{eq:f-bar-residual}
    \overline{r}^{u_i}_{a}
    = - \int\limits_{\Omega_0^e} 
    \left( \frac{\det \vect{F}_0}{\det \vect{F}} \right)^{-2/3}
    \frac{\partial N^{\vect{u}}_a}{\partial X_I}  
     \overline{F}_{iJ} \overline{S}_{JI} \ dV 
    + \int\limits_{\Omega_0^e} \rho_{\R} N^{\vect{u}}_a B_i \ dV
    + \int\limits_{\Gamma_{\vect{T}}^e} N^{\vect{u}}_a T_i \ dS,
\end{equation}
where $\overline{S}_{JI}$ is the second Piola-Kirchhoff stress tensor components evaluated using $\vect{F} = \overline{\vect{F}}$. Accordingly, using the definition of the modified deformation gradient in Eq. \eqref{eq:f-bar-def} and the modified element mechanical residual in Eq. \eqref{eq:f-bar-residual}, the following expressions for the components of the modified element tangent stiffness matrix are obtained after tedious application of the chain rule and tensor algebra,
\begingroup
\allowdisplaybreaks
\begin{align*} \label{eq:elem-tangent-f-bar}
    & k^{u_i u_k}_{ab}
    = - \frac{\partial \overline{r}^{u_i}_a} {\partial u_b^k} \\
    & \qquad = \int\limits_{\Omega_0^e}  \frac{\partial N^{\vect{u}}_a}{\partial X_J} 
    \left[ \left( \frac{\det \vect{F}_0}{\det \vect{F}} \right)^{-1/3}
    \left( \overline{S}_{JL} \delta_{ik} + \overline{F}_{iI}  \left( 2 \frac{ d \overline{S}_{IJ}}{d \overline{C}_{KL}} \right) \overline{F}_{kK} \right) \right] \frac{\partial N^{\vect{u}}_b}{\partial X_L} \ dV \\
    & \qquad \quad
    + \frac{1}{3} \int\limits_{\Omega_0^e} \frac{\partial N^{\vect{u}}_a}{\partial X_J} 
    \left[ \left( \frac{\det \vect{F}_0}{\det \vect{F}} \right)^{-2/3}  \overline{F}_{iI}  \left( 2 \frac{d \overline{S}_{IJ}}{d \overline{C}_{MN}}  \right)  \overline{C}_{MN}  
    \left( F_{0,kL}^{-\T} \frac{\partial N^{\vect{u}}_{0b}}{\partial X_L} - \overline{F}_{kL}^{-\T} \frac{\partial N^{\vect{u}}_b}{\partial X_L} \right) \right] dV \\
    & \qquad \quad
    - \frac{1}{3} \int\limits_{\Omega_0^e} \frac{\partial N^{\vect{u}}_a}{\partial X_J} 
    \left[ \left( \frac{\det \vect{F}_0}{\det \vect{F}} \right)^{-2/3}   \overline{P}_{iJ} 
    \left( F_{0,kL}^{-\T} \frac{\partial N^{\vect{u}}_{0b}}{\partial X_L} - \overline{F}_{kL}^{-\T} \frac{\partial N^{\vect{u}}_b}{\partial X_L} \right) \right]  dV, \\[6pt]
    & k^{u_i \mu}_{ab}
    = - \frac{\partial \overline{r}^{u_i}_a} {\partial \mu^w} 
    =  \int\limits_{\Omega_0^e}  \left( \frac{\det \vect{F}_0}{\det \vect{F}} \right)^{-2/3}  \frac{\partial N^{\vect{u}}_a}{\partial X_I} \left( \overline{F}_{iJ} \frac{d \overline{S}_{IJ}}{d \mu^w} \right) N^{\mu}_b \ dV, 
    \stepcounter{equation}\tag{\theequation} \\
    & k^{u_i \omega_k}_{ab}
    = - \frac{\partial \overline{r}^{u_i}_a} {\partial \omega^{\beta_k}} 
    =  \int\limits_{\Omega_0^e}  \left( \frac{\det \vect{F}_0}{\det \vect{F}} \right)^{-2/3}  \frac{\partial N^{\vect{u}}_a}{\partial X_I} \left( \overline{F}_{iJ} \frac{d \overline{S}_{IJ}}{d \omega^{\beta_k}} \right) N^{\omega}_b \ dV.
\end{align*}
\endgroup
where $\overline{(\bullet)}$ represents the quantity being evaluated using $\vect{F} = \overline{\vect{F}}$. By using the definition of $\overline{\vect{F}}$ for the three-dimensional case in Eq. \eqref{eq:f-bar-def}, the matrix form of the element mechanical residual for the F-bar element in Eq. \eqref{eq:f-bar-residual} can be written as follows,
\begin{equation}
    \overline{\vect{r}}^{\vect{u}}_e
    = - \int\limits_{\Omega_0^e}  \left( \frac{\det \vect{F}_0}{\det \vect{F}} \right)^{-2/3} (\mat{B}_{\vect{u}} \Sigma_{\vect{F}})^{\T} \widetilde{\vect{S}} \ dV 
    + \int\limits_{\Omega_0^e} \rho_{\R} \mat{N}_{\vect{u}}^{\T} \vect{B} \ dV
    + \int\limits_{\Gamma_{\vect{T}}^e} \mat{N}_{\vect{u}}^{\T}  \vect{T} \ dS
\end{equation}

The matrix-vector forms of the components of the element tangent stiffness matrix for the three-dimensional F-bar element in Eq. \eqref{eq:elem-tangent-f-bar} are given by,
\begingroup
\allowdisplaybreaks
\begin{align*} \label{eq:elem-tangent-f-bar-matrx}
    \mat{K}^{\vect{u}\vect{u}}_e
    & = \int\limits_{\Omega_0^e} \left( \frac{\det \vect{F}_0}{\det \vect{F}} \right)^{-1/3} 
    \left[ \mat{G}_{\vect{u}}^{\T} \Sigma_{\vect{S}} \mat{G}_{\vect{u}} 
    + (\mat{B_{\vect{u}} \Sigma_{\vect{F}}})^{\T} \vect{D}_{\mathds{C}} (\mat{B}_{\vect{u}} \Sigma_{\vect{F}}) \right] dV \\
    & \quad
    + \int\limits_{\Omega_0^e} 
    \left( \frac{\det \vect{F}_0}{\det \vect{F}} \right)^{-2/3}  
    \left( \mat{G}_{\vect{u}}^{\T} \mat{Q}_{\mathrm{R}}^0 \mat{G}_{\vect{u}}^0 
    - \mat{G}_{\vect{u}}^{\T} \mat{Q}_{\mathrm{R}} \mat{G}_{\vect{u}} \right) dV, \\
    \mat{K}^{\vect{u} \mu}_e
    & = \int\limits_{\Omega_0^e} \left( \frac{\det \vect{F}_0}{\det \vect{F}} \right)^{-2/3} 
    \mat{G}_{\vect{u}}^{\T} \mat{a}_{\vect{u}\mu} \mat{N}_{\mu} \ dV , 
    \stepcounter{equation}\tag{\theequation} \\
    \mat{K}^{\vect{u}\omega_k}_e 
    & =  \int\limits_{\Omega_0^e} \left( \frac{\det \vect{F}_0}{\det \vect{F}} \right)^{-2/3}
    \mat{G}_{\vect{u}}^{\T} \mat{a}_{\vect{u} \omega_k} \mat{N}_{\omega} \ dV,
\end{align*}
\endgroup
where $\mat{G}^0_{\vect{u}}$ is the non-symmetric gradient matrix of the interpolation functions at the centroid of the 8-node hexahedral element, and $\mat{Q}_{\mathrm{R}}^0$ and $\mat{Q}_{\mathrm{R}}$ are the matrix form of the following fourth-order tensors
\begin{equation}
\begin{aligned}
    (\mathds{Q}_{\mathrm{R}}^0)_{iJkL} 
    = \frac{1}{3} \overline{F}_{iI} \left( 2 \frac{d \overline{S}_{IJ}}{d \overline{C}_{MN}} \right) \overline{C}_{MN} F_{0,kL}^{-\T} 
    - \frac{1}{3} \overline{P}_{iJ}  F_{0,kL}^{-\T}, \\[6pt]
    (\mathds{Q}_{\mathrm{R}})_{iJkL} 
    = \frac{1}{3} \overline{F}_{iI} \left( 2 \frac{d \overline{S}_{IJ}}{d \overline{C}_{MN}} \right) \overline{C}_{MN} \overline{F}_{kL}^{-\T} 
    - \frac{1}{3} \overline{P}_{iJ}  \overline{F}_{kL}^{-\T}.
\end{aligned}
\end{equation}

We used the following convention from Table \ref{tab:tensor_to_matrix_3D} to transform the fourth-order tensors, $\mat{Q}_{\mathrm{R}}^0$ and $\mat{Q}_{\mathrm{R}}$, into matrices for three-dimensional cases, where $(m,n)$ are indices of the matrix and $(i,J,k,L)$ are the indices of the fourth-order tensor.
\begin{table}[ht]
\centering
\begin{tabular}{ c  c  c  c }
    \toprule \toprule
    $m$ &   $(i,J)$   &    $n$    &     $(k,L)$  \\
    \toprule \toprule
    1   &   (1,1)     &     1     &     (1,1) \\ 
    2   &   (2,1)     &     2     &     (2,1) \\ 
    3   &   (3,1)     &     3     &     (3,1) \\ 
    4   &   (1,2)     &     4     &     (1,2) \\ 
    5   &   (2,2)     &     5     &     (2,2) \\
    6   &   (3,2)     &     6     &     (2,3) \\
    7   &   (1,3)     &     7     &     (1,3) \\
    8   &   (2,3)     &     8     &     (2,3) \\
    9   &   (3,3)     &     9     &     (3,3) \\
    \bottomrule
\end{tabular}
\caption{Convention for transforming a fourth-order tensor to a matrix for three-dimensional cases.}
\label{tab:tensor_to_matrix_3D}
\end{table}

For two-dimensional plane strain cases, the definition of $\overline{\vect{F}}$ is slightly different, and consequently, the element formulation is also different. For two-dimensional plane strain case, $\overline{\vect{F}}$, is defined as,
\begin{equation} \label{eq:f-bar-2D-pe}
\overline{\vect{F}} =
\left[
\begin{array}{c c | c}
    \multicolumn{2}{c|}{\multirow{2}{*}{$[\overline{\vect{F}}_{\mathrm{pe}}]$}}  &   0 \\
    \multicolumn{2}{c|}{} &   0 \\
    \hline
    0   &   0    &   1   \\
\end{array} \right],
\end{equation}
where $[\overline{\vect{F}}_{\mathrm{pe}}]_{2 \times 2}$ is defined as the following
\begin{equation}
    \overline{\vect{F}}_{\mathrm{pe}} = \vect{F}^{\mathrm{dev}} \vect{F}_0^{\mathrm{vol}} 
    = \left( \frac{\det \vect{F}_0}{\det \vect{F}} \right)^{1/2} \vect{F}_{\mathrm{pe}},
    \quad \mathrm{where} \ 
    \vect{F}_{\mathrm{pe}}
    =
    \begin{bmatrix}
        F_{11}  &   F_{12} \\
        F_{21}  &   F_{22}
    \end{bmatrix}.
\end{equation}

Using the definition of $\overline{\vect{F}}$ in Eq. \eqref{eq:f-bar-2D-pe}, for plane strain cases, the second Piola-Kirchhoff stress tensor can be computed as
\begin{equation} \label{eq:PK2-stress-f-bar-2D}
    \vect{S}
    = \left( \frac{\det \vect{F}_0}{\det \vect{F}} \right)^{-1/2} \vect{S} \rvert_{\vect{F} = \overline{\vect{F}}},
\end{equation}

Substituting Eq. \eqref{eq:PK2-stress-f-bar-2D} in the standard element mechanical residual Eq. \eqref{eq:elem-residuals}$_1$ leads to the modified residual as follows,
\begin{equation}
\begin{aligned}
    \overline{r}^{u_i}_a
    & = - \int\limits_{\Omega_0^e} 
    \left( \frac{\det \vect{F}_0}{\det \vect{F}} \right)^{-1/2}
    \frac{\partial N_a}{\partial X_I}  
     \overline{F}_{iJ} \overline{S}_{JI} \ dV 
    + \int\limits_{\Omega_0^e} \rho_{\R} N_a B_i \ dV
    + \int\limits_{\Gamma_{\vect{T}}^e} N_a T_i \ dS, 
\end{aligned}
\end{equation}

As is standard, following the Newton-Raphson solution procedure, the element tangent stiffness matrix can be obtained as follows,
\begingroup
\allowdisplaybreaks
\begin{align*}
    & k^{u_i u_k} _{ab}
    && = - \frac{\partial \overline{r}_{u_i}^a} {\partial u^b_k}, \\
    & && = \int\limits_{\Omega_0^e} \frac{\partial N_a}{\partial X_J} 
    \left[ \overline{S}_{JL} \delta_{ik} + \overline{F}_{iI}  \left( 2 \frac{\partial \overline{S}_{IJ}}{\partial \overline{C}_{KL}} \right) \overline{F}_{kK} \right]  \frac{\partial N_b}{\partial X_L}  \ dV \\
    & && \quad 
    + \frac{1}{2}  \int\limits_{\Omega_0^e} \frac{\partial N_a}{\partial X_J}  
    \left[ \left( \frac{\det \vect{F}_0} {\det \vect{F}} \right)^{-1/2}  
    \overline{F}_{iI}  \left( 2 \frac{\partial \overline{S}_{IJ}}{\partial \overline{C}_{MN}} \right) \overline{C}_{MN} 
    \left( F_{0,kL}^{-\T} \frac{\partial N_{0b}}{\partial X_L} - \overline{F}_{kL}^{-\T} \frac{\partial N_b}{\partial X_L} \right) \right] \ dV, \\
    & k^{u_i \mu}_{ab}
    && = - \frac{\partial \overline{r}^{u_i}_{a}} {\partial \mu^w_b}
    = \int\limits_{\Omega_0^e} \left( \frac{\det \vect{F}_0} {\det \vect{F}} \right)^{-1/2} \frac{\partial N^{\vect{u}}_a}{\partial X_I} \left( \overline{F}_{iJ} \frac{d S_{IJ}}{d \mu^w} \right) N^{\mu}_b \ dV, 
    \stepcounter{equation}\tag{\theequation} \\
    & k^{u_i \omega_k}_{ab}
    && = - \frac{\partial \overline{r}^{u_i}_{a}} {\partial \omega^{\beta_k}_b}
    = \int\limits_{\Omega_0^e} \left( \frac{\det \vect{F}_0} {\det \vect{F}} \right)^{-1/2} \frac{\partial N^{\vect{u}}_a}{\partial X_I} \left( \overline{F}_{iJ} \frac{d S_{IJ}}{d \omega^{\beta_k}}  \right)  N^{\omega}_b \ dV.
\end{align*}
\endgroup

The vector form of the mechanical residual for the element is given by,
\begin{equation}
    \overline{\vect{r}}^{\vect{u}}_e
    = - \int\limits_{\Omega_0^e}  \left( \frac{\det \vect{F}_0}{\det \vect{F}} \right)^{-1/2} (\mat{B}_{\vect{u}} \Sigma_{\vect{F}})^{\T} \widetilde{\vect{S}} \ dV 
    + \int\limits_{\Omega_0^e} \rho_{\R} \mat{N}_{\vect{u}}^{\T} \vect{B} \ dV
    + \int\limits_{\Gamma_{\vect{T}}^e} \mat{N}_{\vect{u}}^{\T}  \vect{T} \ dS.
\end{equation}

With the F-bar modification, the matrix forms of the components of the element tangent stiffness matrix can be written as follows,
\begingroup
\allowdisplaybreaks
\begin{align*}
    \mat{K}^{\vect{u}\vect{u}}_e
    & = \int\limits_{\Omega_0^e} \left[ \mat{G}_{\vect{u}}^{\T} \Sigma_{\vect{S}} \mat{G}_{\vect{u}} 
    + (\mat{B_{\vect{u}} \Sigma_{\vect{F}}})^{\T} \vect{D}_{\mathds{C}} (\mat{B}_{\vect{u}} \Sigma_{\vect{F}}) \right] dV \\
    & \quad
    + \int\limits_{\Omega_0^e} 
    \left( \frac{\det \vect{F}_0}{\det \vect{F}} \right)^{-1/2} 
    \left( \mat{G}_{\vect{u}}^{\T} \mat{Q}_{\mathrm{R}}^0 \mat{G}_{\vect{u}}^0 
    -  \mat{G}_{\vect{u}}^{\T} \mat{Q}_{\mathrm{R}} \mat{G}_{\vect{u}} \right)  dV,  \\
    \mat{K}^{\vect{u} \mu}_e
    & = \int\limits_{\Omega_0^e} \left( \frac{\det \vect{F}_0}{\det \vect{F}} \right)^{-1/2} 
    \mat{G}_{\vect{u}}^{\T} \mat{a}_{\vect{u}\mu} \mat{N}_{\mu} \ dV ,
    \stepcounter{equation}\tag{\theequation} \\
    \mat{K}^{\vect{u}\omega_k}_e 
    & =  \int\limits_{\Omega_0^e} \left( \frac{\det \vect{F}_0}{\det \vect{F}} \right)^{-1/2}
    \mat{G}_{\vect{u}}^{\T} \mat{a}_{\vect{u} \omega_k} \mat{N}_{\omega} \ dV
\end{align*}
\endgroup
where $\mat{G}^0_{\vect{u}}$ is the non-symmetric gradient matrix of the interpolation functions at the centroid of the 4-node quadrilateral element, and $\mat{Q}_{\mathrm{R}}^0$ and $\mat{Q}_{\mathrm{R}}$ are the matrix form of the following fourth-order tensors
\begin{equation}
\begin{aligned}
    (\mathds{Q}_{\mathrm{R}}^0)_{iJkL} 
    = \frac{1}{2} \overline{F}_{iI} \left( 2 \frac{d \overline{S}_{IJ}}{d \overline{C}_{MN}} \right) \overline{C}_{MN} F_{0,kL}^{-\T} , 
    \\
    (\mathds{Q}_{\mathrm{R}})_{iJkL} 
    = \frac{1}{2} \overline{F}_{iI} \left( 2 \frac{d \overline{S}_{IJ}}{d \overline{C}_{MN}} \right) \overline{C}_{MN} \overline{F}_{kL}^{-\T} .
\end{aligned}
\end{equation}

We followed the convention listed in Table \ref{tab:tensor_to_matrix_2D} to transform the fourth-order tensors, $\mat{Q}_{\mathrm{R}}^0$ and $\mat{Q}_{\mathrm{R}}$, into matrices to calculate element tangent stiffness matrix,
\begin{table}[H]
\centering
\begin{tabular}{ c  c  c  c }
    \toprule \toprule
    $m$ &   $(i,j)$   &    $n$    &    $(k,l)$  \\
    \toprule \toprule
    1   &   (1,1)     &     1     &     (1,1)   \\ 
    2   &   (2,1)     &     2     &     (2,1)   \\ 
    3   &   (1,2)     &     3     &     (1,2)   \\ 
    4   &   (2,2)     &     4     &     (2,2)   \\ 
    \bottomrule
\end{tabular}
\caption{Convention for transforming a fourth-order tensor to a matrix for the plane strain case.}
\label{tab:tensor_to_matrix_2D}
\end{table}

In our implementation, when F-bar elements are chosen, the constitutive laws and their derivatives presented in Subsection \ref{sec:constitutive-update} and Appendix \ref{sec:appendix-tangents} are evaluated using $\overline{\vect{F}}$ for bilinear quadrilateral and trilinear hexahedral elements in our implementation albeit we do not use the notation $\overline{(\bullet)}$.

\begin{remark}
    F-bar modification for the axisymmetric case is similar to the 3D case except $F_{33} = r/R$ and $F_{3J} = F_{i3} = 0$. The details can be found in \cite{netoDesignSimpleLow1996} and \cite{chesterFiniteElementImplementation2015}; hence, it is omitted here. Additionally, from a strict technical point, we should also perform modifications of the components of the element tangent stiffness matrix, $k_{ab}^{\mu u_k}$, $k_{ab}^{\omega_1 u_k}$, and $k_{ab}^{\omega_2 u_k}$ in Eq. \eqref{eq:elem-tangents}. However, as the F-bar implementation for non-ionic gels in \cite{chesterFiniteElementImplementation2015} demonstrated that the effect is negligible, we ignored these complicated modifications in our current F-bar implementation.
\end{remark}

\begin{landscape}
For a three-dimensional element with two ions, $\beta_1$ and $\beta_2$, the element tangent stiffness matrix and element residual vector can be written as,
\begingroup
\begin{gather*}
\renewcommand{\arraystretch}{1.55}
\mat{K}_e =
\begin{bmatrix}
    k_{11}^{u_1 u_1} & k_{11}^{u_1 u_2} & k_{11}^{u_1 u_3} & k_{11}^{u_1 \mu} & k_{11}^{u_1 \omega_1} & k_{11}^{u_1 \omega_2} 
    & \cdots & \cdots & \cdots & 
    k_{ 1 n_{\mathrm{en}} }^{u_1 u_1} & k_{ 1 n_{\mathrm{en}} }^{u_1 u_2} & k_{ 1 n_{\mathrm{en}} }^{u_1 u_3} & k_{ 1 n_{\mathrm{en}} }^{u_1 \mu} & k_{ 1 n_{\mathrm{en}} }^{u_1 \omega_1} & k_{ 1 n_{\mathrm{en}} }^{u_1 \omega_2} \\
    k_{11}^{u_2 u_1} & k_{11}^{u_2 u_2} & k_{11}^{u_2 u_3} & k_{11}^{u_2 \mu} & k_{11}^{u_2 \omega_1} & k_{11}^{u_2 \omega_2} 
     & \cdots & \cdots & \cdots & 
    k_{ 1 n_{\mathrm{en}} }^{u_2 u_1} & k_{ 1 n_{\mathrm{en}} }^{u_2 u_2} & k_{ 1 n_{\mathrm{en}} }^{u_2 u_3} & k_{ 1 n_{\mathrm{en}} }^{u_2 \mu} & k_{ 1 n_{\mathrm{en}} }^{u_2 \omega_1} & k_{ 1 n_{\mathrm{en}} }^{u_2 \omega_2} \\
    k_{11}^{u_3 u_1} & k_{11}^{u_3 u_2} & k_{11}^{u_3 u_3} & k_{11}^{u_3 \mu} & k_{11}^{u_3 \omega_1} & k_{11}^{u_3 \omega_2} 
     & \cdots & \cdots & \cdots & 
    k_{ 1 n_{\mathrm{en}} }^{u_3 u_1} & k_{ 1 n_{\mathrm{en}} }^{u_3 u_2} & k_{ 1 n_{\mathrm{en}} }^{u_3 u_3} & k_{ 1 n_{\mathrm{en}} }^{u_3 \mu} & k_{ 1 n_{\mathrm{en}} }^{u_3 \omega_1} &  k_{ 1 n_{\mathrm{en}} }^{u_3 \omega_1} \\
    k_{11}^{\mu u_1} & k_{11}^{\mu u_2} & k_{11}^{\mu u_3} & k_{11}^{\mu \mu} & k_{11}^{\mu \omega_1} & k_{11}^{\mu \omega_2} 
     & \cdots & \cdots & \cdots & 
    k_{ 1 n_{\mathrm{en}} }^{\mu u_1} & k_{ 1 n_{\mathrm{en}} }^{\mu u_2} & k_{ 1 n_{\mathrm{en}} }^{\mu u_3} & k_{ 1 n_{\mathrm{en}} }^{\mu \mu} & k_{ 1 n_{\mathrm{en}} }^{\mu \omega_1} & k_{ 1 n_{\mathrm{en}} }^{\mu \omega_2}\\
    k_{11}^{\omega_1 u_1} & k_{11}^{\omega_1 u_2} & k_{11}^{\omega_1 u_3} & k_{11}^{\omega_1 \mu} & k_{11}^{\omega_1 \omega_1} & k_{11}^{\omega_1 \omega_1} 
     & \cdots & \cdots & \cdots & 
    k_{ 1 n_{\mathrm{en}} }^{\omega_1 u_1} & k_{ 1 n_{\mathrm{en}} }^{\omega_1 u_2} & k_{ 1 n_{\mathrm{en}} }^{\omega_1 u_3} & k_{ 1 n_{\mathrm{en}} }^{\omega_1 \mu} & k_{ 1 n_{\mathrm{en}} }^{\omega_1 \omega_1} & k_{ 1 n_{\mathrm{en}} }^{\omega_1 \omega_2} \\
    k_{11}^{\omega_2 u_1} & k_{11}^{\omega_2 u_2} & k_{11}^{\omega_2 u_3} & k_{11}^{\omega_2 \mu} & k_{11}^{\omega_2 \omega_1} & k_{11}^{\omega_2 \omega_2} 
    & \cdots & \cdots & \cdots & 
    k_{ 1 n_{\mathrm{en}} }^{\omega_2 u_1} & k_{ 1 n_{\mathrm{en}} }^{\omega_2 u_2} & k_{ 1 n_{\mathrm{en}} }^{\omega_2 u_3} & k_{ 1 n_{\mathrm{en}} }^{\omega_2 \mu} & k_{ 1 n_{\mathrm{en}} }^{\omega_2 \omega_1} & k_{ 1 n_{\mathrm{en}} }^{\omega_2 \omega_2} \\
    \vdots &  & \vdots  &  & \vdots  & & \ddots & & & & \vdots & & \vdots &  & \vdots \\
    \vdots &  & \vdots  &  & \vdots  & & & \ddots & & & \vdots & & \vdots &  & \vdots \\
    \vdots &  & \vdots  &  & \vdots  & & & & \ddots & & \vdots & & \vdots &  & \vdots \\
    k_{ n_{\mathrm{en}} 1 }^{u_1 u_1} & k_{ n_{\mathrm{en}} 1 }^{u_1 u_2} & k_{ n_{\mathrm{en}} 1 }^{u_1 u_3} & k_{ n_{\mathrm{en}} 1 }^{u_1 \mu} & k_{ n_{\mathrm{en}} 1 }^{u_1 \omega_1} & k_{ n_{\mathrm{en}} 1 }^{u_1 \omega_2} 
    & \cdots & \cdots & \cdots & 
    k_{ n_{\mathrm{en}} n_{\mathrm{en}}}^{u_1 u_1} & k_{ n_{\mathrm{en}} n_{\mathrm{en}}}^{u_1 u_2} & k_{ n_{\mathrm{en}} n_{\mathrm{en}}}^{u_1 u_3} & k_{ n_{\mathrm{en}} n_{\mathrm{en}}}^{u_1 \mu} & k_{ n_{\mathrm{en}} n_{\mathrm{en}}}^{u_1 \omega_1} & k_{ n_{\mathrm{en}} n_{\mathrm{en}}}^{u_1 \omega_2} \\
    k_{ n_{\mathrm{en}} 1 }^{u_2 u_1} & k_{ n_{\mathrm{en}} 1 }^{u_2 u_2} & k_{ n_{\mathrm{en}} 1 }^{u_2 u_3} & k_{ n_{\mathrm{en}} 1 }^{u_2 \mu} & k_{ n_{\mathrm{en}} 1 }^{u_2 \omega_1} & k_{ n_{\mathrm{en}} 1 }^{u_2 \omega_2} 
    & \cdots & \cdots & \cdots & 
    k_{ n_{\mathrm{en}} n_{\mathrm{en}}}^{u_2 u_1} & k_{ n_{\mathrm{en}} n_{\mathrm{en}}}^{u_2 u_2} & k_{ n_{\mathrm{en}} n_{\mathrm{en}}}^{u_2 u_3} & k_{ n_{\mathrm{en}} n_{\mathrm{en}}}^{u_2 \mu} & k_{ n_{\mathrm{en}} n_{\mathrm{en}}}^{u_2 \omega_1} & k_{ n_{\mathrm{en}} n_{\mathrm{en}}}^{u_2 \omega_2}\\
    k_{ n_{\mathrm{en}} 1 }^{u_3 u_1} & k_{ n_{\mathrm{en}} 1 }^{u_3 u_2} & k_{ n_{\mathrm{en}} 1 }^{u_3 u_3} & k_{ n_{\mathrm{en}} 1 }^{u_3 \mu} & k_{ n_{\mathrm{en}} 1 }^{u_3 \omega_1} & k_{ n_{\mathrm{en}} 1 }^{u_3 \omega_2} 
    & \cdots & \cdots & \cdots & 
    k_{ n_{\mathrm{en}} n_{\mathrm{en}}}^{u_3 u_1} & k_{ n_{\mathrm{en}} n_{\mathrm{en}}}^{u_3 u_2} & k_{ n_{\mathrm{en}} n_{\mathrm{en}}}^{u_3 u_3} & k_{ n_{\mathrm{en}} n_{\mathrm{en}}}^{u_3 \mu} & k_{ n_{\mathrm{en}} n_{\mathrm{en}}}^{u_3 \omega_1} & k_{ n_{\mathrm{en}} n_{\mathrm{en}}}^{u_3 \omega_2} \\
    k_{ n_{\mathrm{en}} 1 }^{\mu u_1} & k_{ n_{\mathrm{en}} 1 }^{\mu u_2} & k_{ n_{\mathrm{en}} 1 }^{\mu u_3} & k_{ n_{\mathrm{en}} 1 }^{\mu \mu} & k_{ n_{\mathrm{en}} 1 }^{\mu \omega_1} & k_{ n_{\mathrm{en}} 1 }^{\mu \omega_2} 
    & \cdots & \cdots & \cdots & 
    k_{ n_{\mathrm{en}} n_{\mathrm{en}}}^{\mu u_1} & k_{ n_{\mathrm{en}} n_{\mathrm{en}}}^{\mu u_2} & k_{ n_{\mathrm{en}} n_{\mathrm{en}}}^{\mu u_3} & k_{ n_{\mathrm{en}} n_{\mathrm{en}}}^{\mu \mu} & k_{ n_{\mathrm{en}} n_{\mathrm{en}}}^{\mu \omega_1} & k_{ n_{\mathrm{en}} n_{\mathrm{en}}}^{\mu \omega_2} \\
    k_{ n_{\mathrm{en}} 1 }^{\omega_1 u_1} & k_{ n_{\mathrm{en}} 1 }^{\omega_1 u_2} & k_{ n_{\mathrm{en}} 1 }^{\omega_1 u_3} & k_{ n_{\mathrm{en}} 1 }^{\omega_1 \mu} & k_{ n_{\mathrm{en}} 1 }^{\omega_1 \omega_1} & k_{ n_{\mathrm{en}} 1 }^{\omega_1 \omega_2} 
    & \cdots & \cdots & \cdots & 
    k_{ n_{\mathrm{en}} n_{\mathrm{en}}}^{\omega_1 u_1} & k_{ n_{\mathrm{en}} n_{\mathrm{en}}}^{\omega_1 u_2} & k_{ n_{\mathrm{en}} n_{\mathrm{en}}}^{\omega_1 u_3} & k_{ n_{\mathrm{en}} n_{\mathrm{en}}}^{\omega_1 \mu} & k_{ n_{\mathrm{en}} n_{\mathrm{en}}}^{\omega_1 \omega_1} & k_{ n_{\mathrm{en}} n_{\mathrm{en}}}^{\omega_1 \omega_2} \\
    k_{ n_{\mathrm{en}} 1 }^{\omega_2 u_1} & k_{ n_{\mathrm{en}} 1 }^{\omega_2 u_2} & k_{ n_{\mathrm{en}} 1 }^{\omega_2 u_3} & k_{ n_{\mathrm{en}} 1 }^{\omega_2 \mu} & k_{ n_{\mathrm{en}} 1 }^{\omega_2 \omega_1} & k_{ n_{\mathrm{en}} 1 }^{\omega_2 \omega_2} 
    & \cdots & \cdots & \cdots & 
    k_{ n_{\mathrm{en}} n_{\mathrm{en}}}^{\omega_2 u_1} & k_{ n_{\mathrm{en}} n_{\mathrm{en}}}^{\omega_2 u_2} & k_{ n_{\mathrm{en}} n_{\mathrm{en}}}^{\omega_2 u_3} & k_{ n_{\mathrm{en}} n_{\mathrm{en}}}^{\omega_2 \mu} & k_{ n_{\mathrm{en}} n_{\mathrm{en}}}^{\omega_2 \omega_1} & k_{ n_{\mathrm{en}} n_{\mathrm{en}}}^{\omega_2 \omega_2}
\end{bmatrix}, \\ \\
\text{and} \qquad \qquad \qquad
\vect{r}_e =
\begin{Bmatrix}
    r_1^{u_1} & r_1^{u_2} & r_1^{u_3} & r_1^{\mu} & r_1^{\omega_1} & r_1^{\omega_2} &
    \cdots & 
    r_{n_{\mathrm{en}}}^{u_1} & r_{n_{\mathrm{en}}}^{u_2} & r_{n_{\mathrm{en}}}^{u_3} & r_{n_{\mathrm{en}}}^{\mu} & r_{n_{\mathrm{en}}}^{\omega_1} & r_{n_{\mathrm{en}}}^{\omega_2}
\end{Bmatrix}^{\T}.
\stepcounter{equation}\tag{\theequation} 
\end{gather*}
\endgroup

\end{landscape}

\section{Derivations and computations related to the constitutive model}
\setcounter{table}{0}

It is possible to obtain the components of the free energy density functions by scaling the standard energy potential expressions from the dry state to the hydrated reference state volume. However, we derived them using a statistical mechanics approach to demonstrate that the consistency and low-length-scale physics have been conserved.

\subsection{Derivation of free energy potentials for pre-swollen gels}
\label{sec:appendix-free-energy-derivation}

For a polymer network with $N$ Gaussian chains, the total change in entropy for the deformation from the dry state to the current state is the summation of changes in entropy for all the chains \citep{treloarPhysicsRubberElasticity1975}, \emph{i.e.},
\begin{equation}
    \Delta S'_{\mathrm{net}} = - \frac{N k_B}{2} \left[ (\lambda_1')^2+(\lambda_2')^2+(\lambda_3')^2 - 3 - 2\ln J' \right],
\end{equation}
where the principal stretches $\lambda_1', \lambda_2', \lambda_3'$ are from the dry state to the current state. For elastomers undergoing swelling volume change, the additional dilation term was introduced in \cite{floryStatisticalMechanicsSwelling1950}. For unswollen elastomers, the volumetric dilatation term is zero and hence ignored in most literature on elastomeric networks. Similarly, for isotropic swelling stretch from the dry to the pre-swollen hydrated reference state, $\lambda^d$, we can write the change in entropy as,
\begin{equation}
\begin{aligned}
    \Delta S_{\mathrm{net}}^0 = & - \frac{N k_B}{2} \left[ 3(\lambda^d)^2 - 3 - 2 \ln J^d \right], \\
    = & - \frac{N k_B}{2} \left[ 3(\phi^p_0)^{-2/3} - 3 - 2 \ln (\phi^p_0)^{-1} \right]
\end{aligned}
\end{equation}
Therefore, the change in entropy from the hydrated state to the current state is,
\begin{equation}
\begin{aligned}
    \Delta S_{\mathrm{net,dry}} = & \Delta S'_{\mathrm{net}} - \Delta S_{\mathrm{net}}^d, \\
    = & - \frac{N k_B}{2} \left[ (J^s)^{2/3} ( {\lambda_1^e}^2 + {\lambda_2^e}^2 + {\lambda_3^e}^2) - 3(\phi^p_0)^{-2/3} - 2 \ln \left( \phi^p_0 J' \right) \right], \\
    = & - \frac{N k_B}{2} (\phi^p_0)^{-2/3} \left[ (\phi^p_0)^{2/3}(J^s)^{2/3} ( {\lambda_1^e}^2 + {\lambda_2^e}^2 + {\lambda_3^e}^2) - 3 - 2 (\phi^p_0)^{2/3} \ln \left( J \right) \right], \\
    = & - \frac{N k_B}{2} (\phi^p_0)^{-2/3} \left[ (\phi^p_0 J^s)^{2/3} ( {\lambda_1^e}^2 + {\lambda_2^e}^2 + {\lambda_3^e}^2) - 3 - 2 (\phi^p_0)^{2/3} \ln \left( J \right) \right].
\end{aligned}
\end{equation}
Here, $ \Delta S_{\mathrm{net,dry}} $ is defined as the change in entropy with respect to the dry state. Entropy change in per unit hydrated state volume is given by, 
\begin{equation}
\begin{aligned}
    \Delta S_{\mathrm{net}} = & \Delta S_{\mathrm{net,dry}} \left( \frac{dV_0}{dV_{\mathrm{dry}}} \right) = \phi^p_0 \Delta S_{\mathrm{net,dry}}, \\
    = & \frac{N k_B}{2} (\phi^p_0)^{1/3} \left[ (\phi^p_0 J^s)^{2/3} ( {\lambda_1^e}^2 + {\lambda_2^e}^2 + {\lambda_3^e}^2) - 3 - 2 (\phi^p_0)^{2/3} \ln \left( J \right) \right].
\end{aligned}
\end{equation}

For $\Delta U_{\mathrm{net}}$ change in internal energy from the hydrated state to the current state at any temperature $\theta$, the Helmholtz free energy density for mechanical stretching of the network is now given by,
\begin{equation}    \label{eq:free-energy-neo-hookean-stat-mech}
\begin{aligned}
    \Psi^{\mathrm{mech}} = & \Delta U_{\mathrm{net}} - \theta \Delta S_{\mathrm{net}}, \\
    = & \frac{N k_B \theta }{2} (\phi^p_0)^{1/3} \left[ (\phi^p_0 J^s)^{2/3} ( {\lambda_1^e}^2 + {\lambda_2^e}^2 + {\lambda_3^e}^2) - 3 - 2 (\phi^p_0)^{2/3} \ln \left( J \right) \right].
\end{aligned}
\end{equation}
In polymer networks, we generally take an ideal assumption that there is no energetic contribution to elasticity, hence, $\Delta U_{\mathrm{net}} = 0$. Let define $G_{\mathrm{dry}} = N k_B \theta$ be the shear modulus of the gel at the dry state, and $G = N k_B \theta (\phi^p_0)^{1/3} = G_{\mathrm{dry}} (\phi^p_0)^{1/3}$ be the shear modulus of the gel at the hydrated state \citep{chesterThermomechanicallyCoupledTheory2011}. Additionally, by definition, we know, $I_1 = (\phi^p_0 J^s)^{2/3} \tr (\vect{B}^e) = (\phi^p_0 J^s)^{2/3} \tr (\vect{C}^e) = (\phi^p_0 J^s)^{2/3} \left( {\lambda_1^e}^2 + {\lambda_2^e}^2 + {\lambda_3^e}^2 \right)$ is the first invariant of Cauchy deformation tensor. Hence, we can write,
\begin{equation} \label{eq:psi_r_nh_final}
\begin{aligned}
    \Psi^{\mathrm{mech}} = \frac{G}{2} \left[ I_1 - 3 - 2 (\phi^p_0)^{2/3} \ln \left( J \right) \right].
\end{aligned}
\end{equation}

Now, as is standard, the Helmholtz free energy for $N$ lattice sites in polymer-solvent mixture is given by \citep{floryThermodynamicsHighPolymer1942,doiIntroductionPolymerPhysics1996},
\begin{equation}
    \Delta A^{\mathrm{mix,pol}} = N k_B \theta \left[ (1-\phi^p)  \ln (1-\phi^p) +  \chi \phi^p (1-\phi^p) \right] 
\end{equation}
Thus, we can write the Helmholtz free energy density of the polymer-solvent mixture as,
\begin{equation} \label{eq:psi_r_2poly-mix}
\begin{aligned}
    \Psi^{\mathrm{mix,pol}} = \frac{\Delta A^{\mathrm{mix}}_{\mathrm{pol}}}{V_0}
    = & \frac{N}{V_0} k_B  \theta (1-\phi^p) \left[  \ln (1-\phi^p) +  \chi \phi^p  \right], \\
    = & \frac{N v_0}{V_0} \frac{k_B \theta}{v_0} (1-\phi^p)  \left[   \ln (1-\phi^p) +  \chi \phi^p  \right], \\
    = & \frac{v}{V_0} \frac{R \theta}{\mathcal{V}^w} (1-\phi^p)   \left[   \ln (1-\phi^p) +  \chi \phi^p  \right], \\
    = & \frac{\phi^p_0}{\phi^p} \frac{R \theta}{\mathcal{V}^w} (1-\phi^p)  \left[   \ln (1-\phi^p) + \chi \phi^p  \right], \\
    = & \phi^p_0 \left[\frac{R\theta}{\mathcal{V}^{w}}\frac{1-\phi^{p}}{\phi^{p}} \left(\ln \left( 1-\phi^{p} \right) + \chi\phi^{p} \right) \right],
\end{aligned}
\end{equation}
where $v_0$ is the volume of each lattice site, such that $v = N v_0$ represents the volume of the polymer-solvent mixture in the current configuration. We further used the incompressibility assumption so that $\frac{v}{V_0} = J = \phi^p_0 J^e J^s \approx \frac{\phi^p_0}{\phi^p}$.

The free energy density expression for the dilute ionic mixture in the hydrated reference state remains the same as that of the dry reference state. The derivation of the particular form used in this article is given by \cite{narayanCoupledElectrochemomechanicalTheory2022}; hence, we did not include it here.

\subsection{Local Newton-Raphson procedure for constitutive evaluation}    \label{sec:appendix-local-newton}

Now, by using the definition of polymer volume fraction, $\phi^p$, from Eq. \eqref{eq:polymer-vol-frac-def}, the mean pressure, $p$, from Eq. \eqref{eq:mean-pressure-def-2}, the Lagrange multiplier, $\mathcal{P}$, from Eq. \eqref{eq:lagrange-mult-def}, and the kinematic relation, $J^s= 1/\phi^p$, we can compute,
\begingroup
\begin{align*} 
    \frac{d \phi^p}{d C^w} 
    & = - \frac{\mathcal{V}^w (\phi^p)^2 }{\phi^p_0}, \\
    \frac{\partial p}{\partial \phi^p} 
    & = - \left[ \frac{G}{3\phi^p_0} \left( I_1 - 3 (\phi^p_0)^{2/3} \right) + \frac{\kappa}{\phi^p} \right], \\
    \frac{\partial p}{\partial C^w} 
    & = \left( \frac{\partial p}{\partial \phi^p}  \right) \left( \frac{d \phi^p}{d C^w}  \right), \\
    \frac{\partial p}{\partial \vect{F}} 
    & = - \left( \frac{2G}{3 \phi^p_0 J^s} \vect{F}  + \kappa  \vect{F}^{-\T} \right), 
    \stepcounter{equation}\tag{\theequation} \\
    \frac{\partial p}{\partial \vect{C}} 
    & = - \left( \frac{G}{3 \phi^p_0 J^s} \mathds{1} + \kappa \frac{ \vect{C}^{-1} } {2} \right), \\
    \frac{\partial \mathcal{P}}{\partial \phi^p} 
    & = \frac{\kappa}{\phi^p} \left( \ln J^e - 1\right).
\end{align*}
\endgroup

Using these derivatives, the components of the Jacobian matrix of the local residual vector in Eq. \eqref{eq:local-residual-system} can be computed as follows,
\begingroup
\allowdisplaybreaks
\begin{align*}          \label{eq:local-jacobian}
    \frac{\partial \mathcal{G}_1}{\partial C^w}  
    & = \frac{d \phi^p}{d C^w} \left[ R \theta \left( 1 - \frac{1}{1-\phi^{p}} + 2 \chi \phi^{p} \right) 
    + \left( \frac{\partial \mathcal{P}}{\partial \phi^p} \right) \mathcal{V}^w \right]
    + \frac{R\theta}{{(C^w)}^2} \sum_{\beta_k} C^{\beta_k}, \\
    \frac{\partial \mathcal{G}_1}{\partial C^{\beta_1}}
    & =  - \frac{R \theta}{C^w}, \\
    \vdots \\
    \frac{\partial \mathcal{G}_1}{\partial C^{\beta_n}}
    & = - \frac{R \theta}{C^w},   \\
    \frac{\partial \mathcal{G}_1}{\partial \psi} 
    & = 0, \\
    \frac{\partial \mathcal{G}_2}{\partial C^w} 
    & = - \frac{R\theta}{C^w} 
    + \left( \frac{\partial p}{\partial C^w} \right) \mathcal{V}^{\beta_1}, \\
    \frac{\partial \mathcal{G}_2}{\partial C^{\beta_1}} & = \frac{R \theta}{C^{\beta_1}}, \\
    \frac{\partial \mathcal{G}_2}{\partial C^{\beta_2}} & = 0,  \\
    \vdots \\
    \frac{\partial \mathcal{G}_2}{\partial C^{\beta_n}} & = 0, \\
    \frac{\partial \mathcal{G}_2}{\partial \psi} & = F z^{\beta_1} 
    \stepcounter{equation}\tag{\theequation} \\
    \vdots \\ 
    \frac{\partial \mathcal{G}_{n+1}}{\partial C^w} 
    & = - \frac{R\theta}{C^w} 
    + \left( \frac{\partial p}{\partial C^w} \right) \mathcal{V}^{\beta_n}, \\
    \frac{\partial \mathcal{G}_{n+1}}{\partial C^{\beta_1}} & = 0, \\
    \vdots \\
    \frac{\partial \mathcal{G}_{n+1}}{\partial C^{\beta_n}} & = \frac{R \theta}{C^{\beta_n}}, \\
    \frac{\partial \mathcal{G}_{n+1}}{\partial \psi} & = F z^{\beta_n} \\
    \frac{\partial \mathcal{G}_{n+2}}{\partial C^w} 
    & = \sum_{\beta_k} z^{\beta_k} \exp \left( \frac{\omega^{\beta_k} -  p \mathcal{V}^{\beta_k} - F\psi z^{\beta_k} - \omega^{0,{\beta_k}} }{R \theta} \right)  \left( 1 - \left( \frac{C^w}{R \theta} \right) \left( \frac{\partial p}{\partial C^w} \right) \mathcal{V}^{\beta_k} \right), \\
    \frac{\partial \mathcal{G}_{n+2}}{\partial C^{\beta_1}}
    & = 0, \\
    \vdots \\
    \frac{\partial \mathcal{G}_{n+2}}{\partial C^{\beta_n}} & = 0, \\
    \frac{\partial \mathcal{G}_{n+2}}{\partial \psi}
    & =  - \frac{F  C^w}{R \theta}  \sum_{\beta_k} \left( z^{\beta_k} \right)^2 \exp \left( \frac{\omega^{\beta_k} - p \mathcal{V}^{\beta_k} - F\psi z^{\beta_k} - \omega^{0,{\beta_k}} }{R \theta} \right).
\end{align*}
\endgroup
We should note that, to keep the presentation concise, we did not include the derivative terms that are zero and used the continuation symbol $(\vdots)$ to skip them.

\subsection{Computation of consistent material tangent moduli} 
\label{sec:appendix-tangents}

Recall the following equations \eqref{eq:internal-vars-derivatives} and \eqref{eq:material-tangent-moduli}, 
\begin{equation}
\begin{gathered}
    \frac{d \vect{q}}{d (\bullet)} 
    = - \left( \frac{\partial \boldsymbol{\mathcal{G}}}{\partial \vect{q}} \right)^{-1} \frac{\partial \boldsymbol{\mathcal{G}}}{\partial (\bullet)}, \\
    \frac{d \boldsymbol{\mathcal{F}}}{d (\bullet)}
    = \frac{\partial \boldsymbol{\mathcal{F}}}{\partial (\bullet)}
    + \frac{\partial \boldsymbol{\mathcal{F}}}{d \vect{q}} \frac{d \vect{q}}{\partial (\bullet)}.
\end{gathered}
\end{equation}
Thus, the first step is to compute the derivatives of the algorithmic internal variables with respect to the degrees of freedom or their gradients, and then to compute the consistent material tangent moduli. $\frac{\partial \boldsymbol{\mathcal{G}}}{\partial \vect{q}}$ is already given in Eq. \eqref{eq:local-jacobian} for our prescribed constitutive model. Thus we first need to calculate $\frac{\partial \boldsymbol{\mathcal{G}}}{\partial (\bullet)}$ where $ (\bullet) = \{ \vect{F}, \ \vect{C}, \ \mu^w, \ \omega^{\beta_k} \} $

\begin{enumerate} [label=(\alph*),itemsep=0pt,topsep=0pt,leftmargin=0.75cm]

\item Derivative of the local residual vector, $\boldsymbol{\mathcal{G}}$, with respect to the deformation gradient, $\vect{F}$, are as follows,
\begingroup
\begin{align*}
    \frac{\partial \mathcal{G}_1} {\partial \vect{F}} 
    & = \kappa \mathcal{V}^w \left( \ln J^e - 1 \right) \vect{F}^{-\T}, \\
    \frac{\partial \mathcal{G}_2} {\partial \vect{F}} 
    & =  \frac{\partial p}{\partial \vect{F}} \mathcal{V}^{\beta_1}, \\
    \vdots  \stepcounter{equation} \tag{\theequation} \\
    \frac{\partial \mathcal{G}_{n+1}} {\partial \vect{F}} 
    & =  \frac{\partial p}{\partial \vect{F}} \mathcal{V}^{\beta_{n}}, \\
     \frac{\partial \mathcal{G}_{n+2}} {\partial \vect{F}} 
    & =  - \frac{C^w}{R \theta}  \left( \frac{\partial p}{\partial \vect{F}} \right)  
    \sum_{\beta_k} z^{\beta_k} \mathcal{V}^{\beta_k}
    \exp \left( \frac{\omega^{\beta_k} 
    - F \psi z^{\beta_k} - p \mathcal{V}^{\beta_k} - \omega^{0,{\beta_k}} } {R \theta} \right).
\end{align*}
\endgroup
$\vect{c}_{\mu \vect{u}}$ and $\vect{c}_{\omega_k \vect{u}}$ in Eq. \eqref{eq:elem-tangent-matrix} are vectors of dimension $\left[ \bullet \right]_{\mathrm{1} \times n_{\mathrm{dim}}^2}$ obtained by reshaping $\frac{\partial C^w}{\partial \vect{F}}$ and $\frac{\partial C^{\beta_k}}{\partial \vect{F}}$, respectively.

Similarly, the derivative of the local residual vector, $\boldsymbol{\mathcal{G}}$, with respect to the Cauchy-Green tensor, $\vect{C}$, are given by,
\begingroup
\begin{align*}
    \frac{\partial \mathcal{G}_1} {\partial \vect{C}} 
    & = \kappa \mathcal{V}^w  \left( \ln J^e - 1 \right) \frac{\vect{C}^{-1}}{2}, \\
    \frac{\partial \mathcal{G}_2} {\partial \vect{C}} 
    & =  \frac{\partial p}{\partial \vect{C}} \mathcal{V}^{\beta_1}, \\
    \vdots  \stepcounter{equation} \tag{\theequation} \\
    \frac{\partial \mathcal{G}_{n+1}} {\partial \vect{C}} 
    & =  \frac{\partial p}{\partial \vect{C}} \mathcal{V}^{\beta_{n}}, \\
    \frac{\partial \mathcal{G}_{n+2}} {\partial \vect{C}} 
    & = - \frac{C^w}{R \theta}  \left( \frac{\partial p}{\partial \vect{C}} \right)  
    \sum_{\beta_k} z^{\beta_k} \mathcal{V}^{\beta_k}
    \exp \left( \frac{\omega^{\beta_k} 
    - F \psi z^{\beta_k} - p \mathcal{V}^{\beta_k} - \omega^0_{\beta_k} } {R \theta} \right).
\end{align*}
\endgroup

The derivative of the local residual vector, $\boldsymbol{\mathcal{G}}$, with respect to $\mu^w$ and $\omega^{\beta_k}$ are given by,
\begingroup
\begin{align*}
    \frac{d \boldsymbol{\mathcal{G}}}{d \mu^w} & = \{ -1 \ 0 \ 0 \ \cdots \ 0 \}^{\T}, \\
    \text{and} \quad
    \frac{\partial \mathcal{G}_1}{\partial \omega^{\beta_k}} & = 0, \\
    \frac{\partial \mathcal{G}_2}{\partial \omega^{\beta_k}} & = 0, \\
    \vdots  \stepcounter{equation} \tag{\theequation} \\
    \frac{\partial \mathcal{G}_{k+1}}{\partial \omega^{\beta_k}} & = -1, \\
    \vdots \\
    \frac{\partial \mathcal{G}_{n+2}}{\partial \omega^{\beta_k}} 
    & =  \frac{C^w}{R \theta}  \exp \left( \frac{\omega^{\beta_k} 
    - F \psi z^{\beta_k} - p \mathcal{V}^{\beta_k} - \omega{^{0,\beta_k}} } {R \theta} \right) z^{\beta_k}.
\end{align*}
\endgroup

\item With derivatives of algorithmic internal variables at hand, we can now compute the material tangent moduli using Eq. \eqref{eq:material-tangent-moduli}. First, $\mat{D}_{\mathds{C}}$ in Eq. \eqref{eq:elem-tangent-matrix} is the matrix form of the following fourth-order tensor:
\begin{equation}
    \mathds{C}_{IJKL} 
    = 2 \frac{dS_{IJ}}{dC_{KL}} 
    = 2 \left( \frac{\partial S_{IJ}}{\partial C_{KL}} + \frac{\partial S_{IJ}}{\partial C^w} \frac{d C^w}{d C_{KL}} \right).
\end{equation}
Based on our proposed constitutive model, $\mathds{C}_{IJKL}$ can be computed as,
\begin{equation}
\begin{gathered}
    \mathds{C}_{IJKL} =  \kappa \phi^p_0 J^s C^{-1}_{IJ} C^{-1}_{KL} 
    + \left( G (\phi^p_0)^{2/3} - \kappa \phi^p_0 J^s  \ln J^e \right) \left( C^{-1}_{IK} C^{-1}_{JL}  + C^{-1}_{JK} C^{-1}_{IL} \right) \\
    + 2 \left( \frac{\partial S_{IJ}}{\partial C^w} \right) 
    \left( \frac{d C^w}{d C_{KL}} \right),
\end{gathered}
\end{equation}
where 
\begin{equation}
    \frac{\partial S_{IJ}}{\partial C^w} = \kappa \mathcal{V}^w ( \ln J^e - 1) C^{-1}_{IJ}.
\end{equation}

$\mathds{C}_{IJKL}$ is then transformed into its matrix form, $\mat{D}_{\mathds{C}}$, using the Voigt notation convention listed in Table \ref{tab:voigt_conevntion_3D} for three-dimensional cases. Similar conventions can be defined for axisymmetry and two-dimensional plane strain cases.
\begin{table}[ht]
\centering
\begin{tabular}{ c  c  c  c }
    \toprule \toprule
    $m$ &   $(I,J)$   &    $n$    &    $(K,L)$  \\
    \toprule \toprule
    1   &   (1,1)     &     1     &     (1,1) \\ 
    2   &   (2,2)     &     2     &     (2,2) \\ 
    3   &   (3,3)     &     3     &     (3,3) \\ 
    4   &   (2,3)     &     4     &     (2,3) \\ 
    5   &   (1,3)     &     5     &     (1,3) \\
    6   &   (1,2)     &     6     &     (1,2) \\
    \bottomrule
\end{tabular}
\caption{Voigt notation convention for transforming a symmetric fourth-order tensor to a matrix for three-dimensional cases. Here, $(m,n)$ are indices of the matrix, and $(I,J,K,L)$ are the indices of the fourth-order tensor.}
\label{tab:voigt_conevntion_3D}
\end{table}

$\vect{a}_{\vect{u}\mu}$ and $\vect{a}_{\vect{u} \omega_k}$ in Eq. \eqref{eq:elem-tangent-matrix} are the vector forms of the following second-order tensor:
\begin{equation}
    F_{iJ} \frac{\partial S_{JI}}{\partial C^w}  \frac{d C^w}{d \mu^w} 
    \qquad \text{and} \qquad
    F_{iJ} \frac{\partial S_{JI}}{\partial C^w}   \frac{d C^w}{d \omega^{\beta_k}}.
\end{equation}

$\mat{D}_{\mu \vect{u}}$ and $\mat{D}_{\omega_k \vect{u}}$ in Eq. \eqref{eq:elem-tangent-matrix} are the matrix forms of the third-order tensors defined as follows, respectively:
\begin{equation}
\begin{aligned}
    \frac{d J^w_I}{d F_{kL}}  
    & = \frac{\partial J^w_I}{\partial F_{kL}}  
    + \frac{\partial J^w_I}{\partial C^w} \frac{d C^w}{d F_{kL}}, \\
    & = \frac{ D^w C^{w} }{ R \theta }  \left[ F^{-1}_{Ik} C^{-1}_{LJ} + F^{-1}_{Jk} C^{-1}_{LI} \right]   (\Grad \mu^w)_J
    - \left[\frac{ D^w }{ R \theta } C^{-1}_{IJ} (\Grad \mu^w)_J \right] 
    \left( \frac{d C^w}{d F_{kL}} \right), \\
    \text{and }  
    \frac{d J^{\beta_k}_I}{d F_{kL}}  
    & = \frac{ D^{\beta_k} C^{\beta_k} }{ R \theta } \left[ F^{-1}_{Ik} C^{-1}_{LJ} + F^{-1}_{Jk} C^{-1}_{LI} \right] (\Grad \omega^{\beta_k})_J 
    - \left[\frac{D^{\beta_k}}{ R \theta } C^{-1}_{IJ} (\Grad \omega^{\beta_k})_J \right]  \left( \frac{d C^{\beta_k}}{d F_{kL}} \right).
\end{aligned}
\end{equation}
The matrix forms are obtained by a column-order reshaping on $kL$ indices.

$\mat{m}_{\mu \mu}$, $\mat{m}_{\mu \omega_k}$, $\mat{m}_{\omega_k \mu}$, $\mat{m}_{\omega_i \omega_j}$ in Eq. \eqref{eq:elem-tangent-matrix} are vectors defined as,
\begingroup
\begin{align*} \label{eq:mobility-tangents}
    (\mat{m}_{\mu \mu})_I 
    & = \frac{d J^w_I}{d \mu^w}
    = \frac{\partial J^w_I}{\partial C^w} \frac{d C^w}{d \mu^w}
    = \left[ - \frac{D^w}{R \theta} C^{-1}_{IJ} (\Grad \mu^w)_J \right] \left( \frac{d C^w}{d \mu^w} \right),  \\
    (\mat{m}_{\mu \omega_k})_I 
    & = \frac{d J^w_I}{d \omega^{\beta_k}}
    = \frac{\partial J^w_I}{\partial C^w} \frac{d C^w}{d \omega^{\beta_k}}
    = \left[ - \frac{D^w}{R \theta} C^{-1}_{IJ} (\Grad \mu^w)_J \right] \left( \frac{d C^w}{d \omega^{\beta_k}} \right), 
    \stepcounter{equation} \tag{\theequation}   \\
    (\mat{m}_{\omega_k \mu})_I 
    & = \frac{d J^{\omega^{\beta_k}}_I}{d \mu^w}
    = \frac{\partial J^{\beta_k}_I}{\partial C^{\beta_k}} \frac{d C^{\beta_k}}{d \mu^w}
    = \left[ - \frac{D^{\beta_k}}{R \theta} C^{-1}_{IJ} (\Grad \omega^{\beta_k})_J \right] \left( \frac{d C^{\beta_k}}{d \mu^w} \right), \\
    (\mat{m}_{\omega_i \omega_j})_I 
    & = \frac{d J^{\omega^{\beta_i}}_I}{d \omega^{\beta_j}}
    = \frac{\partial J^{\beta_i}_I}{d C^{\beta_i}} \frac{d C^{\beta_i}}{d \omega^{\beta_j}}
    = \left[ - \frac{D^{\beta_i}}{R \theta} C^{-1}_{IJ} (\Grad \omega^{\beta_i})_J \right] \left( \frac{d C^{\beta_i}}{d \omega^{\beta_j}} \right).
\end{align*}
\endgroup

Finally, $\mat{M}^{w w}$ and $\mat{M}^{\beta_k \beta_k}$ in Eq. \eqref{eq:elem-tangent-matrix} are the matrix forms of the second-order mobility tensors for corresponding species, $w$ and $\beta_k$, respectively, as defined in Eq. \eqref{eq:flux-law-solvent}$_2$ and Eq. \eqref{eq:flux-law-ion}$_2$. Since we did not adopt cross-diffusion-based diffusion kinetic laws in this work, the following cross-diffusion mobility tensor terms are zero, \emph{i.e.},
\begin{equation}
    \mat{M}^{w \beta_k} = \mat{M}^{\beta_i \beta_j}  = 0.
\end{equation}
Although the direct cross-gradient mobility tensors vanish, $m_{\omega_i\omega_j}$ in Eq. \eqref{eq:mobility-tangents} is generally nonzero for $i \ne j$, because the locally determined concentration $C^{\beta_i}$ depends on all electrochemical potentials through the coupled constitutive equations and electroneutrality constraint.

\end{enumerate}

\section{Details of Abaqus user element subroutine implementation} 
\label{sec:abaqus-implementation}
\setcounter{table}{0}

Our Abaqus/Standard user element subroutine (UEL) implementation includes standard isoparametric Lagrange elements in conjunction with Gaussian quadrature to evaluate the element tangent stiffness matrix, $\vect{K}_e$, and the element residual vector, $\vect{r}_e$. As previously mentioned, we restricted our implementation to first-order Lagrangian elements, considering the computational cost associated with the solution procedure of multi-field finite element procedures. Specifically, we implemented 4-node tetrahedral (TET4) and 8-node hexahedral (HEX8) PE gel elements in three dimensions and 3-node triangular (TRI3) and 4-node quadrilateral (QUAD4) elements under axisymmetric and plane strain conditions in two dimensions. For TRI3 and TET4 elements, only the exact full Gaussian integration scheme is available. On the other hand, for QUAD4 and HEX8 elements, both reduced integration and full integration schemes are available. For fully-integrated QUAD4 and HEX8 elements, we optionally included the F-bar element formulation \citep{netoDesignSimpleLow1996,chesterCoupledTheoryFluid2010}, as presented in Appendix \ref{sec:appendix-f-bar-element}, to alleviate possible volumetric locking. Table \ref{tab:hydrogel-element_types} lists the available elements and their corresponding Gaussian integration schemes used in our implementation. For the sake of generality, we included the reduced integration scheme for several elements; however, we never used reduced integration in performing analyses. Figure S1 depicts the topology and integration scheme of the implemented elements alongside the nodal degrees of freedom and algorithmic internal variables.
\begin{table}[htbp]
\centering
\begin{tabular}{c c c}
    \toprule \toprule
    \textbf{Element key}    &   \textbf{Element description}   & \textbf{No. of integration points} \\
    \toprule  \toprule
    U1  &   4-node tetrahedral                  & 1         \\
    U2  &   8-node hexahedral                   & 8 and 1   \\
    U3  &   3-node triangular (axisymmetric)    & 1         \\
    U4  &   4-node quadrilateral (axisymmetric) & 4 and 1   \\
    U5  &   3-node triangular (plane strain)    & 1         \\
    U6  &   4-node quadrilateral (plane strain) & 4 and 1   \\
    \bottomrule
\end{tabular}
\caption{Types of PE hydrogel elements and integration schemes implemented as Abaqus/Standard user elements.}
\label{tab:hydrogel-element_types}
\end{table}

Since Abaqus does not have a specific load step procedure available for PE gels, we chose the coupled temperature-displacement analysis step in Abaqus/Standard. In this procedure (step), besides the displacement degrees of freedom, twenty additional scalar degrees of freedom can be specified for any user-defined element. At every time step, for each iteration, the Abaqus/Standard solver calls the UEL subroutine for all the user-defined PE gel elements in the model. In its UEL subroutine interface, Abaqus/Standard requires the user to program the element residual vector, $\vect{r}_e$, in \texttt{RHS} variable and the element tangent stiffness matrix, $\mat{K}_e$, in \texttt{AMATRX} variable to perform the implicit nonlinear solution procedure. We additionally used the Abaqus variable \texttt{SVARS} from the UEL subroutine to store the history of the algorithmic internal variables, $\vect{q}$ in Eq. \eqref{eq:internal-vars}. The internal state variable vector (\texttt{SVARS}) provided by Abaqus is from the previously converged time, $t$, and the degrees of freedom are from the last iteration. Once the residual vector and element tangent stiffness matrices are formed, Abaqus performs the global element assembly in Eq. \eqref{eq:global-system} and iterative solution procedure in Eq. \eqref{eq:global-iteration} for all the user-defined elements.

At the global level, we let Abaqus control the time-stepping scheme based on its convergence criteria, which can be specified by the user. However, we used the Abaqus-provided UEL variable, \texttt{PNEWDT}$=\Delta t_{\mathrm{new}} / \Delta t_{\mathrm{current}}$, to cut back on the time step size in the case of convergence issues with the local iterative nonlinear solver or any large changes in the solution of degrees of freedom. During our numerical simulations, we observed a significant enhancement of robustness at critical time steps because of this programmed time step adjustment. Consequently, this technique allowed us to use a larger maximum time step size for a specific analysis step. To ensure further robustness of our implementation, the local nonlinear solver was augmented with a backtracking line search approach with the convergence criterion set based on the absolute or relative norm of the local residuals being less than $\varepsilon_{\mathrm{tol}} = 10^{-9}$. If needed, this criterion can be easily changed in our source code by the user. Following the iterative solution of algorithmic internal variables, all the material tangents as presented in Appendix \ref{sec:appendix-tangents} are computed, and then the element residual vectors and tangent stiffness matrices are formed. At this stage, Abaqus performs the global assembly of the element tangent stiffness matrices and residual vectors, followed by the nonlinear solution procedure. In all of our simulations, we used the standard Newton solver available in Abaqus/Standard. Since both of our standard and F-bar element formulations are unsymmetric, we chose the unsymmetric matrix storage scheme and unsymmetric monolithic solver to perform the analyses. However, it is also possible to use the staggered (or separated) solver through the solution technique interface of Abaqus without making any modifications to our current implementation. This technique can be advantageous in analyses with severe convergence issues; however, it is beyond the scope of this work.

\end{appendices}

\clearpage
\setcounter{page}{1}
\setcounter{section}{0}
\setcounter{figure}{0}
\resetlinenumber
\makeatletter
\renewcommand \thesection{S\@arabic\c@section}
\renewcommand\thetable{S\@arabic\c@table}
\renewcommand \thefigure{S\@arabic\c@figure}
\makeatother

\begin{center}
    {\Large \textbf{Supplementary Information}} 

    \vspace{18pt}

    {\Large A transient nonlinear finite element framework and implementation of coupled electro-chemo-mechanics of polyelectrolyte hydrogels}
    
    \vspace{12pt}

    Bibekananda Datta\textsuperscript{a}, 
    Brandon K. Zimmerman\textsuperscript{b}, 
    Thao D. Nguyen\textsuperscript{a}

    \vspace{6pt}
    
    {\small
    \textit{\textsuperscript{a}Department of Mechanical Engineering, Johns Hopkins University, Baltimore, MD 21218, USA \\
    \textsuperscript{b}Lawrence Livermore National Laboratory, Livermore, CA 944550, USA}
    }
    \vspace{6pt}
    \hrule
\end{center}

\section{Topology of the implemented PE gel elements}

\begin{figure}[ht]
\begin{center}
    \includegraphics[width=\textwidth, trim={1cm 6.5cm 1cm 6.5cm}, clip] {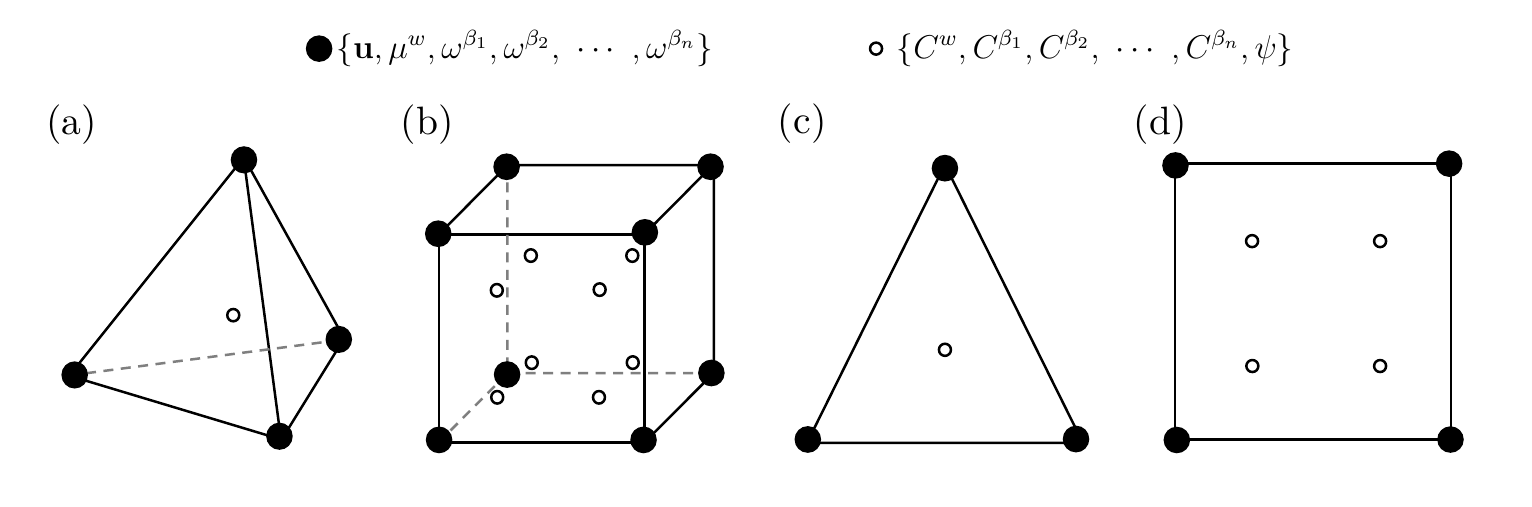}
    \caption{Topology of implemented polyelectrolyte gel elements with the list of degrees of freedom at the nodes and internal variables at the Gauss points: (a) 4-node tetrahedron, (b) 8-node hexahedron, (c) 3-node triangle, (4) 4-node quadrilateral.}
    \label{fig:pe-gel-elements}
\end{center}
\end{figure}

As described in the main article, we implemented first-order tetrahedral and hexahedral elements in three dimensions and first-order triangular and quadrilateral elements in two-dimensional axisymmetric and plane strain conditions. The topology of the elements is standard, with their interpolation functions being described by Lagrangian polynomials. We used the standard Gaussian quadrature rule to perform numerical integration. In Figure \ref{fig:pe-gel-elements}, we denote the nodal degrees of freedom and internal variables computed at the integration points.

\clearpage
\section{Tutorial for the Abaqus user element subroutine}

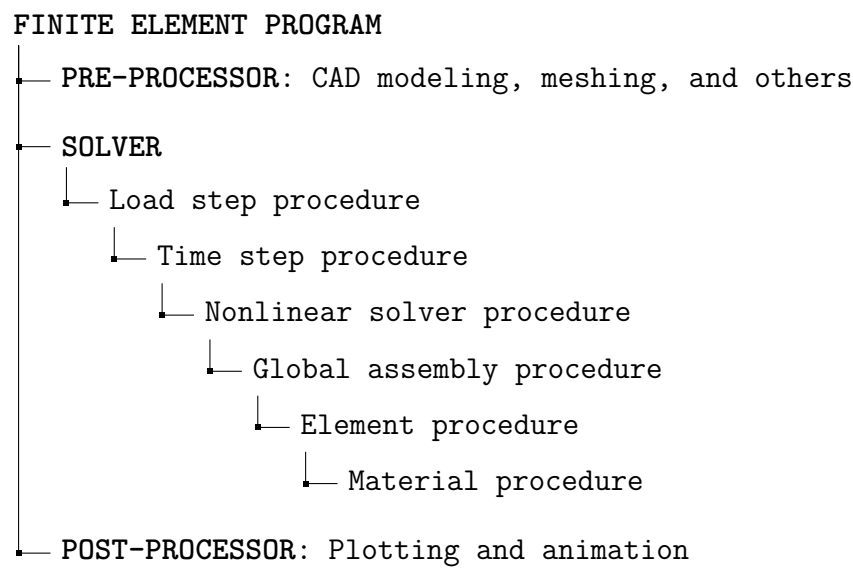
\begin{figure}[ht] 
\begin{center}
\begin{forest}
  for tree={
    font=\ttfamily,
    grow'=0,
    child anchor=west,
    parent anchor=south,
    anchor=west,
    calign=first,
    edge path={
      \noexpand\path [draw, \forestoption{edge}]
      (!u.south west) +(6pt,0) |- node[fill,inner sep=1pt] {} (.child anchor)\forestoption{edge label};
    },
    before typesetting nodes={
      if n=1
        {insert before={[,phantom]}}
        {}
    },
    fit=band,
    before computing xy={l=18pt},
  }
[\textbf{FINITE ELEMENT PROGRAM}
  [\textbf{PRE-PROCESSOR}: CAD modeling{,} meshing{,} and others]
  [\textbf{SOLVER}
    [Load step procedure
        [Time step procedure
            [Nonlinear solver procedure
                [Global assembly procedure
                    [Element procedure
                        [Material procedure]
                    ]
                ]
            ]
        ]
    ]
  ]
  [\textbf{POST-PROCESSOR}: Plotting and animation]
]
\end{forest}
\caption{Procedural organization of a generic nonlinear finite element program.}
\end{center}
\end{figure} \label{fig:fem-program}

\subsection{Configuring Abaqus for user subroutines}

To execute Abaqus subroutines, we will need to have the Intel Fortran compiler (included in the Intel oneAPI packages) and Microsoft Visual Studio installed and linked to Abaqus. If we are installing a newer version of Microsoft Visual Studio, please make sure it is compatible with the Intel oneAPI version that we have. Typically, Intel oneAPI versions that are released after the Microsoft Visual Studio version are compatible with each other. The following guidelines are from 2021, but it was executed with the same steps in 2023 with the latest versions. Abaqus can be installed in any order, but Microsoft Visual Studio and Intel oneAPI must be installed in that order for this to work.

\subsubsection{Installing Microsoft Visual Studio}

First, we have to install Microsoft Visual Studio. Intel oneAPI 2021 supports Visual Studio 2017 and Visual Studio 2019. I used VS 2019 Community Edition, which is available for free. Once the installer is downloaded, click on it to start the installation process. On the installation page, please select \textbf{Desktop Development with C++} and keep the rest of it as the default. It will take a few minutes to install.

\subsubsection{Installing Intel oneAPI}

Intel oneAPI is split into two packages: Intel oneAPI Base Toolkit and Intel oneAPI HPC Toolkit. To be able to use Intel Fortran and C++ compilers, both toolkits (especially for MKL libraries and parallel processing) are required.

\begin{enumerate} [leftmargin=0.55cm,itemsep=0pt,topsep=0pt]

    \item Once both of the packages are downloaded, first install the Intel oneAPI Base Toolkit (I chose the default installation). It will recognize the existence of Visual Studio. Please click on the appropriate version and move forward with installation. It might take a few minutes to install the Base Toolkit. The next step is to install the HPC toolkit in the same procedure. This usually takes a shorter time than the Base Toolkit to get installed.

    \item For default installation of Intel oneAPI, its components should be located in \textcolor{RoyalBlue}{\texttt{C:\textbackslash Program Files (x86)\textbackslash Intel\textbackslash oneAPI}}. Navigate to this directory and then navigate further to ensure the ifort executable is available in the following (or similar) directory. Copy the installation directory into a text file for later use. The directory should look like: \\
    \textcolor{RoyalBlue}{\texttt{C:\textbackslash Program Files (x86)\textbackslash Intel\textbackslash oneAPI\textbackslash compiler\textbackslash 2021.1.1\textbackslash windows\textbackslash bin\textbackslash intel64}}.
 
\end{enumerate}

\subsubsection{Linking Intel Fortran compiler to Abaqus}

To link the Intel Fortran compiler from the Intel oneAPI package to Abaqus to compile user subroutines, the Abaqus batch file needs to be modified. The file needs to be saved as administrator after the modification. To do that, open the file with Notepad++, which allows the file to be saved as administrator, or run Windows Notepad from the program menu as administrator, and then open the file from the application.

\begin{enumerate} [leftmargin=0.55cm,itemsep=0pt,topsep=0pt]

    \item Navigate to \textcolor{RoyalBlue}{\texttt{C:\textbackslash Program Files (x86)\textbackslash Intel\textbackslash oneAPI\textbackslash compiler\textbackslash 2021.1.1\textbackslash env}} (or similar directory depending on the installation and version) and ensure the \texttt{vars.bat} file is available. If the directory name is different for the installation, we will need to search for the \texttt{vars.bat} file through Windows Explorer and navigate to the directory. Copy the file directory into a text file, which should look like this: \\
    \textcolor{RoyalBlue}{\texttt{C:\textbackslash Program Files (x86)\textbackslash Intel\textbackslash oneAPI\textbackslash compiler\textbackslash 2021.1.1\textbackslash env\textbackslash vars.bat}}.

    \item Navigate to \textcolor{RoyalBlue}{\texttt{C:\textbackslash SIMULIA\textbackslash Commands}} directory and open the \texttt{abq2020.bat} (or the Abaqus version installed) file with any text editor. At the beginning of the file, add the following lines to the file and \textbf{save it as administrator}.
    
\begin{lstlisting} [basicstyle=\ttfamily\small, frame=single, backgroundcolor=\color{lightgray!20}, columns=fullflexible, aboveskip=12pt, belowskip=0pt, breaklines=true, xleftmargin=.2cm]
SET PATH=%PATH%; C:\Program Files (x86)\Intel\oneAPI\compiler\2021.1.1\windows\bin\intel64; call "C:\Program Files (x86)\Intel\oneAPI\compiler\2021.1.1\env\vars.bat" intel64
\end{lstlisting}

    As we can see, the \texttt{PATH} in the first line is the directory for \texttt{ifort} compiler executable, which was copied during installation. The second line calls the batch file for \texttt{ifort} compiler, which sets the environment variables when Abaqus is invoked from the command line.

    \item To verify if the linking process was successful, open the Abaqus command line terminal from the Windows Program menu and type \texttt{abaqus verify -all} on the terminal. Upon successful installation and linking, it should print and show \texttt{PASS} on the terminal. To get information about different packages, type \texttt{abaqus info=system} on the terminal screen, which will print out all the necessary information.

\begin{lstlisting} [basicstyle=\ttfamily\small, frame=single, backgroundcolor=\color{lightgray!20}, aboveskip=12pt, belowskip=0pt, breaklines=true, xleftmargin=.2cm]
$ abaqus verify -all
$ abaqus info=system
\end{lstlisting}
 
\end{enumerate}

\subsubsection{Linking Intel MKL libraries with Abaqus compiler}

\begin{enumerate} [leftmargin=0.55cm,itemsep=0pt,topsep=-6pt]

    \item Navigate to \textcolor{RoyalBlue}{\texttt{C:\textbackslash Program Files (x86)\textbackslash Intel\textbackslash oneAPI\textbackslash mkl\textbackslash 2021.1.1\textbackslash env}}, and ensure the \texttt{vars.bat} file is available for oneMKL package. This batch file is for the Intel oneMKL library. Copy the file directory as usual.

    \item Navigate to \textcolor{RoyalBlue}{\texttt{C:\textbackslash SIMULIA\textbackslash Commands}} and open the \texttt{abq2020.bat} (or corresponding version) file. Add the following line below the two lines previously added for the ifort compiler. Save the file as administrator using the same approach as above; either run Notepad as administrator from the Windows program menu or use Notepad++.
\begin{lstlisting} [basicstyle=\ttfamily\small, frame=single, backgroundcolor=\color{lightgray!20}, columns=fullflexible, aboveskip=12pt, belowskip=0pt, breaklines=true, xleftmargin=.2cm]
call "C:\Program Files (x86)\Intel\oneAPI\mkl\2021.1.1\env\vars.bat" intel64
\end{lstlisting}

    \item In the previous step, providing the Intel oneMKL batch file location in the ABAQUS batch file will source the environment variables when ABAQUS is called to run simulations. However, to compile code with the Intel oneMKL library, you will need to do one more step. Locate to the \textcolor{RoyalBlue}{\texttt{C:\textbackslash SIMULIA\textbackslash EstProducts\textbackslash 2020\textbackslash win\_b64\textbackslash SMA\textbackslash site}} directory and open the \texttt{win86\_64.env} file. Find the following line in the file and add the \texttt{/Qmkl:sequential} compiler option as shown below. The rest of the file should be the same. Save the file as administrator.

\begin{lstlisting} [basicstyle=\ttfamily\small, frame=single, backgroundcolor=\color{lightgray!20}, columns=fullflexible, aboveskip=12pt, belowskip=0pt, breaklines=true, xleftmargin=.2cm]
compile_fortran=['ifort','/Qmkl:sequential', ... ... ]
\end{lstlisting}

\end{enumerate}

\subsection{Brief description of the main source code}

The main source code has a few subroutines. The following are brief descriptions of the subroutines within the main source code.

\begin{description}[topsep = 0pt, style=multiline, leftmargin= 4cm, itemsep = 0pt, font=\normalfont\ttfamily]

    \item[UEL] 
    This is the Abaqus-provided subroutine to program any user element. We do not perform any finite element calculations in this subroutine; rather perform all the initial checks, determine some additional variables related to the element (and material) type, and pass these variables to a subsequent subroutine in which all the main calculations are performed.

    \item[UVARM]
    This is an Abaqus-provided subroutine that allows saving user-defined variables at each integration point as an output of the \texttt{.odb} file. We will use this subroutine to visualize element output (stress and strain tensor components) in Abaqus/Viewer by storing this variable using a global variable approach. Since we will use standard Lagrangian elements in this implementation, we can use this Abaqus subroutine to overlay the element output on standard Abaqus elements for visualization. More details about this technique are given below.

    \item[UEXTERNALDB]
    This is also an Abaqus user subroutine, which is used to open an error logging file and log the errors and warning messages from the analysis to a \texttt{.dat} file.

\end{description}

\subsection{Brief description of the element and material modules}

\begin{description}[topsep = 0pt, style=multiline, leftmargin= 4cm, itemsep = 0pt, font=\normalfont\ttfamily]

    \item[pegel\_element] 
    This is the element formulation module that contains two main subroutines, \texttt{pegel\_general} and \texttt{pegel\_axisymmetric}, and an additional subroutine, \texttt{assembleElement}. Subroutine \texttt{pegel\_general} has the element formulation for 3D elements and 2D plane strain elements. Formulation for axisymmetric element is different and difficult to generalize with other 3D and 2D elements; hence, it was done in a separate subroutine called \texttt{pegel\_axisymmetric}. Based on the type of elements being used in the analysis,  \texttt{UEL} subroutine calls either of these subroutines to perform element operations. These subroutines make calls to the material model subroutine at each integration point to compute constitutive laws and tangent moduli. With these quantities at the integration points, they form the components of the element residual vector and element tangent matrix. Once the components are formed, they call \texttt{assembleElement} to perform the assembly procedure. Remember, this is not a global assembly; rather, it performs assembly for a single element since PE gel elements have a large number of degrees of freedom arranged in a specific way. Any changes made to the element formulation will have to be done in these subroutines.

    \item[pegel\_material]
    This module contains the material model subroutine. Currently, it only has one material model, as demonstrated in the article. The existing material model has \texttt{matID = 1} (see Table \ref{tab:pe_gel_elem_props}). The material model subroutine \texttt{neohookean\_flory2} represents the material model that can be described using a compressible Neo-Hookean potential, binary Flory-Huggins mixture potential, and dilute ionic mixture potential. The main subroutine contains another subroutine to compute the internal variables at each integration point. To add new material models, users will need to program the material model within this subroutine. Users can use the \texttt{matID} flag to call additional material model subroutines they will develop. If there are multiple material models, it would be easier to break up the module into multiple submodules and put them in different files.
    
\end{description}

\subsection{Brief description of the additional modules}

There are a few other modules used in programming the element formulation. The name of the source file for those modules is the same as the module name.

\begin{description}[topsep = 0 pt, style=multiline, leftmargin=4.2cm, itemsep = 0pt, font=\normalfont\ttfamily]

    \item[global\_parameters] 
    This module includes the definition of the default floating-point variable type and some other common global variables that are used in writing subroutines in other modules, as well as the Abaqus \texttt{UEL} and \texttt{UVARM} subroutines.

    \item[error\_logging]
    This module includes a derived data type and a few type-bound procedures (also known as a class in object-oriented programming) which can be used to log general information about simulation, any warning and error messages, and debugging information in \texttt{.dat} files and on the terminal if the \texttt{interactive} option is used to run the \texttt{abaqus} command.

    \item[lagrange\_element]
    This module defines a simple derived data type to define element type and defines the interpolation functions and their derivatives for the most common first and second-order Lagrangian elements, \emph{i.e.}, 1D (bar), 2D (triangular and quadrilateral), and 3D (tetrahedral and hexahedral) used in solid mechanics. In the future, first and second-order 3D wedge elements will be included. Additionally, there are plans to include a few more subroutines that can return element operator matrices for scalar (heat transfer/ diffusion) and vector elements (displacement) based on the element definition.

    \item[gauss\_quadrature]
    This module includes a few subroutines that can return weights and coordinates for integration points in standard 1D, 2D, and 3D Lagrange elements to perform full and reduced integration. As of now, this is sufficient to perform numerical integration related to quasi-static or static problems. In the future, these subroutines will be extended to return weights and coordinates of integration points to perform higher-order integration required for dynamic problems.

    \item[linear\_algebra]
    This module is subdivided into three different submodules, named \texttt{utility}, \texttt{computation}, and \texttt{lapack}. These submodules contain procedures (functions and subroutines) to perform standard linear algebraic operations on vectors and matrices with simplified APIs similar to NumPy and MATLAB. The \texttt{lapack} submodule contains simplified interfaces for some of the common LAPACK subroutines to perform matrix computations on double-precision rank-2 arrays.

    \item[nonlinear\_solver]
    This module contains subroutines and relevant settings for solving nonlinear linear algebraic equations. We utilized the module to solve for the system of nonlinear equations formed by the set of internal variables. There are two main subroutine interfaces, \texttt{fzero} and \texttt{fsolve}. The first one is used for a single variable equation, and either Newton or the hybrid fail-safe Newton method can be used. \texttt{fsolve} is used for solving a system of nonlinear equations using the Newton method, but it can be augmented with a backtracking line search algorithm. For both subroutines, the user can specify options to control the tolerance, algorithm, number of maximum iterations, to use an analytical or numerical Jacobian, etc.

    \item[solid\_mechanics] 
    This module contains a few simple subroutines to perform standard tasks of reshaping vectors and matrices based on different types of analysis used in solid mechanics, \emph{i.e.}, plane strain, plane stress, axisymmetric, three-dimensional, etc.

    \item[post\_processing]
    This subroutine includes three global variables related to building the Abaqus model and storing user-defined element output for post-processing in Abaqus/Viewer.

    \item[surface\_integration]
    This module contains a few subroutines to perform surface integration on hexahedral and quadrilateral elements, which might be necessary to apply traction or flux-type boundary conditions. As of now, this module is not in use, but it can be used in the future to implement these natural boundary conditions to the elements.

\end{description}

\subsection{Modeling in Abaqus}

As mentioned previously, we can use Abaqus/CAE to build the model for the user subroutine if the nodal connectivity (topology of the element) of the user element is the same as one of the standard elements available in Abaqus. This means we can still do most of the pre-processing work in Abaqus/CAE even if the interpolation functions and integration schemes are different than those of the standard Abaqus user elements. If the user elements have a different topology than the standard Abaqus elements, we will have to use different pre-processing software to generate the mesh, such as using Cubit to generate an isogeometric mesh based on splines. Here, we will focus on standard Lagrangian elements and will use Abaqus/CAE as the primary pre-processing.

\subsubsection{Visualization of element output using \texttt{UVARM} subroutine}

If the user elements have similar nodal connectivity as the standard Abaqus elements, we can use Abaqus/CAE to build most of the finite element model (meshing, element and node sets, surfaces, contact interaction, load step, time steps, solver settings, etc.). We will discuss the details in the next section. Even if standard Lagrangian interpolation functions and integration schemes are used for the user elements, visualization of the element output is not natively supported in Abaqus/Viewer after the solution is obtained. 

According to Abaqus documentation, if the user element shares the same nodal connectivity and integration points with any standard Abaqus element (especially displacement-based mechanical elements), users can overlay the standard Abaqus elements with negligible elastic properties on the user elements. This will allow users to visualize the node output in the Abaqus/Viewer. To view element output, users can use the \texttt{UVARM(...)} subroutine from Abaqus. This subroutine is called at each integration point for the standard Abaqus element. A user can store the results from the user element using the global variable technique and assign them to the \texttt{uvar} variable of the standard elements in the subroutine. In the source code, a three-dimensional array named \texttt{globalPostVars(:,:,:)} has been declared as a global variable. The first index of this global variable represents the element number, the second index represents the number of integration points, and the third index represents the post-processing quantity. This approach has been adopted from \cite{chesterFiniteElementImplementation2015}.

We will discuss the steps for adding the layer of standard elements to the user elements in the next section. Briefly, a similar set of elements with the same nodal connectivity is added in the Abaqus input file after generating the primary model in Abaqus/CAE. This new element set has a different element numbering from the user elements. For example, if the model has 100 user elements numbered from 1-100, the overlaying standard elements can have element numbers from 100001-100100, with an offset number between the two sets being 100000. In the source code, the element offset number has been defined as \texttt{elemOffset}. When building a model, users have to be careful about matching this number for the overlaying set of elements (see next section). Since two sets of elements (user elements and standard elements) have been defined with the same element connectivity, the Abaqus solver will call UEL for the user elements and then call its internal element procedure subroutine for the standard Abaqus elements to calculate the element stiffness matrix and residual vectors. Since they share the same connectivity, during the assembly process, they will be added. However, if the overlaying standard elements have an insignificantly small elastic modulus, then their contribution to the element stiffness matrix (\texttt{AMATRX}), consequently, to the global stiffness matrix will be negligible to alter the results. Using these same overlaying elements, we can prescribe traction and pressure-type boundary conditions, and these contributions will be added to the \texttt{RHS} vector and will influence the result. Here, we will follow this method since programming traction or pressure-type boundary conditions is an intensive task.

This technique is applicable if and only if the user elements share the same nodal connectivity and integration points as one of the Abaqus standard elements. If the element connectivity and the location of integration points are different, the user may need to write a Python script to post-process the output.

Currently, the \texttt{numElem = 50000} variable has been defined within the \texttt{post\_processing} module, which means \texttt{globalPostVars(:,:,:)} can contain data for a maximum of 50000 elements. Abaqus will throw an \texttt{ELEMENT LOOP} error if there are more user-defined elements in the model. So, if the Abaqus model has more than 50000 user-defined elements, then the user will need to change it. This number has to be at least the same as the number of user elements or higher. If needed, the user will also need to update the variable \texttt{elemOffset} in the same module.

\subsubsection{Building primary model in Abaqus/CAE}

Building an Abaqus model to use with the UEL subroutine in Abaqus/CAE is not a straightforward task. It involves the following steps:

\begin{enumerate} [leftmargin=0.55cm, itemsep=0pt, topsep=0pt]

    \item Create the CAD model in the \texttt{Part} module. Define material properties, section, etc., and insert the part into the \texttt{Assembly} module.
    
    \item Now mesh of the desired model in Abaqus/CAE using standard built-in elements in the \texttt{Mesh} module that is \textbf{similar} to the user element, \emph{i.e.}, has the same number of nodes and topology. For example, if we would like to model the domain using 8-node hexahedral elements (U2), then create the mesh using C3D8 elements from Abaqus. Similarly, for 4-node axisymmetric quadrilaterals (U4), use CAX4 elements, and for 4-node plane strain quadrilaterals (U6), use CPE4 elements.

    \item In the assembly module, define all the node sets and elements. We can do it before creating the mesh by using the geometry feature.

    \item Now we can create all the amplitudes using the graphical user interface of Abaqus/CAE. However, it is easier to create using keywords in the input file by using parameters, so it is easier to change from case to case.

    \item Before exporting the input file from Abaqus/CAE, on the model tree on the left-hand side of the window, right-click on the model name and then click on \textbf{Edit Attributes}. Check the box that reads \textbf{Do not use parts and assemblies in input files}. This will simplify the format of the input file being exported for further modification.
    
    \item The final task in Abaqus/CAE is to create a \texttt{job} from the model and export the Abaqus input file (\texttt{.inp}) for further modification. If it gives any warning about section assignment, ignore that and export the input file anyway.

\end{enumerate}

Open the input file in a text editor such as Notepad++ or VS Code. It will start with some basic information and the description of the nodal coordinates and element connectivity of the generated mesh. The main section of the as-exported input file should look somewhat like the following:

\begin{lstlisting} [basicstyle=\ttfamily\small, language=bash, frame=single, backgroundcolor=\color{lightgray!20}, aboveskip=18pt, belowskip=6pt, breaklines=true,  xleftmargin=0.1cm, xrightmargin=0.1cm]
*Node
1,           0.,           0.
2, 0.000210268525,         0.
...
...
...
*Element, type=CAX4
1,   1,   2,  17,  16
2,   2,   3,  18,  17
...
...
...
*Nset, nset=all, generate
1,  420,    1
*Elset, elset=all, generate
1,  378,    1
...
...
...
\end{lstlisting}

\subsubsection{Preparing the input file for a simple PE gel analysis}

In this section, we will look into different sections of an Abaqus input file for a PE gel model and discuss how it is different than a standard input file. To demonstrate the structure of an input file, we take the first example from our article, \emph{i.e.}, free swelling of cylindrical cationic PE gel reported by \cite{sunMultiresponsiveToughHydrogels2015}.

Before we start making modifications to the input file, let us first look at the properties in Table \ref{tab:pe_gel_elem_props} that need to be specified. There are 15 properties to be defined for the current material model of the gel. 3 of these properties (density, pKa, and molar volume of polymer) are not currently used but kept to extend the code in the future. Additionally, the gel and external solution have 2 different types of ions, with each ion having 5 properties. The default material has \texttt{matID = 1}, and more material-specific constitutive models can be added to the code. We also need to specify the number of integration points, the flag for F-bar element formulation, the number of properties for each ion, and the number of post-processed variables.

\begin{table}[ht]
\centering
\begin{tabular}{c c c}
    \toprule \toprule
    \textbf{Property Name} & \textbf{Symbol}  & \textbf{Variable in UEL} \\
    \toprule  \toprule
    Universal gas constant  & $R$ & \texttt{Rgas} \\
    Faraday's constant      & $F$ & \texttt{Fcon} \\
    Absolute temperature    & $\theta$ & \texttt{theta} \\
    Initial polymer volume fraction & $\phi_0^p$    & \texttt{phi0} \\
    Density of polymer                  & $\rho$    & \texttt{rho} \\
    Shear modulus at reference state    & $G$       & \texttt{Gshear} \\
    Bulk modulus at reference state     & $\kappa$  & \texttt{Kappa} \\
    Log of acid disassociation constant & pKa       & \texttt{pKa}  \\
    Initial fixed charge of polymer     & $C_0^{\mathrm{fix}}$ & \texttt{C0\_fix} \\
    Molar volume of polymer             & $\mathcal{V}^p$     & \texttt{Vp} \\
    Charge number of polymer & $z^p$ & \texttt{Zfix} \\
    Chemical potential of pure solvent & $\mu^0_w$ & \texttt{mu0} \\
    Molar volume of the solvent & $\mathcal{V}^w$ & \texttt{Vw} \\ 
    Flory-Huggins interaction parameter & $\chi$ & \texttt{chi} \\
    Diffusion coefficient of the solvent & $D^w$ & \texttt{Dw} \\
    Initial concentration of ion & $C^{\beta_k}_0$ & \texttt{Cion0} \\
    Electrochemical potential of pure ion, & $\omega^{0,\beta_k}$ & \texttt{Omg0} \\
    Molar volume of ion & $V^{\beta_k}$ & \texttt{Vion} \\
    Charge number of ion & $z^{\beta_k}$ & \texttt{Zion} \\
    Diffusion coefficient of ion & $D^{\beta_k}$ & \texttt{Dion} \\
    \midrule
    No. of integration points & & \texttt{nInt} \\
    Fbar formulation flag & & \texttt{fbarFlag} \\
    Elastomeric material ID & & \texttt{matID} \\
    No of properties for each ion & & \texttt{nIonProps} \\
    No. of post-processed variables & & \texttt{nPostVars} \\
    \bottomrule
\end{tabular}
\caption{List of properties used in element and material definition.}
\label{tab:pe_gel_elem_props}
\end{table}

First, we will define a set of parameters at the beginning of the input file (before nodal coordinates and element connectivity) so that we can make changes to the input file quickly. Look at the following list of parameters. Since we are using an axisymmetric element, our \texttt{nDim = 2} and \texttt{nStress = 4}; see Table \ref{tab:elem_parameters} for details. Besides the Cauchy stress and Euler strain, we are storing all the internal state variables for post-processing, so we defined all these parameters accordingly. We then define all the properties required for the gel and the ions. Here, we took Na$^+$ as our first ion and Cl$^-$ as our second ion. Notice that we assigned arbitrary but somewhat low initial concentration values for the ions, and they maintain the \textbf{electroneutrality condition}. Assigning any large ionic concentration value for the ions can result in wrong results, as our constitutive model can only accommodate dilute cases.

\begin{table}[ht]
\centering
\begin{tabular}{c c c c}
    \toprule \toprule
    Element & Description  & \texttt{nDim} & \texttt{nStress}         \\
    \toprule  \toprule
    U1, U2 & 3D elements                & 3 & 6     \\
    U3, U4 & 2D axisymmetric elements   & 2 & 4     \\
    U5, U6 & 2D plane strain elements   & 2 & 3     \\
    \bottomrule
\end{tabular}
\caption{\texttt{nDim} and \texttt{nStress} parameters for implemented elements.}
\label{tab:elem_parameters}
\end{table}

\begin{lstlisting} [basicstyle=\ttfamily\small, language=bash, frame=single, backgroundcolor=\color{lightgray!20}, aboveskip=18pt, belowskip=6pt, breaklines=true,  xleftmargin=0.1cm, xrightmargin=0.1cm]
*Heading
***********************************************************
** PARAMETERS TO MAKE THE FILE MORE ADAPTIVE
***********************************************************
*Parameter
nDim        = 2
nStress     = 4
nNode       = 4
nInt        = 4
fbar        = 1
matID       = 1
nIons       = 2
nIonProps   = 5
matProps    = 15 + nIons*nIonProps
intProps    = 5
nstatev     = (nIons + 2)
nsvars      = nstatev*nInt
nPostVars   = 2*nStress + nstatev
***********************************************************
** PHYSICAL CONSTANTS AND PROPERTIES
***********************************************************
Rgas        = 8.3145
Fcon        = 96485.0
theta       = 298.0
phi0        = 0.312
rho         = 1100.0
Gshear      = 48.0e3
kappa       = 50.0*Gshear
pKa         = 5.0
Cp_fix      = 460.0
Vp          = 8.928e-3
Zp          = 1.0
mu0         = 0.0
Vw          = 1.8e-5
chi         = 0.40
Dw          = 9.0e-7
** properties of ions in gel
Cion1_gel   = 340.0
Cion2_gel   = 800.0
Omg0_1      = 0.
Omg0_2      = 0.
Vion_1      = 2.38e-6
Vion_2      = 2.24e-6
Zion_1      = 1.0
Zion_2      = -1.0
Dion_1      = 4.0e-8
Dion_2      = 4.0e-8
\end{lstlisting}

Once all the parameters and properties are defined, our next step is to calculate the parameters to prescribe the initial and boundary conditions. We then define all the time parameters, such as the simulation time for each Abaqus load step and the initial, minimum, and maximum time step size for each load step.
\begin{lstlisting} [basicstyle=\ttfamily\small, language=bash, frame=single, backgroundcolor=\color{lightgray!20}, aboveskip=18pt, belowskip=6pt, breaklines=true,  xleftmargin=0.1cm, xrightmargin=0.1cm]
***********************************************************
** CALCULATE PARAMETERS RELATED TO IC AND BC
***********************************************************
RT          = Rgas*theta
Cw0_gel     = (1.0 - phi0)/Vw
**
initMU      = mu0 + RT*(phi0+log(1-phi0)+chi*phi0**2) - RT/Cw0_gel*(Cion1_gel + Cion2_gel)
initOmg1    = Omg0_1 + RT*log(Cion1_gel/Cw0_gel)
initOmg2    = Omg0_2 + RT*log(Cion2_gel/Cw0_gel)
**
** properties of ions in low salt solution
**
Cw_sol            = 55000.0
**
Cion1_sol_low     = 50.0
Cion2_sol_low     = 50.0
**
mu_sol_low        = mu0 - RT/Cw_sol*(Cion1_sol_low + Cion2_sol_low)
omg1_sol_low      = Omg0_1 + RT*log(Cion1_sol_low/Cw_sol)
omg2_sol_low      = Omg0_2 + RT*log(Cion2_sol_low/Cw_sol)
***********************************************************
** TIME PARAMETERS
***********************************************************
t_eql           = 1.0
tRamp_eql       = 0.1
dtInit_eql      = 1.0e-3
dtMax_eql       = 0.25
**
t_swell         = 3600.0*24.0
tRamp_swl       = 180.0
dtInit_swl      = 1.0e-3
dtMax_swl       = 300.0
**
dtMin           = 1.0e-15
\end{lstlisting}

This is one of the most important sections of the input file that needs appropriate modification. The nodal coordinates are as it was exported from Abaqus/CAE. However, we now change the definition of the standard Abaqus element to our user element with all the necessary attributes.

In the first line of the user element definition, we will need to specify an element key (U4), number of nodes, number of coordinates, number of real (floating point) properties, number of integer properties, number of state variables within the element, etc. Additionally, our element has an unsymmetric element tangent matrix, so we specify that as well.

In the second line, we define the degrees of freedom. In our code, the element tangent matrix and element residual vector were assembled in the order of displacement, solvent chemical potential, and ion electrochemical potentials. Since it is an axisymmetric analysis, we use DOF 1 and 2 for displacement, 11 for solvent chemical potential, and 12 and 13 for two different ions. If the model had more ions, we would have added more degrees of freedom for them. The current version of Abaqus/Standard allows up to 30 degrees of freedom. Since we are using the temperature-displacement analysis, Abaqus will perform the convergence check for mechanical degrees of freedom (1--6) and temperature degrees of freedom (11--30). Hence, our additional degrees of freedom (non-mechanical) start from 11. We do not make any changes to the element connectivity and keep the element and node set definition as they are.

\begin{lstlisting} [basicstyle=\ttfamily\small, language=bash, frame=single, backgroundcolor=\color{lightgray!20}, aboveskip=18pt, belowskip=6pt, breaklines=true,  xleftmargin=0.1cm, xrightmargin=0.1cm]
***********************************************************
** NODAL COORDINATES
***********************************************************
*Node
1,           0.,           0.
2, 0.000210268525,         0.
...
...
...
***********************************************************
** USER-DEFINED PE GEL ELEMENT AND ELEMENT CONNECTIVITY
***********************************************************
*User Element,Type=U4,Nodes=<nNode>,Coordinates=<nDim>,Properties=<matProps>,Iproperties=<intProps>,Variables=<nsvars>,Unsymm
1,2,11,12,13
*Element, type=U4
1,   1,   2,  17,  16
2,   2,   3,  18,  17
...
...
...
***********************************************************
** ALL THE NODE SETS AND ELEMENT SETS
***********************************************************
*Nset, nset=all, generate
1,  420,    1
*Elset, elset=all, generate
1,  378,    1
...
...
...
\end{lstlisting}

As described before, we now create dummy elements for visualization. These dummy elements have the same element connectivity as the user-defined elements, but the element numbering has an offset of 10000. This offset numbering will have to match the \texttt{elemOffset} variable in the \texttt{post\_processing} module. The best way to create the dummy element connectivity is to use a MATLAB or Python script to read the as-exported input file and add the offset number to the original element connectivity, and paste them into the input file.

Our next step is to define a single element to make coupled temperature-displacement analysis available through the input file. It does not matter what topology of the elements we choose, as long as it has the suffix \texttt{T} for temperature, we should be fine. Define the nodal coordinates and element connectivity, then create a node set and an element set for this element as well.
\begin{lstlisting} [basicstyle=\ttfamily\small, language=bash, frame=single, backgroundcolor=\color{lightgray!20}, aboveskip=18pt, belowskip=6pt, breaklines=true,  xleftmargin=0.1cm, xrightmargin=0.1cm]
***********************************************************
** DUMMY ELEMENT CONNECTIVITY
***********************************************************
*Element, type=CAX4
100001, 1, 2, 17, 16
100002, 2, 3, 18, 17
...
...
...
*Elset, elset=ElDummy, generate
100001,100378,1
***********************************************************
** EXTRA ELEMENTS
***********************************************************
*Node
999996, 0.0, 0.0
999997, .00001, 0.0
999998, .00001, .00001
999999, 0.0, .00001
*Element, Type=CPE4T
999999,999996,999997,999998,999999
*Nset, nset=extraElement
999996,999997,999998,999999
*Elset,elset=extraElement
999999
\end{lstlisting}

Now we define the properties of the PE gel element. Unlike standard Abaqus elements, we do not need to specify any section for user-defined elements. Follow the list of properties in Table \ref{tab:pe_gel_elem_props}. Make sure not to define more than 8 properties per line, and first define the real (floating point) properties, followed by the integer properties.

Next, we need to define material properties for the dummy elements we created for visualization. Notice that the Young's modulus for the dummy elements is negligible, and we defined the number of user-defined outputs for these elements as well for visualization. Finally, we need to define properties for the extra element we created to make the temperature-displacement analysis available in Abaqus. This element also has very negligible properties which will not affect the analysis.

\begin{lstlisting} [basicstyle=\ttfamily\small, language=bash, frame=single, backgroundcolor=\color{lightgray!20}, aboveskip=18pt, belowskip=6pt, breaklines=true,  xleftmargin=0.1cm, xrightmargin=0.1cm]
***********************************************************
** MATERIAL DEFINITION OF GEL ELEMENTS
***********************************************************
*uel property,elset=all
*uel property,elset=all
** Rgas Fcon theta phi0 Gshear kappa Cp_fix Zp
<Rgas>, ... ... <pKa>,
<Cp_fix>, ... ... , <Cion1_gel>,
<Omg0_1>, ... ... , <Zion_2>,
<Dion_2>, ... ... <nPostVars>

**
***********************************************************
** MATERIAL PROPERTIES OF THE DUMMY ELEMENTS
***********************************************************
*Solid section, elset=ElDummy, material=DummyMaterial 
*Material, name=DummyMaterial
*Elastic
1.e-20
*User output variables
<nPostVars>
***********************************************************
** MATERIAL PROPERTIES OF THE EXTRA ELEMENT
***********************************************************
*Solid section, elset=extraElement, material=extraMaterial
*Material, name=extraMaterial
*Elastic
1.e-20
*Conductivity
1.0
*Density
1.0
*Specific heat
1.0
\end{lstlisting}

Now we define the amplitudes for the solvent chemical potential and electrochemical potential of the ions. We will use these amplitudes to define boundary conditions.
\begin{lstlisting} [basicstyle=\ttfamily\small, language=bash, frame=single, backgroundcolor=\color{lightgray!20}, aboveskip=18pt, belowskip=6pt, breaklines=true,  xleftmargin=0.1cm, xrightmargin=0.1cm]
***********************************************************
** AMPLITUDE DEFINITION
***********************************************************
** as-prepared to low salt solution
*Amplitude, name=mu_prep_to_low, definition = smooth step
0.0, <initMU>, <tRamp_swl>, <mu_sol_low>
*Amplitude, name=omg1_prep_to_low, definition = smooth step
0.0, <initOmg1>, <tRamp_swl>, <omg1_sol_low>
*Amplitude, name=omg2_prep_to_low, definition = smooth step
0.0, <initOmg2>, <tRamp_swl>, <omg2_sol_low>
\end{lstlisting}

Our final step is to define all the initial and boundary conditions necessary to perform Abaqus/Standard analysis. We define initial conditions for the solvent chemical potential and ion electrochemical potentials. We also define an arbitrary initial condition for the extra element we created for the temperature-displacement analysis.

For the boundary conditions, we prescribed those as discussed in the article. To avoid any numerical issues with rigid body motion, we fixed the extra element in the load steps and prescribed the same temperature as its initial state. To make the user-defined output available, we will need to request the element output (uvarm) explicitly in each step.
\begin{lstlisting} [basicstyle=\ttfamily\small, language=bash, frame=single, backgroundcolor=\color{lightgray!20}, aboveskip=18pt, belowskip=6pt, breaklines=true,  xleftmargin=0.1cm, xrightmargin=0.1cm]
***********************************************************
** INITIAL CONDITIONS
***********************************************************
*Initial conditions, type=temperature
all, <initMU>, <initOmg1>, <initOmg2>
*Initial conditions, type=temperature
extraElement, 0.0
***********************************************************
** STEP DEFINITION
***********************************************************
** this step equilibrates the gel to its as-prepared state
*Step, Name=prep, nlgeom=yes, inc=50000, UNSYMM=YES
*Coupled temperature-displacement, stabilize, deltmx=1000.0
<dtInit_eql>, <t_eql>, <dtMin>, <dtMax_eql>
*CONTROLS, PARAMETERS=TIME INCREMENTATION
,,,,,,,30,,,,,,
*CONTROLS, PARAMETERS=LINE SEARCH
10,1.0,0.0001,0.25,0.10
*Boundary
left,1,1,0.0
bottom,2,2,0.0
all, 11, 11, <initMU>
all, 12, 12, <initOmg1>
all, 13, 13, <initOmg2>
extraElement,encastre
extraElement,11,11,0.0
**
*Restart, write, overlay
*Output, field, frequency=5, time marks=no
*node output, nset=all
u, nt, rf   
** specify the user-defined output for the dummy element set
*element output, elset=elDummy
uvarm, le
*End Step
***********************************************************
***********************************************************
*Step, Name=swell, nlgeom=yes, inc=50000, UNSYMM=YES
*Coupled temperature-displacement, stabilize, deltmx=100.0
<dtInit_swl>, <t_swell>, <dtMin>, <dtMax_swl>
*CONTROLS, PARAMETERS=TIME INCREMENTATION
,,,,,,,30,,,,,,
*CONTROLS, PARAMETERS=LINE SEARCH
10,1.0,0.0001,0.25,0.10
**
*Boundary, op=new
left,1,1,0.0
bottom,2,2,0.0
** direct amplitude BC
*Boundary, amplitude=mu_prep_to_low, op=new
top, 11, 11, 1.0
right, 11, 11, 1.0
*Boundary, amplitude=omg1_prep_to_low, op=new
top, 12, 12, 1.0
right, 12, 12, 1.0
*Boundary, amplitude=omg2_prep_to_low, op=new
top, 13, 13, 1.0
right, 13, 13, 1.0
**
*Boundary, op=new
extraElement,encastre
extraElement,11,11,0.0
**
*Restart, write, overlay
*Output, field, frequency = 5, time marks=no
*node output, nset=all
u, nt, rf   
** specify the user-defined output for the dummy element set
*element output, elset=elDummy
uvarm, le
*End Step
***********************************************************
\end{lstlisting}

\subsubsection{Running Abaqus/Standard job}

The Abaqus user element subroutine can not be executed from Abaqus/CAE; it has to be executed from the terminal. We can use the Abaqus command line terminal, Windows cmd, or Windows PowerShell to execute the following command to run an Abaqus/Standard analysis.

\begin{lstlisting} [language=bash, basicstyle=\ttfamily\small, frame=single, backgroundcolor=\color{lightgray!20}, aboveskip=12pt, belowskip=0pt, breaklines=true,  xleftmargin=0.1cm, xrightmargin=0.1cm]
abaqus interactive double analysis job=<job_name> inp=<input_file_name> user=<source_code>
\end{lstlisting}

Make sure to change the variables (job name, input file name, user subroutine name, etc.) inside the angle brackets \texttt{< >} to run the Abaqus job. The Abaqus user manual contains additional arguments that can be used while invoking Abaqus from the command-line terminal.

\end{document}